\documentclass[useAMS]{mn2e}
\usepackage{amssymb,amsmath,graphicx,times} 
\usepackage[flushleft]{threeparttable}
\title[The ALHAMBRA Survey: Bayesian Photometric Redshifts.]{The ALHAMBRA Survey: Bayesian Photometric Redshifts \\ 
with 23 bands for 3 squared degrees.}
\author[Molino et al.]{A. Molino$^1$, N. Ben\'itez$^1$,  M. Moles$^2$, A. Fern\'andez-Soto$^{3,4}$, D. Crist\'obal-Hornillos$^2$,   
\newauthor{B. Ascaso$^1$, Y. Jim\'enez-Teja$^1$, W. Schoenell$^1$, P. Arnalte-Mur$^5$, M. Povi\'c$^1$, D. Coe$^6$,}
\newauthor{C. L\'opez-Sanjuan$^2$, L. A. D\'iaz-Garc\'ia$^2$, J. Varela$^2$, M. Stefanon$^7$, J. Cenarro$^2$,}
\newauthor{I. Matute$^1$, J. Masegosa$^1$, I. M\'arquez$^1$, J. Perea$^1$, A. Del Olmo$^1$, C. Husillos$^1$,} 
\newauthor{E. Alfaro$^1$, T. Aparicio-Villegas$^{1,8}$, M. Cervi\~no$^{1,9}$, M. Huertas-Company$^{10,11}$,}
\newauthor{J. A. L. Aguerri$^{9}$, T. Broadhurst$^{12}$, J. Cabrera-Ca\~no$^{13}$, J. Cepa$^{9,14}$, R. M. Gonz\'alez$^1$,}
\newauthor{L. Infante$^{15}$, V. J. Mart\'\i nez$^{4,16,17}$,  F. Prada$^1$, J. M. Quintana$^1$}
\\
$^1$IAA-CSIC, Glorieta de la astronom\'ia S/N. 18008, Granada, Spain \\
$^2$Centro de Estudios de F\'isica del Cosmos de Arag\'on (CEFCA), Plaza San Juan 1, 44001 Teruel, Spain \\
$^3$Instituto de F\'isica de Cantabria (CSIC-UC), E--39005, Santander, Spain \\
$^4$Unidad Asociada Observatori Astron\'omic (IFCA - Universitat de Val\'encia), Valencia, Spain \\
$^5$Institute for Computational Cosmology, Department of Physics, Durham University, South Road, Durham DH1 3LE, UK \\ 
$^6$Space Telescope Science Institute, Baltimore, MD, USA \\
$^7$Physics and Astronomy Department, University of Missouri, Columbia, MO 65211, USA \\
$^8$Observat\'orio Nacional-MCT, Rua Jos\'e Cristino, 77. CEP 20921-400, Rio de Janeiro-RJ, Brazil \\
$^9$Instituto de Astrof\'isica de Canarias, V\'ia L\'actea s/n, La Laguna, Tenerife 38200, Spain \\
$^{10}$GEPI, Paris Observatory, 77 av. Denfert Rochereau, 75014 Paris, France \\
$^{11}$University Denis Diderot, 4 Rue Thomas Mann, 75205 Paris, France \\
$^{12}$Department of Theoretical Physics, University of the Basque Country UPV/EHU, Bilbao, Spain \\
$^{13}$Departamento de F\'i sica At\'omica, Molecular y Nuclear, Facultad de F\'i sica, Universidad de Sevilla, Spain \\
$^{14}$Departamento de Astrof\'isica, Facultad de F\'isica, Universidad de la Laguna, Spain \\
$^{15}$Departamento de Astronom\'ia, Pontificia Universidad Cat\'olica. Santiago, Chile  \\
$^{16}$Departament d'Astronom\'ia i Astrof\'isica, Universitat de Valencia, Spain  \\
$^{17}$Observatori Astronomic de la Universitat de Valencia, Spain}

\begin{document}
\maketitle
\label{firstpage}

\begin{abstract}
 The ALHAMBRA (Advance Large Homogeneous Area Medium Band Redshift Astronomical) survey has observed 8 different regions of the sky, including sections of the COSMOS, DEEP2, ELAIS, GOODS-N, SDSS and Groth fields using a new photometric system with 20 optical, contiguous $\sim$300$\AA$ filters plus the $JHKs$ bands. The filter system is designed to optimize the effective photometric redshift depth of the survey, while having enough wavelength resolution for the identification of faint emission lines. The observations, carried out with the Calar Alto $3.5$m telescope using the wide field optical camera LAICA and the NIR instrument Omega-2000, represent a total of $\sim$700hrs of on-target science images. Here we present multicolor PSF-corrected photometry and photometric redshifts for $\sim$438,000 galaxies, detected in synthetic $F814W$ images. The catalogs are complete down to a magnitude I$\sim$24.5AB and cover an effective area of $2.79$ deg$^{2}$. Photometric zeropoints were calibrated using stellar transformation equations and refined internally, using a new technique based on the highly robust photometric redshifts measured for emission line galaxies. We calculate Bayesian photometric redshifts with the BPZ2.0 code, obtaining a precision of $\delta_{z}$/(1+$z_{s}$)=1$\%$ for I$<$22.5 and $\delta_{z}$/(1+$z_{s}$)=1.4$\%$ for 22.5$<$I$<$24.5. The global $n(z)$ distribution shows a mean redshift $<$z$>$=0.56 for I$<$22.5 AB and $<$z$>$=0.86 for I$<$24.5 AB. Given its depth and small cosmic variance, ALHAMBRA is a unique dataset for galaxy evolution studies. 
\end{abstract}


\section{INTRODUCTION}
\label{intro}

\begin{figure}
\begin{center}
\includegraphics[width=8.5cm]{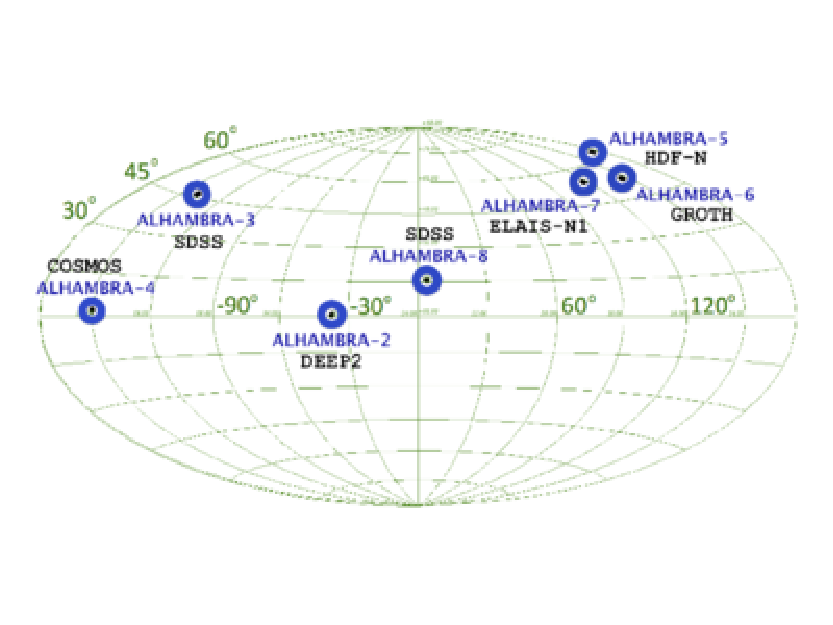}
\caption{The figure shows the different fields observed by the ALHAMBRA survey along with their correspondence with other existing surveys. The mean galactic coordinates are specified in Table \ref{campos}.}
\label{sphere}
\end{center}
\end{figure}

  Photometric redshifts (Baum 1962, Lanzetta, Fern\'andez-Soto \& Yahil 1997, Ben\'itez 2000) have become a powerful tool for cosmology and galaxy evolution studies. Several medium-band photometric surveys have been carried out in the last years:  the UBC-NASA survey (Hickson \& Mulrooney 1998), CADIS (Wolf et al. 2001b), COMBO-17 (Wolf et al. 2001a) and most recently, COSMOS-21 (Taniguchi 2004, Ilbert et al. 2009), NEWFIRM (van Dokkum et al. 2009) or SHARDs (P\'erez-Gonz\'alez et al. 2013). The ALHAMBRA (Advance Large Homogeneous Area Medium Band Redshift Astronomical) survey (Moles et al. 2008) has been optimized to detect and measure precise and reliable photometric redshifts for a large population of galaxies over 8 different fields. Broadband photometric surveys can be significantly shallower, in terms of photometric redshift depth, than well designed, medium band imaging (see Wolf et al. (2001a) and Ben\'itez et al. (2009b) for a systematic study). ALHAMBRA uses a especially designed filter system (see also Aparicio-Villegas et al. 2010) which covers the whole optical range (3500$\AA$ to 9700$\AA$) with 20 contiguous, equal-width, non overlapping, medium-band filters along with the standard JHKs near-infrared bands. The initial goal of the project was covering a total area of 4 deg$^{2}$ on the sky divided into 8 non-contiguous regions (Fig. \ref{sphere}). 

\vspace{0.2cm}

 The photometric system has been specifically designed to optimize photometric redshift depth and accuracy (Ben\'itez et al. 2009b), while having enough sensitivity for the detection and identification of faint emission lines (Bongiovanni et al. 2010, Matute et al. 2012, Matute et al. 2013). The observations presented here were carried out with the Calar Alto 3.5m telescope using both the wide field camera LAICA in the optical range and the Omega-2000 camera in the Near Infrared (NIR) from 2005 to 2012. In order to define a constant and homogeneous window for all the ALHAMBRA fields, we generated synthetic F814W detection images (corresponding to HST/ACS F814W). These images are photometrically complete down to a magnitude $m_{F814W}\le24.5$, have a much better $S/N$ than individual filters, and allow direct comparisons with other space-based surveys as COSMOS. Hereafter, $m_{F814W}$ magnitudes correspond to the magnitudes derived on the synthetic F814W images.

\vspace{0.2cm}

 In this paper we summarize the ALHAMBRA survey observations and the data reduction in Section \ref{general}. We describe in detail the photometric catalogues used to derive photometric redshifts in Section \ref{photom}. This includes the aperture-matched PSF-corrected photometry, the generation of synthetic F814W detection images and their corresponding star masking treatment, a statistical star/galaxy classification and an empirical estimation of the photometric uncertainties. We discuss several photometric checks in Section \ref{phintchecks}. In Section \ref{bpz} we analyze the methodology used to derive the photometric redshifts, the different methods used to compute photometric zeropoint calibrations and the photometric redshift accuracy quantification as a function of F814W magnitude, redshift and $Odds$. Finally we present the description of the ALHAMBRA photometric redshift catalogues in Section \ref{catalogs}.       
 
\vspace{0.3cm} 
 
All optical and near-IR magnitudes in this paper are on the AB system. Cosmological parameters of $H_{0}$ = 70 km$s^{-1}$ Mpc$^{-1}$, $\Omega_{M}$ = 0.3, $\Omega_{\Lambda}$ = 0.7 are assumed throughout.

\section{OBSERVATIONS AND DATA REDUCTION}
\label{general}

\subsection{Observations}

 The ALHAMBRA survey has imaged a total area of 4.0 deg$^{2}$ in eight separated regions of the sky during a seven-year period (2005-2012). Observations were carried out with the 3.5m telescope at the Calar Alto Observatory (CAHA, Spain) making use of the two wide-field imagers in the optical (LAICA) and in the NIR (Omega-2000). The ALHAMBRA fields have been observed whenever the conditions were good (seeing$<$1.6", airmass$<$1.8), for a total on-target exposure time of $\sim$700hrs for the whole survey, corresponding to $\sim 32$ hrs for each field. The integration time was split into $\sim$27.8 hrs for medium-band filters and $\sim$4.2 hrs for broad-band Near Infrared (NIR) filters, as explained in Cristobal-Hornillos et al. 2013, (in prep.). 

 Although ALHAMBRA-01 has already been observed, its analysis has not been included in this paper due to issues with its primary photometric calibration at the time. For a detailed description of the NIR observations, we refer the reader to Crist\'obal-Hornillos et al. (2009). The description of the optical observations will be available in Cristobal-Hornillos et al. 2013 (in prep.).

\begin{figure*}
\includegraphics[width=17.cm]{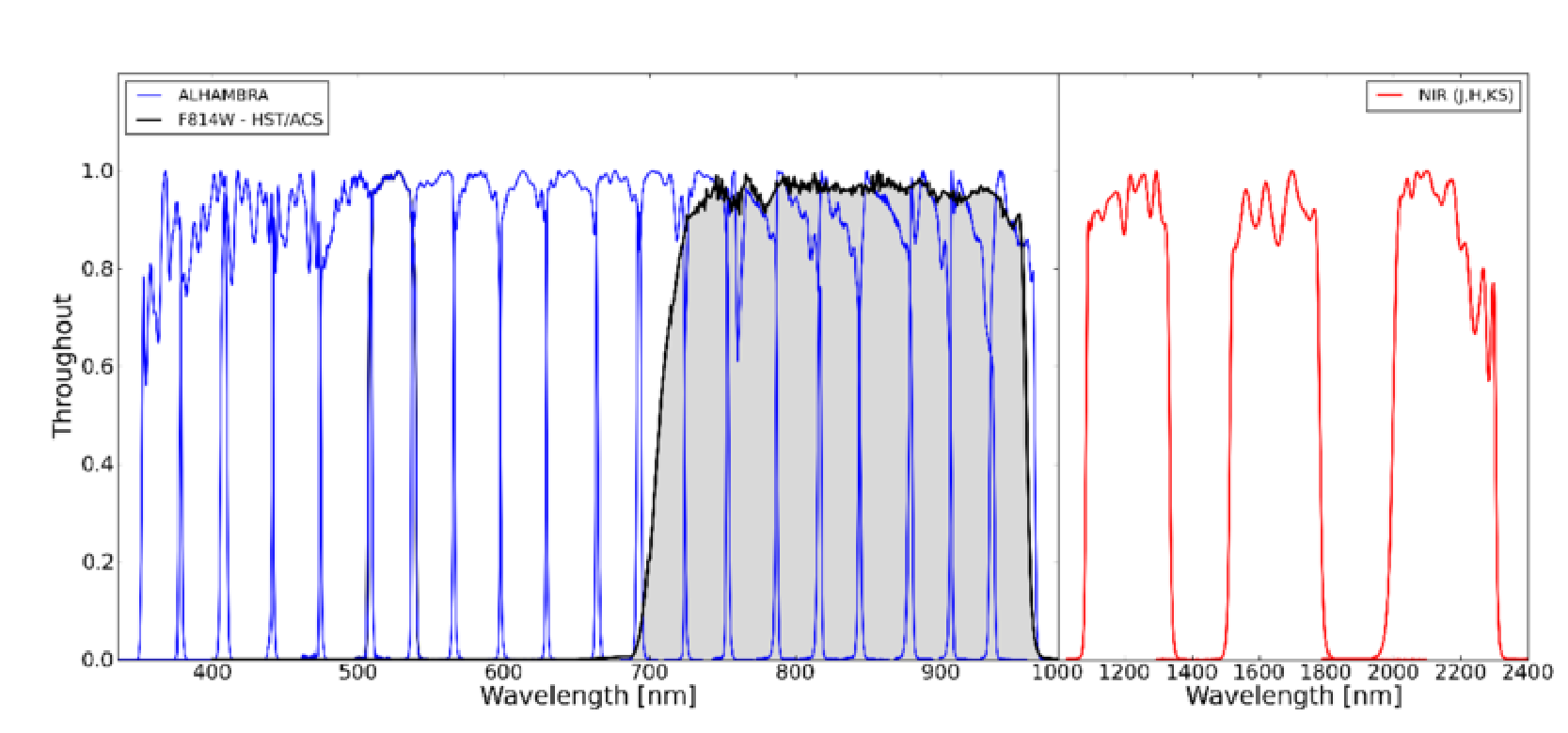}
\caption{The ALHAMBRA survey filter set. On the left-hand side, solid blue lines represent the optical filter system composed by 20 contiguous, equal-width, non overlapping, medium-band ($\sim$300$\AA$) filters. The solid black line corresponds to the synthetic F814W filter used to define a constant observational window across fields. On the right-hand side, solid red lines represent the standard $JHKs$ near-infrared broad bands. All transmission curves are normalized to its maximum value.}
\label{ALHAMBRA_optfilt}
\end{figure*}

\begin{figure*}
\begin{center}
\includegraphics[width=17.6cm]{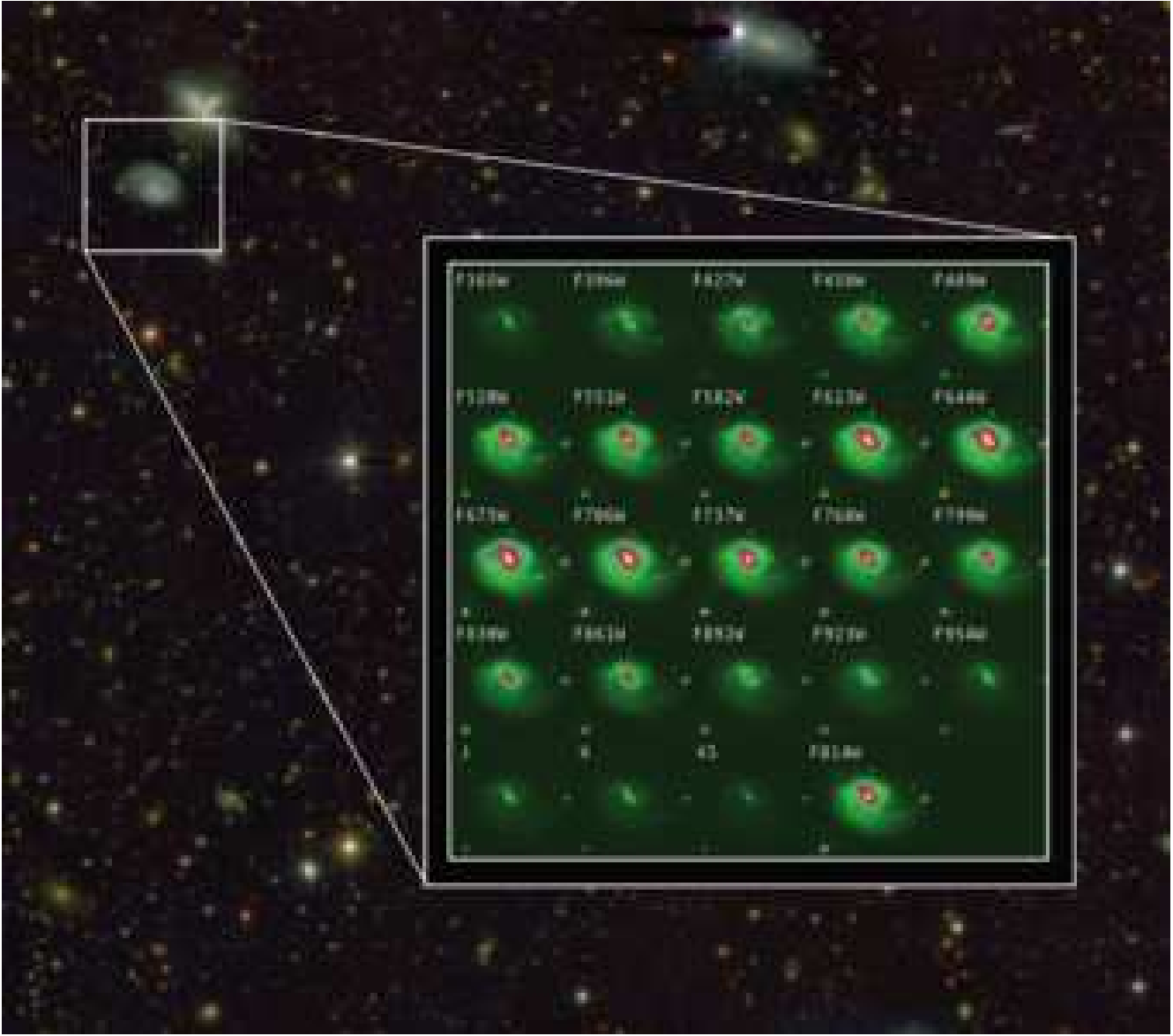}
\caption{The ALHAMBRA survey. The figures show how a galaxy looks like when observed through the ALHAMBRA filter system. The optical range is covered horizontally from top to bottom and left to right, the last row corresponds to the $J$, $H$ and $Ks$ NIR filters along with the synthetic F814W detection image. The background color image was generated using the $Trilogy$ software (www-int.stsci.edu/$^{\sim}$dcoe/trilogy/Intro.html).}
\label{colorimageexample}
\end{center}
\end{figure*} 

\begin{table*}
\caption{{\small The ALHAMBRA survey selected fields}}
\begin{center}
\label{campos}
\begin{tabular}{|l|c|c|c|c|c|c|c|clclclcl}
\hline
Field   &   Overlapping    &  RA        &   DEC      &    Area / Effective & Number  & Science & Detected($^{1}$)  &   Density ($^{1,2}$) \\
Name &    Survey           & (J2000)  &  (J2000)  &         [deg$^{2}$]          &  catalogs    &  Images &   Sources  & [\#/deg$^{2}$]  \\
\hline
\hline
ALHAMBRA-1 &     ---          & 00 29 46.0 & +05 25 30 & 0.50 / ---     & --- & 192 & --- & --- \\
\hline
ALHAMBRA-2 & DEEP2      & 01 30 16.0 & +04 15 40 & 0.50 / 0.45  & 8 & 192 & 67.791 & 77.144 \\
\hline
ALHAMBRA-3 & SDSS         & 09 16 20.0 & +46 02 20 & 0.50 / 0.47 & 8 & 192 & 68.015 & 75.000 \\
\hline
ALHAMBRA-4 & COSMOS   & 10 00 00.0 & +02 05 11 & 0.25 / 0.23 & 4 & 96   & 38.464 & 93.261 \\
\hline
ALHAMBRA-5 & HDF-N     & 12 35 00.0 & +61 57 00   & 0.25 / 0.24 & 4 & 96    & 42.618 & 82.300 \\
\hline
ALHAMBRA-6 & GROTH    & 14 16 38.0 & +52 24 50 & 0.50 / 0.47  & 8 & 192  & 66.906 & 77.740 \\
\hline
ALHAMBRA-7 & ELAIS-N1 & 16 12 10.0 & +54 30 15 & 0.50 / 0.47 & 8 &  192 & 79.453 & 82.185 \\
\hline
ALHAMBRA-8 & SDSS        & 23 45 50.0 & +15 35 05 & 0.50 / 0.46 & 8 & 192 & 75.109 & 82.452 \\
\hline
&                    &                   &                  & 3.00 / 2.79 & 48 & 1344 &  438.356 &  $<$81.440$>$ \\
\hline
\hline
$^{1}$ w/o duplications &                    &                   & &&&&& \\
$^{2} m_{F814W}<$ 24. &                    &                   & &&&&& \\
\end{tabular}
\end{center}
\end{table*}

\subsection{Data Reduction}
\label{dataredu}

 In order to homogenize the data sets from both imagers, NIR images from the OMEGA-2000 detector were converted from their original pixel size, 0.45 "/pix, to 0.221 "/pix to match the pixel size of the LAICA images. As explained in Crist\'obal-Hornillos et al. (2009), individual images from each run have been dark current corrected, flat fielded and sky subtracted. Bad pixels, cosmic rays, linear patterns and ghost images have also been masked out. Processed images have been finally combined using SWARP (Bertin et al. 2002) software and the applied geometrical transformations have been incorporated in WCS headers.

\vspace{0.2cm}

 The total 2.8 deg$^{2}$ consideted in this work are divided into 7 non-contiguous regions of the sky (as summarized in Table \ref{campos}), each of which split into 2 non-overlapping strips composed by 4 individual CCDs, as schematically illustrated in Appendix \ref{LAICAscheme}. Each one of the 48 CCDs represents the minimum area (15.5'$\times$15.5') covered by all the 23 individual filters. To quantify the survey effective area (Section \ref{Flagimages}), FLAG images have been created where pixels not satisfying established photometry quality criteria have been flagged. Meanwhile both RMS-map and WEIGHT-maps have been generated accounting for the level of photometric uncertainties present across individual images.

\subsection{Filter set}
\label{filters}

 Once the instrumental setup and exposure time are fixed, the filter set has a powerful effect on the photo-z performance (see Wolf et al.(2001a) and Ben\'itez et al. (2009b)). Table \ref{surveyaccuracy} summarizes a small list of different photometric filter systems and their photometric redshift accuracy. The ALHAMBRA survey designed its own photometric system (Ben\'itez et al. 2009b) optimizing both photometric depth and accuracy. As seen in Fig. \ref{ALHAMBRA_optfilt}, the system covers the 3500-9700$\AA$ optical window with 20 constant-width ($\sim$300$\AA$), non overlapping filters. We also use the $J$, $H$ and $Ks$ NIR bands. Including  both Optical + NIR observations helps to break color-redshift degeneracies, reducing the fraction of catastrophic outliers and increasing the ALHAMBRA photometric redshift depth. In Fig. \ref{colorimageexample} we show a galaxy observed through the whole ALHAMBRA filter system. The main properties of each individual filter are summarized in Table \ref{tablefilter}.

\begin{table}
\caption{{\small Photometric Redshift Surveys. Since for narrow/medium-band photometric surveys (*) the photometric redshift accuracy is strongly dependent on the signal-to-noise, we compared the performance from both surveys applying a similar cut in magnitude (R$<$23AB for COMBO-17 and I$<$23AB for ALHAMBRA). For the brightest sources, both surveys reach a performance similar to COSMOS's or MUSYC's ($\delta_{z}$/(1+z) $<$ 0.01).}}
\begin{center}
\label{surveyaccuracy}
\begin{tabular}{|l|c|c|c|}
\hline
\hline
survey  &  Reference & Bands  & $\delta_{z}$/(1+z) \\
\hline
HDF & Sawicki (1997) & 4 & 0.080 \\
SDSS/DR6 & Csabai (2003) & 5 & 0.035 \\
SWIRE & Rowan-Robinson (2008) & 5 & 0.035 \\
HUDF  &  Coe (2006)   &  6 & 0.040 \\
HDF  & Fern{\'a}ndez-Soto (1999) & 7 & 0.060 \\
CFHTLS & Ilbert (2006) & 9 & 0.030 \\
GOODS & Dahlen (2010) & 12 & 0.040 \\
CLASH & Molino (2014, prep.) & 16 & 0.025 \\
COMBO-17(*) & Wolf (2008) & 17 & \textbf{0.020} \\
ALHAMBRA(*) & Molino (this work) & 23 & \textbf{0.010} \\
COSMOS & Ilbert (2009) & 30 & 0.007 \\
MUSYC & Cardamone (2010) & 32 & 0.007 \\
JPAS & Ben\'itez (2009a, 2014) & 59 & 0.003 \\
\hline
\hline
\end{tabular}
\end{center}
\end{table} 

\vspace{0.2cm}
\begin{table}
\caption{{\small Summary of the multiwavelength filter set for ALHAMBRA. The FWHM, the exposure time and the limiting magnitude (measured on 3" diameter aperture) correspond to the average value among the 48 CCDs.}}
\begin{center}
\label{tablefilter}
\begin{tabular}{|l|c|c|c|c|c|c|c|c|c|c|c|c|c|c|c|c|}
\hline
\hline
CAMERA   &   FILTER    & $\lambda_{eff} $ & FWHM  & $\langle$t$_{exp}$$\rangle$ & $\langle$m$_{lim}^{(3")}$$\rangle$ \\
                  &                     &   $[\AA]$           & $[\AA]$ &    [sec]      &  (5-$\sigma$)    \\
\hline
Optical      &                  &            &          &            &               \\
\hline  
LAICA       &  F365W    &  365  &   279  &  3918  &   23.7 \\
LAICA       &  F396W    &  396  &   330  &  2896  &   23.8 \\
LAICA       &  F427W    &  427  &   342  &  2774  &   23.8  \\
LAICA       &  F458W    &  458  &   332  &  3079  &   23.8  \\
LAICA       &  F489W    &  489  &   356  &  2904  &   24.2 \\ 
LAICA       &  F520W    &  520  &   326  &  2664  &   24.1  \\
LAICA       &  F551W    &  551  &   297  &  2687  &   23.7  \\
LAICA       &  F582W    &  582  &   324  &  2936  &   23.8   \\
LAICA       &  F613W    &  613  &   320  &  2940  &   23.9  \\
LAICA       &  F644W    &  644  &   357  &  4043  &   23.8  \\
LAICA       &  F675W    &  675  &   314  &  4575  &   23.5  \\
LAICA       &  F706W    &  706  &   332  &  5668  &   23.7   \\
LAICA       &  F737W    &  737  &   304  &  7095  &   23.5  \\
LAICA       &  F768W    &  768  &   354  &  8824  &   23.5  \\
LAICA       &  F799W    &  799  &   312  &  8992  &   23.2   \\
LAICA       &  F830W    &  830  &   296  &  11436 &  23.2  \\
LAICA       &  F861W    &  861  &   369  &  10505 &  22.9 \\
LAICA       &  F892W    &  892  &   303  &  9044  &  22.5 \\
LAICA       &  F923W    &  923  &   308  &  6338 &   22.1  \\
LAICA       &  F954W    &  954  &   319  &  5620 &   21.5  \\
\hline
NIR  &            &            &          &            &               \\
\hline
OMEGA  &  $J$     &  1216  &  2163  &  5169 &   22.6  \\
OMEGA  &  $H$      &  1655  &  2191  &  5055 &   21.9  \\
OMEGA  &  $Ks$   &  2146  &  2412  &  5050 &   21.4  \\
\hline
Detection      &            &                &             &            &               \\
\hline
SYNTH     &  F814W    &  845  &  2366  &  73522 &  24.5 \\
\hline
\hline
\end{tabular}
\end{center}
\end{table}

\subsection{Primary photometric zeropoint calibration.}

 A set of transformation equations between the ALHAMBRA and the Sloan Digital Sky Survey (SDSS, York et al. 2000) was derived based on a collection of primary standard stars from the Next Generation Spectral Library (HST/STIS NGSL, Gregg et al. 2004), as explained in Aparicio-Villegas et al. (2010).   

\vspace{0.2cm} 

  We applied these equations to all stars with good photometry in both ALHAMBRA and the SDSS/DR7 data, and derived photometric zeropoints from the mean difference between instrumental and synthetic magnitudes, yielding an internal error no larger than a few hundredths of a magnitude for stars in each CCD and filter combination. For an in-depth discussion of the calibration of the ALHAMBRA optical photometric system we refer the reader to Crist\'obal-Hornillos et al. 2013 (in prep.). 

 The ALHAMBRA survey has used the 2MASS catalogue (Cutri et al. 2003) to calibrate its NIR images. As explained in Crist\'obal-Hornillos et al. (2009) several dozens of stars common to both datasets with high $S/N$ were selected in each field, yielding photometric zeropoint offsets with uncertainties of $\sim$0.03 mag. We will discuss the procedure to refine the photometric zeropoints calibrations via SED-fitting techniques in Section \ref{PZPR}. 


\section{Photometry}
\label{photom}
\subsection{Multi-wavelength Photometry}
\label{MWP}

\begin{figure}
\begin{center}
\includegraphics[width=8.5cm]{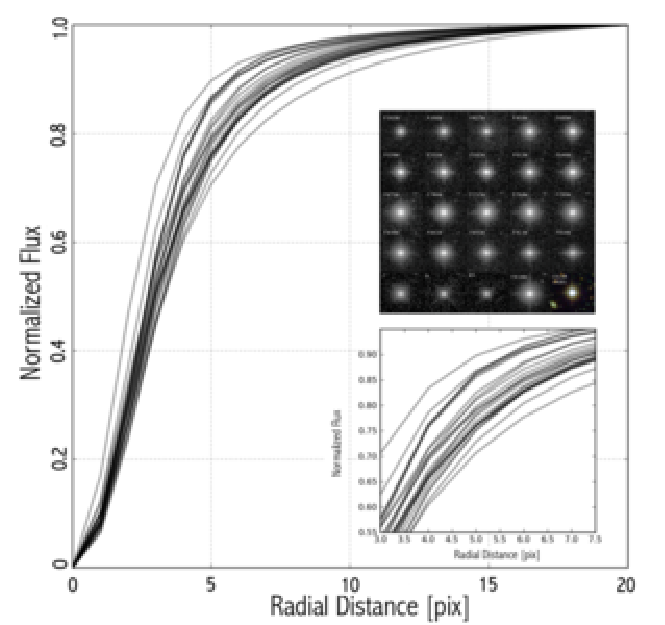}
\caption{Seeing variability across photometric bands. For a single star, solid black lines represent the scatter in the normalized stellar growth curve as a consequence of the varying  PSF across filters (inset top panel). This effect has to be corrected in order to estimate accurate colors. }
\label{stellarPSFefect}
\end{center}
\end{figure}

\begin{figure}
\includegraphics[width=9.0cm]{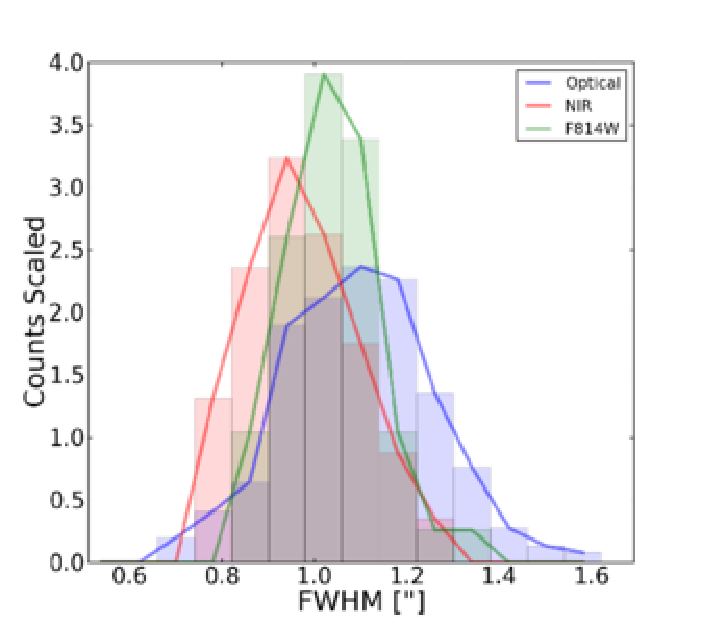}
\caption{Distribution of $seeing$ conditions for the ALHAMBRA fields. The figure shows the distribution of PSFs (measured as the FWHM in arcsec) for optical images (blue), NIR (red) and synthetic F814W detection images (green). The PSFs range from 0.7" to 1.6", with the optical images having  $<$FWHM$>$$\sim$1.1", the NIR images $<$FWHM$>$$\sim$0.9" and the synthetic F814W detection images $<$FWHM$>$$\sim$1.0".}
\label{PSF_var_histo}
\end{figure}

 As it was thoroughly described in Coe et al. (2005), measuring multicolor photometry in images with different PSFs is not a trivial task. To perform good quality multi-color photometry, it is necessary to sample the same physical region of the galaxy taking into account the smearing produced by different PSFs as seen in Fig. \ref{stellarPSFefect}. We show the distribution of optical, NIR \& synthetic F814W PSFs in the ALHAMBRA survey in Fig. \ref{PSF_var_histo}, for values between 0.7" and 1.6".   

\vspace{0.2cm}

 One of the methods to correct for PSF effects is to smooth the whole dataset to the worst seeing condition, making all images look as they had been taken under the same seeing conditions (Loh \& Spillar 1986, Labb\'e et al. 2003, Capak et al. 2007). This methodology produces consistent apertures across the filters, but strongly degrades the best observations  to the level of the worst. Here we have used \textit{ColorPro} (Coe et al. 2006) which accurately corrects for PSF effects without degrading image quality; Laidler et al. 2007; De Santis et al. 2007; Kuijken 2008; Wolf et al. 2008 have also developed similar approaches.  

\vspace{0.2cm}

 To improve the photometric depth and homogeneity, we relied on deep synthetic F814W images (section $\ref{f814wimages}$) which are the best option for photometric aperture definitions (given its enhanced $S/N$) and for galaxy morphology estimations.  

\subsection{PSF-Matched Aperture-Corrected photometry.}
\label{ColorPro}

\textit{ColorPro} derives accurate PSF-corrected photometry without degrading high quality images. Initially the software defines every photometric aperture based on the selected detection image. Then it goes filter by filter and estimates how much flux a galaxy has missed within that aperture as a consequence of the difference between the PSF of the image in that filter and the detection PSF. This correction is applied to each filter, yielding PSF-corrected magnitudes with as little PSF degradation as possible.  

 SExtractor (Bertin \& Arnouts 1996) ISOphotal apertures produce the most robust colors for faint objects (Ben\'itez et al. 2004) while SExtractor AUTO apertures provide better estimations of galaxy total magnitudes. To encompass the usefulness of both measurements, \textit{ColorPro} defines a photometric transformation which provides both SExtractor ISOphotal colors and total magnitudes. 

\vspace{0.2cm}

Total magnitudes are defined as:

\begin{equation}
M_{i} = M^{ISO}_{i} + (M_{det}^{AUTO} - M_{det,i}^{ISO})
\end{equation} while the first term corresponds to the standard SExtractor ISOphotal magnitude for sources detected on the i$^{th}$-band, the second term incorporates the PSF-correction (by applying the photometric differences when degrading the detection image ($M_{det}$) to the i$^{th}$-PSFs condition ($M_{det,i}$)). Hence, the second term  extends SExtractor ISOphotal magnitudes into total magnitudes.

\vspace{0.2cm}

Meanwhile ISOphotal colors are derived as:

\vspace{0.1cm}

\begin{equation}
M_{j}   = M^{ISO}_{j}   + (M_{det}^{AUTO} - M_{det,j}^{ISO})
\end{equation}
\begin{equation}
{M_{i} - M_{j} = M^{ISO}_{i} - M^{ISO}_{j} + (M_{det,j}^{ISO} - M_{det,i}^{ISO})}
\end{equation}

\vspace{0.2cm}

where resulting $M_{i}-M_{j}$ colors are just the combination of their SExtractor ISOphotal magnitudes plus a second term including their relative PSF-corrections. As expected, in those cases with equal PSF the second term might be cancelled out providing colors directly from the SExtractor ISOphotal magnitudes. For a more detailed explanation, we refer the reader to Coe et al. (2006). 

\subsection{PSF models}
\label{psfmodels}

  As required by \textit{ColorPro}, it was necessary to generate PSF models for each individual image. We used the package $DAOPHOT$ from IRAF (Stetson 1987) which uses an hybrid method to compute PSF models. It first fits the central region of the stars by using an analytical function (Gaussian, Lorentzian, Moffat or Penny), and then the outermost parts (regions connected with the background) are empirical fitted point by point; typical residuals between stars and models are around $\sim3\%$.    

\vspace{0.2cm}

  We initially ran SExtractor on each image using a very high threshold ($\sim$100$\times\sigma_{Background}$) to detect only bright sources. We kept detections with SExtractor $CLASS\_STAR>$0.9. When plotting the magnitude vs the FWHM for those selected objects, we find that they are located in the region of the brightest and most compact sources (Section \ref{geomag}). To avoid both saturated stars and misclassified galaxies, we culled objects outside the range 16$<$m$<$22.5, yielding a final sample of several hundred of stars per image.

\begin{figure}
\begin{center}
\includegraphics[width=8.5cm]{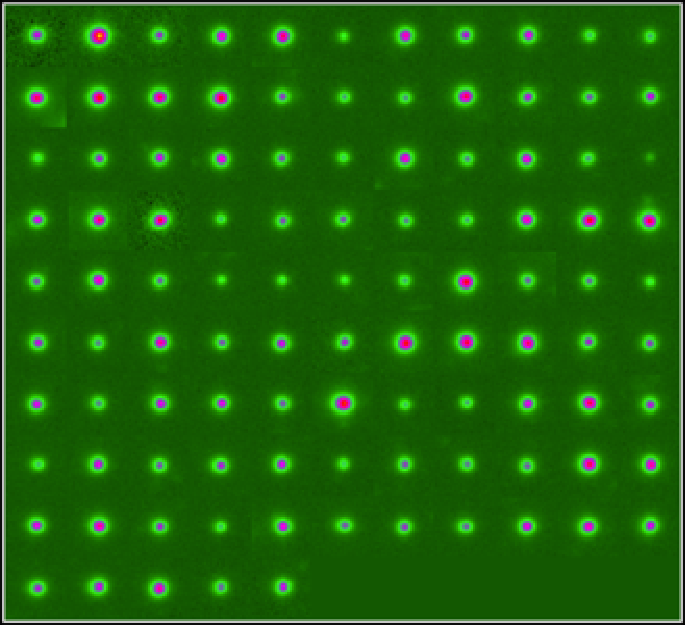}
\caption{Stars selection. Several hundred of non saturated and well-isolated stars were interactively selected across the images, to derive the PSF models. A careful selection is essential to ensure the accuracy of the PSF-corrected magnitudes.}
\label{Starsselection}
\end{center}
\end{figure}

\vspace{0.2cm}
 Afterwards, we visually rejected stars with contaminating neighbors and generated mosaic-like images (Fig. \ref{Starsselection}). These images decreased the computational time required by $DAOPHOT$ to model the PSF and generated much higher quality results. Finally the PSF models were normalized to the same flux. 
\vspace{0.2cm}

 Among the different analytical models considered by $DAOPHOT$, the most recurrent one was the Penny2 profile. This model consists of a Gaussian-like function but with Lorentzian wings. Although typical residuals for PSF models from CCD1, CCD2 and CCD4 are around $3\%$, CCD3 shows a different behavior with systematically larger residuals of $5-10\%$. This different behavior was probably due to the differences in the efficiency of this detector (CCD3), which was not science grade and significantly worse than CCD1, CCD2 and CCD4 (Cristobal-Hornillos et al. 2013, in prep.). 

\subsubsection{PSF Model verification.}
\label{psfverification}

  We systematically verified each PSF model. First we compared its FWHM with the registered seeing (from the image header) and with the mean FWHM value for the stars used to derive the model. The observed scatter among PSFs does not exceed $3-5\%$, ensuring that stars and models are well in agreement.

\vspace{0.2cm}

  The PSF stability among different CCDs was also checked. As introduced in Section \ref{dataredu}, given the spatial configuration of the LAICA optical system, the four CCDs simultaneously imaged similar parts of the sky under the same atmospheric conditions and with the same passbands. This fact made it possible to perform statistical  comparisons among detectors. Once again we observed good agreements among CCD1, CCD2 and CCD4 but a larger deviation for CCD3 close to a $5-10\%$.

\begin{figure}
\includegraphics[width=8.25cm]{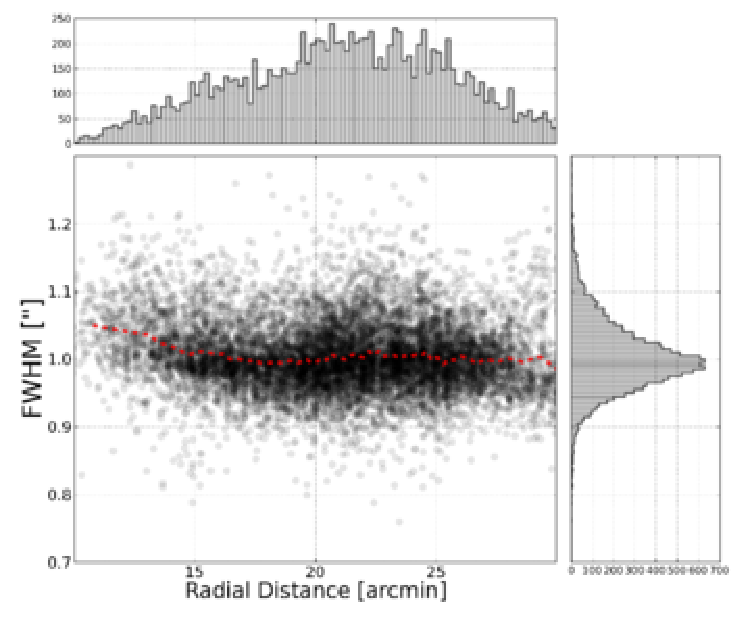}
\caption{Radial PSF variability across images. The figure shows the radial dependence of the PSF (expressed in arcsec and referred to the primary mirror telescope) for the compilation of stars used to derive the PSF models. The mean value of the distribution (dashed red line) has a scatter smaller than $5\%$ enabling the usage of a single PSF model per image.}
\label{radialpsf}
\end{figure}  

\vspace{0.2cm}

 Finally we studied the radial PSF variability across images to ascertain the usage of a single PSF model per image. We defined a new reference system linking every detection (from each CCD) to the center of the telescope's focal plane. In Fig. \ref{radialpsf} we show the dependence of the FWHM as a function of the radial distance for $\sim$20.000 stars, finding a variation smaller than $5\%$. 
 
\subsection{Simulations.}
\label{simulations}

 We carried out several simulations to test the accuracy of \textit{ColorPro}. To do so we degrade a much better resolution image to the typical ALHAMBRA conditions (in terms of PSF and background noise) and run  \textit{ColorPro} on it expecting to retrieve null colors (equal magnitudes) for galaxies observed under different PSFs.
 
\vspace{0.2cm} 
 
\begin{figure}
\includegraphics[width=8.5cm]{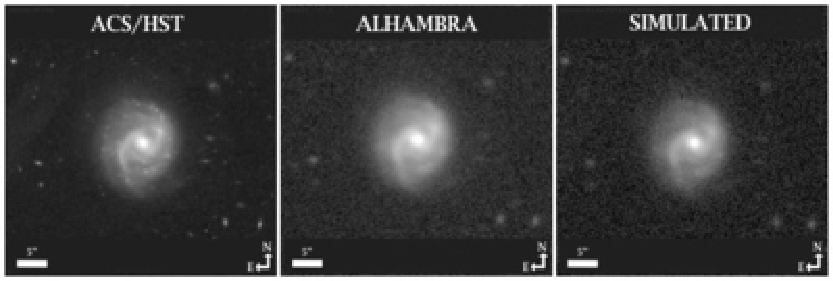}
\caption{Example of the simulated dataset reproducing the typical observational conditions of the ALHAMBRA images (section \ref{simulations}). From left to right, we show a galaxy in the ACS/HST image, in the ALHAMBRA synthetic F814W detection image and in the rescaled + PSF degraded + background reapplied ACS/HST image.}
\label{ALHmosaico}
\end{figure}

 We created a mosaic image by rearranging four HST/ACS F814W images from the COSMOS-survey (Scoville et al. 2007) which overlap the ALHAMBRA fields. We rescaled the mosaic to the ALHAMBRA pixel size (from the ACS 0.065 "/pix to the LAICA 0.221 "/pix), convolved with $\sim$200 PSFs randomly drawn from our models and reapplied background noise using typical values for the ALHAMBRA images (empirically measured as explained in Section \ref{photoerr}). An example of the simulated images is shown in Fig. \ref{ALHmosaico} which compares the ACS/HST image of a galaxy (left panel) with the ALHAMBRA image (middle panel) and the simulated image (right panel). 

\subsubsection{Reliability.}
\label{reliability}

  We ran \textit{ColorPro} on the new set of $\sim$200 mosaics, using the same configuration used for the ALHAMBRA catalogs. We excluded all the detections with photometric problems reported by SExtractor  (\textit{SExtractor\_Flag}$>$1) to eliminate several ghosts and other artifacts (trails) within the original images. 

\vspace{0.2cm} 

\begin{figure}
\includegraphics[width=8.3cm]{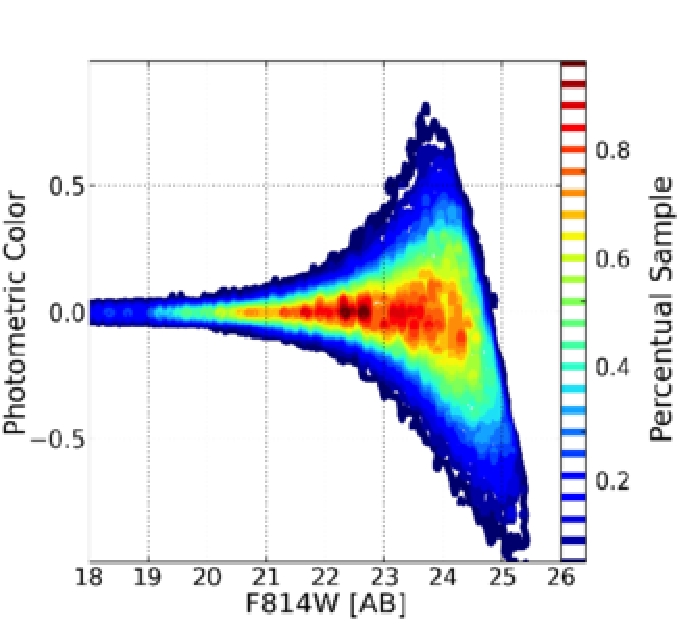}
\includegraphics[width=8.5cm]{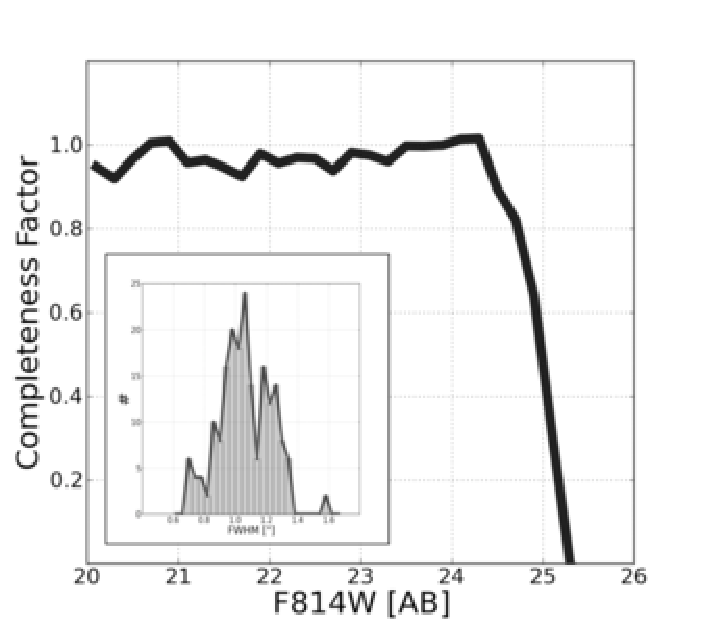}
\caption{PSF-corrected photometry verifications. We designed a set of simulations (Section \ref{simulations}) to estimate both the reliability of ColorPro deriving PSF-corrected photometry and the expected completeness in our images. Top panel shows how ColorPro successfully retrieved null colors (same magnitudes) across simulated images, with a dispersion below $3\%$ for magnitudes brighter than $m_{F814W}=23.0$ and an increasing error, as expected from the added photometric noise, at fainter magnitudes. Bottom panel shows the expected completeness as a function of F814W magnitude.}
\label{simulcompleteness}
\end{figure}

  We found that the simulated colors showed a dispersion of $\sigma$$\sim$0.03, which marks a photometric precision floor, for sources brighter than magnitude $m_{F814W}=23$.0 and, as expected from the uncertainties arising from the photometric noise, an increasing error for fainter magnitudes. For most of the magnitude range, there are negligible biases, as seen in the top panel of Fig. \ref{simulcompleteness}. This shows that \textit{ColorPro} is capable of performing accurate PSF-corrections for ALHAMBRA-like data.

\subsubsection{Completeness.}

  We studied the expected photometric completeness for the ALHAMBRA fields in terms of PSF and background level. For this purpose, we used the previous simulations to derive the statistical probability of detecting a sample of faint galaxies when observed under the typical ALHAMBRA conditions. In the bottom panel of Fig. \ref{simulcompleteness} we show the expected fraction of missed galaxies per magnitude range and square degree. The result indicates that ALHAMBRA is photometrically complete down to a magnitude of $m_{F814W}\sim$24. At fainter magnitudes, the number of detections decreases rapidly, with $\sim40\%$ of the galaxies lost at $m_{F814W}\sim 25$. 

\subsection{Synthetic F814W detection images.}
\label{f814wimages}
 
 In photometric surveys it is a common practice to stack the best quality images in order to generate detection images which are deeper than the images obtained in any individual filter. To define an homogeneous detection image for all ALHAMBRA fields, we generated synthetic F814W images as the properly weighted combination of several individual bands. To calculate the weights we generated, using the empirically calibrated template library of Ben\'\i tez (2014), a realistic galaxy mock catalog up to the typical ALHAMBRA depth, and solved, using least-squares, the system of equations between the corresponding synthetic colors (eq. \ref{transfeq}). 

\vspace{0.2cm} 

The system of equations among filters for $N_g$ galaxies is defined as follows:

\begin{equation}
\label{transfeq}
m_{F814W,i}=\sum_{j=1}^{N_f} a_{i,j} \times m_{i,j}
\end{equation}

\begin{figure*}
\includegraphics[width=8.5cm]{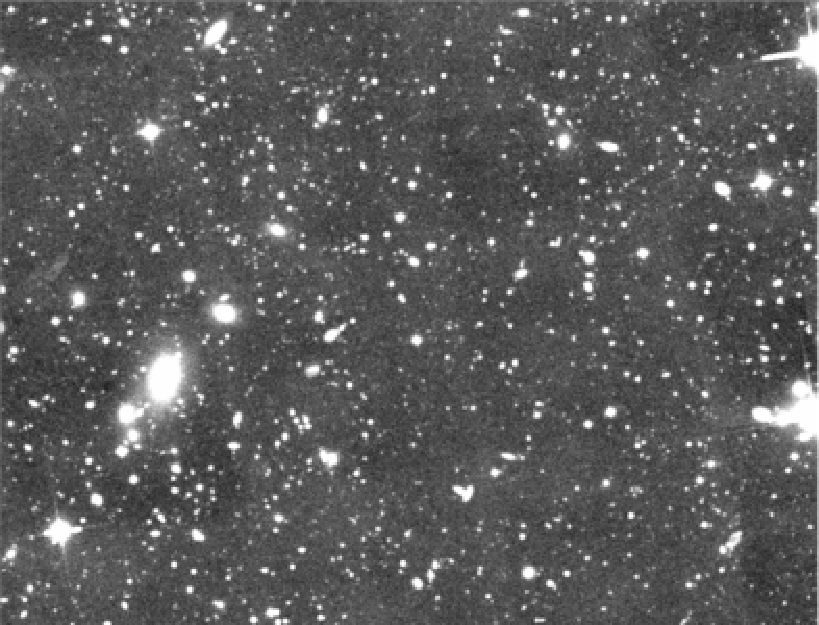}
\includegraphics[width=8.5cm]{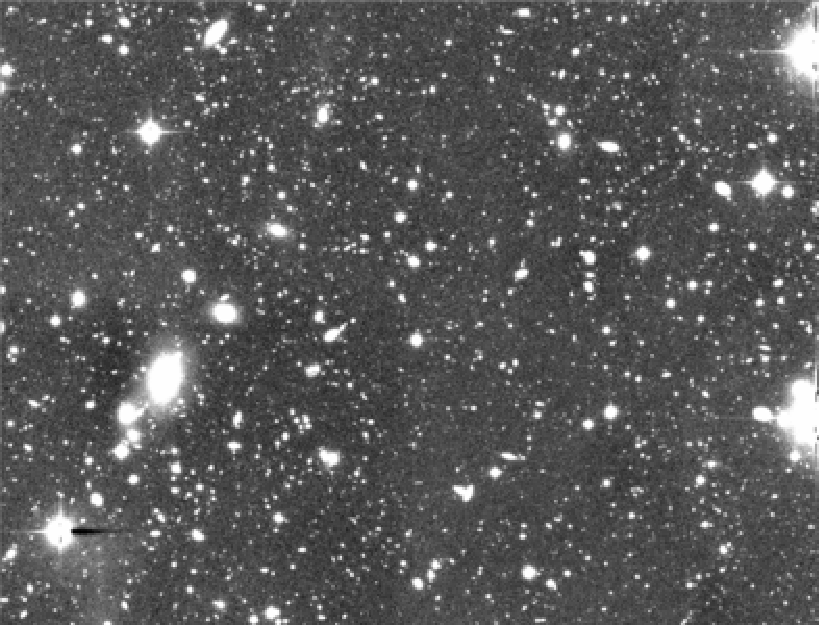}
\caption{Example of the synthetic F814W images derived for the ALHAMBRA fields. Left panel shows how the original HST/ACS F814W image looks like after been scaled to the ALHAMBRA pixel size, convolved with the ALHAMBRA PSF and photometric noise reapplied. Right panel shows the synthetic ALHAMBRA F814W detection image.}
\label{f814w_acs_alh}
\end{figure*}

  An example of the so-derived synthetic F814W images is shown in Fig. \ref{f814w_acs_alh}. The least square solution, with an rms error smaller than $1\%$ is:    

\[
\begin{array}{c}
F814W  =   \\
   0.105\times\textbf{F706W} + 0.178\times\textbf{F737W} + 0.179\times\textbf{F768W} + \\
+ 0.142\times\textbf{F799W} + 0.115\times\textbf{F830W} + 0.119\times\textbf{F861W} +     (5) \\
+ 0.073\times\textbf{F892W} + 0.049\times\textbf{F923W} + 0.040\times\textbf{F954W}  \\ 
\label{f814wtranseq}
\end{array}
\]

\begin{figure*}
\begin{center}
\includegraphics[width=17.5cm]{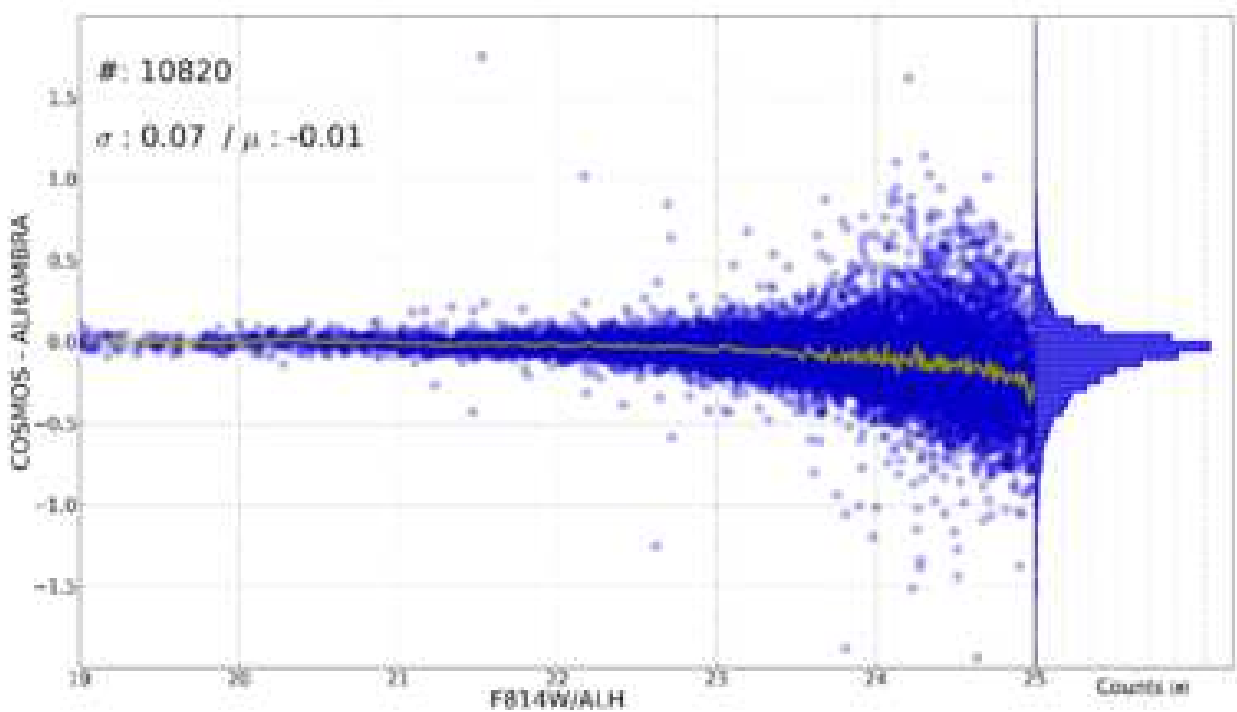}
\caption{Photometric comparison between the F814W/COSMOS and the synthetic ALHAMBRA/F814W images. In order to  reproduce the same photometric measurements as done by Ilbert et al (2009), we ran SExtractor on the synthetic  ALHAMBRA/F814W images using fixed circular apertures of 3". Selecting a common sample of $\sim$10800 detections  between ALHAMBRA and COSMOS, we did not find neither photometric zeropoint offsets nor significant bias for detections with magnitudes $19<m_{F814W}<23$. For sources fainter than $m_{F814W}=23.0$, an increasingly dependence on the  magnitude is observed as a consequence of the rapidly decreasing $S/N$ for the ALHAMBRA detections. To match the  ALHAMBRA and COSMOS F814W magnitudes for fixed $3\arcsec$ apertures, we derived a magnitude-dependent correction which is included in the final catalogues, as explained in Section \ref{catalogs}.}
\label{cosmos2alhambra}
\end{center}
\end{figure*}

 Given that the typical error in the individual bands is $2-3\%$, total zeropoint error in the F814W image is quite  small, providing high homogeneity. To verify the calibration of the synthetic F814W images, we performed a photometric comparison with the COSMOS field. To reproduce the same photometric measurements as in Ilbert et al. (2009), we ran SExtractor using fixed circular apertures of $3\arcsec$. We retrieved $\sim$10800 common sources with ALHAMBRA in the $19 < m_{F814W}< 25.5$ range. The photometric comparison is shown in Fig. \ref{cosmos2alhambra}.

\vspace{0.2cm}
 
 No photometric zeropoint offsets or trends are apparent up to magnitudes $m_{F814W} = 23.5$. For the faintest sample, the retrieved photometric color (COSMOS/F814W magnitude - ALHAMBRA/F814W magnitude) becomes progressively negative, indicating that COSMOS magnitudes are brighter than ALHAMBRA's. The trend observed seems to be caused by a combination of an aperture and filter shape effect. On the one hand, as explained in Sect. \ref{psfmodels}, the ALHAMBRA PSF-models seemed to prefer a Penny2 profile, which consists of a Gaussian-like function but with Lorentzian wings, in addition the PSF is almost an order of magnitude larger than the HST PSF. This much more extended PSF spreads flux outside the aperture diameter, an effect which is much more important for the faintest sources. This is also observed in the simulation in Sect. \ref{simulations}. In Fig.\ref{simulcompleteness} (top panel) where the same trend is observed at faint magnitudes. On the other hand, the ALHAMBRA/F814W images were created combining the last optical individual filters as shown in equation \ref{f814wtranseq}. Unfortunately, the last F892W, F923W and F954W optical filters have relatively low S/N. Therefore our ``F184'' filter becomes progressively bluer with magnitude, therefore having a much wider PSF, which intensifies the above described effect. 
 
 In an effort  to extend the accuracy of the ALHAMBRA photometric measurements, we derived a magnitude-dependent correction to make ALHAMBRA magnitudes reproduce the COSMOS estimations for fixed apertures of $3\arcsec$. These corrected magnitudes  are included in the final catalogues as explained in the appendix \ref{descriptcatalogues}.    
 
\vspace{0.2cm}

  We also ran SExtractor on both ACS/F814W and synthetic ALHAMBRA/F814W images using the same SExtractor configuration. This analysis provided a characterization of the differences in the detections between both images. For detection magnitudes $19<m_{F814W}<23.5$ only a few dozens of sources per CCD were missing from the synthetic ALHAMBRA/F814W images. Detections fainter than magnitudes $m_{F814W}=23.5$ showed a increasing distribution of undetected sources peaking at a magnitude $m_{F814W}\sim 25.5$, well beyond the ALHAMBRA photometric completeness limit.

\subsubsection{Masks.}
\label{masks}

\begin{figure}
\begin{center}
\includegraphics[width=4.1cm]{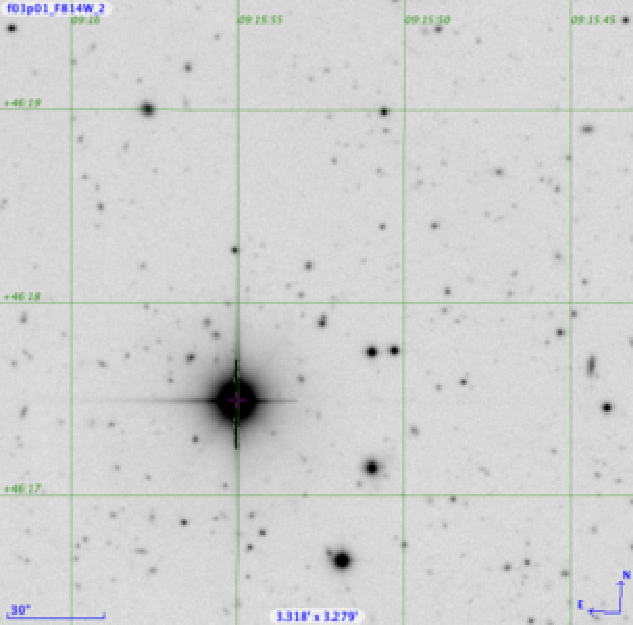}
\includegraphics[width=4.1cm]{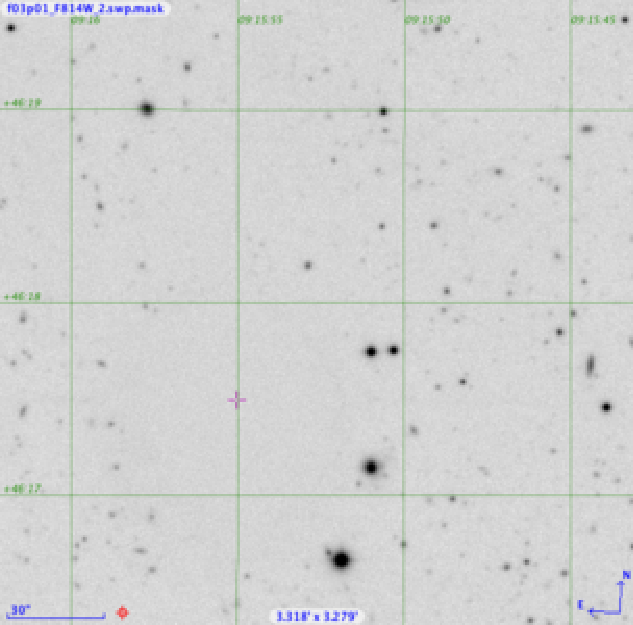}
\caption{Star masking. In order to improve both the photometric depth and the photometric measurements, F814W detection images were masked out, purging saturated stars, spikes, ghosts, negative areas and other artifacts. This  figure shows an example of how a saturated star from the original image (left panel) disappears after replacing all its pixels with background signal (right panel).}
\label{maskexample}
\end{center}
\end{figure}

  In order to improve the source detection efficiency, we masked every saturated star, stellar spike, ghost and damaged area. Initially we ran SExtractor on each synthetic F814W detection image with a special configuration to detect just very bright and extended sources. We visually checked the extracted sources to exclude any possible nearby galaxy. Then we convolved the resulting SExtractor segmentation maps with a Gaussian function to broaden the previously defined apertures and so remove residual contributions from stellar halos. We repeated the same procedure on the inverse image to deal with negative extended regions generated by saturated stars. We combined both positive and negative segmentation maps, defining the total region to be masked out. Finally, we replaced all flagged pixels with background noise to minimize the variation of the image RMS. An example of the masking procedure for a saturated star is shown in Fig. \ref{maskexample}. 

\vspace{0.2cm}

\subsubsection{SExtractor configuration.}
\label{SExconf}

 Assuming an expected variability in terms of PSF (and therefore in photometric depth) among the F814W detection images, we explored the optimal SExtractor configuration which maximizes the number of real detections. For non crowded fields, the most relevant parameters are the minimum number of contiguous pixels $DETECT\_MINAREA$ and the threshold which the signal has to exceed to be considered a real detection $DETECT\_THRESHOLD$.

\vspace{0.2cm} 

  In the Gaussian limit, the sky noise should in principle have a symmetric structure, and a similar amount of spurious objects is expected to be found on both sides of the image. Using this assumption, we looked for the SExtractor detection threshold with produced no more than $3\%$ contamination by spurious detections. We fixed the  $DETECT\_MINAREA$ at twice the image FWHM, and run SExtractor on both the image and its inverted negative side. The results are shown in Fig. \ref{thranalysisplot}. In the Appendix \ref{SExconfApp}, we show an example of the typical configuration used to perform source detection. 

\begin{figure}
\includegraphics[width=8.5cm]{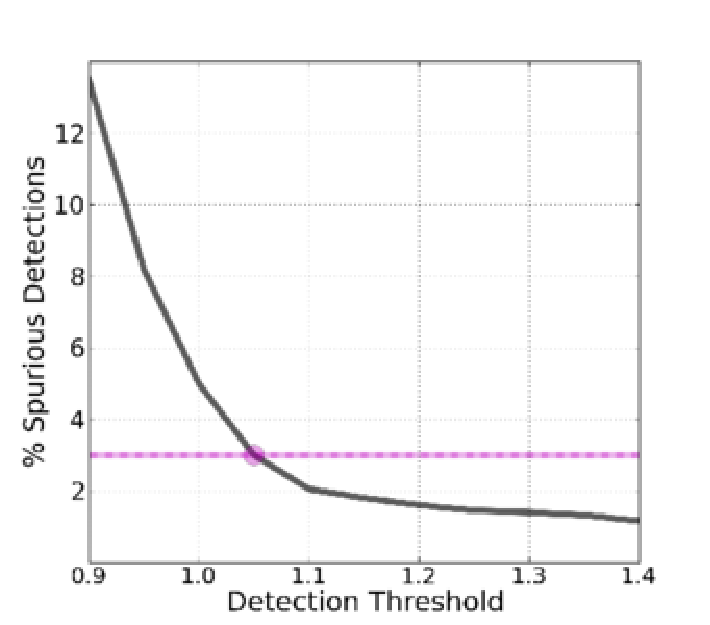}
\caption{Detection threshold. We fix the $DETECT\_MINAREA$ at twice the image FWHMW and then ran SExtractor on the positive and inverted negative image, obtaining the fraction of spurious over real detections as a function of $DETECT\_THRESHOLD$. We set the threshold to the value that reported no more than 3\% of spurious detections.}
\label{thranalysisplot}
\end{figure} 

\subsubsection{Flag Images.}
\label{Flagimages}

 In order to be able to quantify the survey effective area, we generated $FLAG$ images for each individual CCD where all problematic pixels were set to 0.  As the effective exposure time rapidly decreases when approaching to the image edges, we defined homogeneous areas where all sources have adequate exposure in all the 23 bands. We normalized individual weight maps to the maximum exposure time and then flagged the regions with a relative exposure time below  $60\%$ (mostly near the image edges). 

\vspace{0.2cm}

 The flag images also incorporate the stellar mask information (Section \ref{masks}); masked out regions are  replaced by background noise in the science images, to avoid interfering with SExtractor background determination. 
We compute the effective area for each F814W detection image as the direct conversion of the total number of non-flagged pixels into deg$^{2}$, as shown in Table \ref{effectivearea}. Including all the forty-eight F814W detection images yields a total surveyed area of 2.79 deg$^{2}$. 

\subsubsection{RMS Images.}
\label{rmsimages}

 As the effective exposure time on an image is position-dependent, detected sources on the edges will have shorter exposures than sources on the center, generating $S/N$ gradients. From a source-detection point of view, as synthetic F814W images are generated as the combination of many filters, occasional inhomogeneities registered on individual WEIGHT maps became averaged out. However, on individual filters (especially for the case of NIR images) we found occasional inhomogeneities across the images (Crist\'obal-Hornillos 2013, prep.), which affects the photometric depth. 

\vspace{0.2cm}

  To help disentangle whether a galaxy may be missed in a given filter as a consequence of its intrinsic luminosity (below the detection threshold) or due to an insufficient  photometric depth, we used the WEIGHT maps  (Crist\'obal-Hornillos 2013, prep.) to generate a new set of inverseRMS images and define two additional photometric Flags using the following expression: 

\begin{equation}
1/RMS = \sqrt{Weight}
\end{equation}

 Hence, the $irms\_OPT\_Flag$ and the $irms\_NIR\_Flag$ Flags indicates the number of individual bands in which an object has a signal in its inverseRMS below 80\% of the maximum value. Therefore, galaxies with large values in these photometric flags (indicating a large fraction of filters photometrically flagged) may provide unreliable photometric redshift estimations.

\subsection{Star/galaxy separation.}
\label{SG}
 The star/galaxy classification is a necessary step for accurate extragalactic surveys. Stars as real point-like sources (PLS) are observed as the most compact objects in an astronomical image. However, as objects get fainter (decreasing its $S/N$) it becomes progressively harder discerning their real morphologies. 

\vspace{0.2cm}

 We followed a statistical approach to perform star/galaxy separation. We assigned a probability to every detection given its apparent geometry, $F814W$ magnitude, optical $F489W$ - $F814W$ and NIR $J$-$Ks$ colors. For each variable we derived the corresponding probability distribution function (PDF) based on the typical distribution of stars and galaxies. Therefore, every detection is classified in terms of the probability of being a star or a galaxy, as follows:

\vspace{0.2cm}

\begin{equation}
P_{Star} = P_{Star}^{FWHM} \times P_{Star}^{m_{F814W}} \times P_{Star}^{Opt} \times P_{Star}^{NIR}
\end{equation}

\begin{equation}
P_{Gal} = P_{Gal}^{FWHM} \times P_{Gal}^{m_{F814W}} \times P_{Gal}^{Opt} \times P_{Gal}^{NIR}
\end{equation}

\begin{equation}
where \,\,\,\,\,\, P = P_{Star} + P_{Gal} = 1
\end{equation}

 Final probabilities are stored on the statistical variable $Stellar\_Flag$ included in the catalogues. The derivation of each of the four independent PDFs is described below. 

\subsubsection{Geometry and Magnitude.}
\label{geomag}
We used the COSMOS HST/ACS images to explore the star/galaxy selection algorithms, since they are considerably deeper and with an obviously much narrower PSF than the ALHAMBRA dataset.   

\vspace{0.2cm}

\begin{figure*}
\begin{center}3
\includegraphics[width=18.cm]{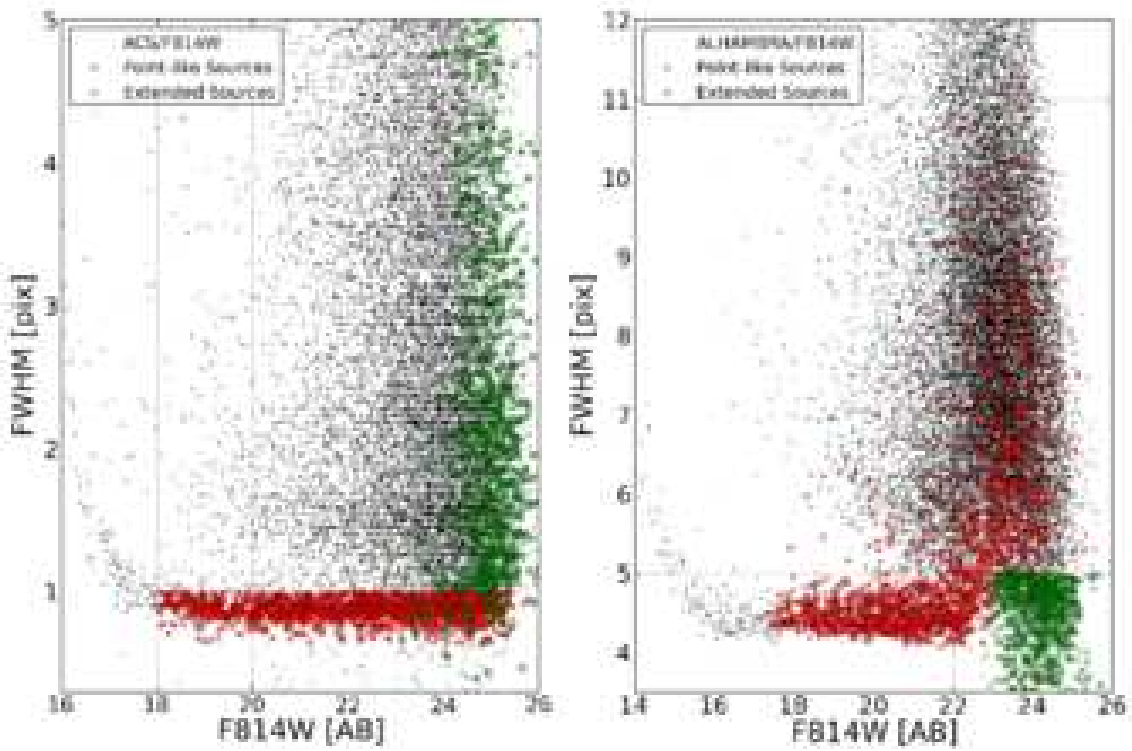}
\caption{We studied the shape degradation of point-like sources observed in our fields as a function of the apparent magnitude. We used the ACS/F814W images from COSMOS to select real point-like sources (narrowest FWHM) within the ALHAMBRA fields and so understand how these sources might look like when observed under the ALHAMBRA PSF conditions. As it can be seen in the figure, $m_{ACS/F814W}<22.5$ objects classified as point-like sources on the ACS/F814W images (red circles) were also classified as point-like sources on the ALHAMBRA/F814W images. However, point-like sources on the ACS/F814W images with $m_{ACS/F814W}>22.5$ show an increasingly wider FWHM on the ALHAMBRA/F814W images. This fact illustrates the degradation of purely ``geometrical'' information with decreasing $S/N$. Inverting the procedure, we also find that point-like sources in the ALHAMBRA images below the $m_{F814W}$ threshold are often the central regions of extended, faint objects in the ACS images.}
\label{PLS}
\end{center}
\end{figure*}

 We ran SExtractor twice, first on the ACS/F814W images and then on the ALHAMBRA/F814W images in single-image mode, plotting the detected sources in a FWHM vs $m_{ACS/F814W}$ diagram as shown in Fig. \ref{PLS}. We selected detections classified as PLS (point-like sources) in the ACS/F814W images and used them to match up the ALHAMBRA/F814W detections.  

\vspace{0.2cm}

 As seen in Fig. \ref{PLS}, sources brighter than $m_{ACS/F814W}=22.5$ classified as PLS on the ACS/F814W images were equally classified as PLS on the ALHAMBRA/F814W (open red circles). However, PLS fainter than $m_{ACS/F814W}=22.5$  showed progressively larger FWHM values on the ALHAMBRA/F814W. Therefore, ALHAMBRA images cannot be used reliably 
for morphological information fainter than $m_{F814W}=22.5$. 

\vspace{0.2cm}
 
 We also investigated the nature of the faint detections appearing as PLS in the ALHAMBRA/F814W images but clearly not belonging to the ACS/F814W PLS sample (green dots). We inverted the procedure, selecting faint $m_{F814W}>22.5$ PLS in ALHAMBRA and matching them to the ACS/F814W detections. The result showed that some of those detections were actually very faint extended sources in the ACS images, and where classified as PLS in the ALHAMBRA/F814W images because only the central, compact region of the source was detected above the detection threshold.  

\subsubsection{Photometric colors.}

\begin{figure*}
\begin{center}
\includegraphics[width=8.5cm]{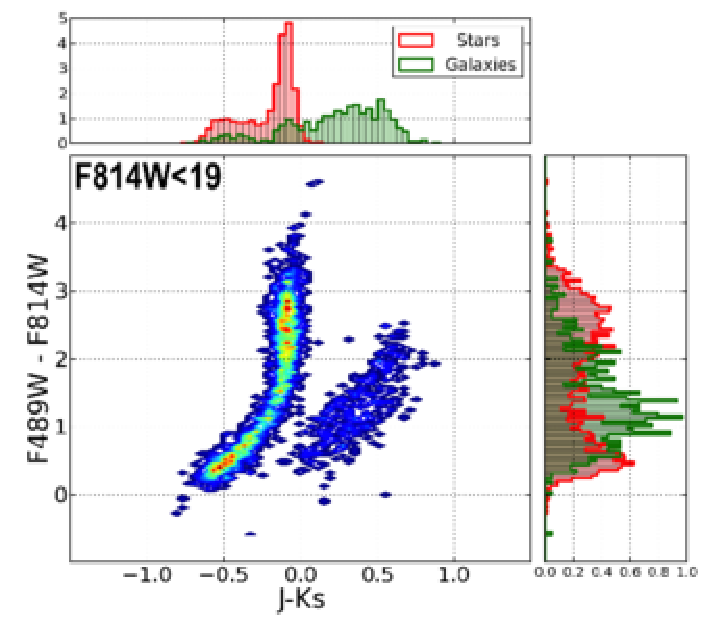}
\includegraphics[width=8.5cm]{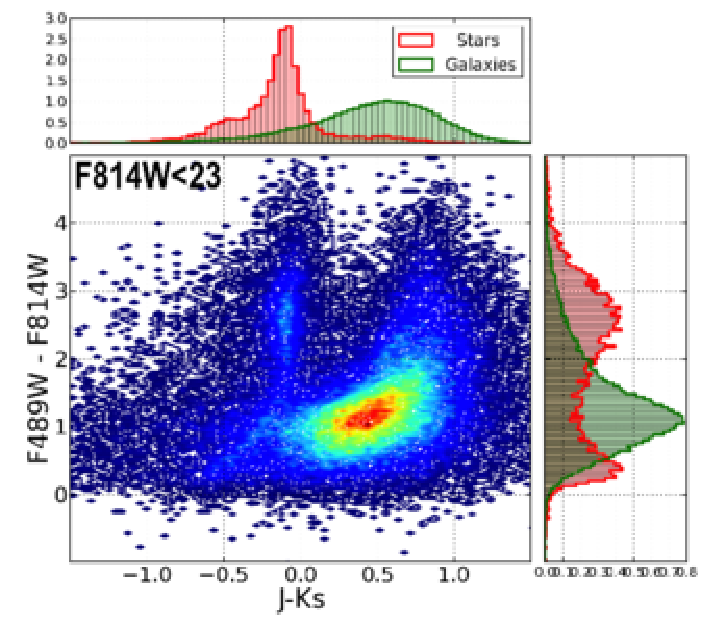}
\caption{Effect of the photometric uncertainties in the separation between stars and galaxies. The $F489W-F814W/J-K_S$ colors of objects classified as point-like sources in the ACS/F814W images are plotted on the left, those of extended objects on the right. This diagram is a very useful discriminating tool for $m_{F814W}<22.5$, but becomes almost useless at fainter magnitudes}
\label{starsgalaxy3Dm21}
\end{center}
\end{figure*}

 Stars can be told apart from galaxies based on their spectral differences (Daddi et al. 2004). By combining two  photometric colors (one in the optical, one in the NIR) is possible to identify two separated regions where stars and galaxies are typically located, as shown in the left panel of Fig. \ref{starsgalaxy3Dm21}. We use the optical $F489W-F814W$ and NIR $J-K_s$ to study how well the color-color method works with increasing photometric depth in the ALHAMBRA images.  

 In order to generate a control sample, we assumed that real PLS (as classified by ACS/F814W images) were all "stars" whereas well extended sources (ES) were assumed to be "galaxies". Considering the resolution of the ACS/HST images and the magnitude range involved in this analysis, the so-derived sample of stars/galaxies represented a good approximation as the expected fraction of misclassified galaxies or QSOs is actually negligible.

\vspace{0.2cm}
 
 We tested the reliability of this methodology by gradually decreasing the S/N of the sample. Initially we selected only sources with very high S/N ($m_{F814W}<19$) as shown on the left-hand side of Fig. \ref{starsgalaxy3Dm21}. However, as sources get fainter ($m_{F814W}<23$, on the righ-hand side) separating both two classes becomes progressively complicated with ES and PLS spreading into, respectively, the stellar and galactic loci. 

\subsubsection{Stellar flag}
\label{stFlag}

\begin{figure}
\begin{center}
\includegraphics[width=4.15cm]{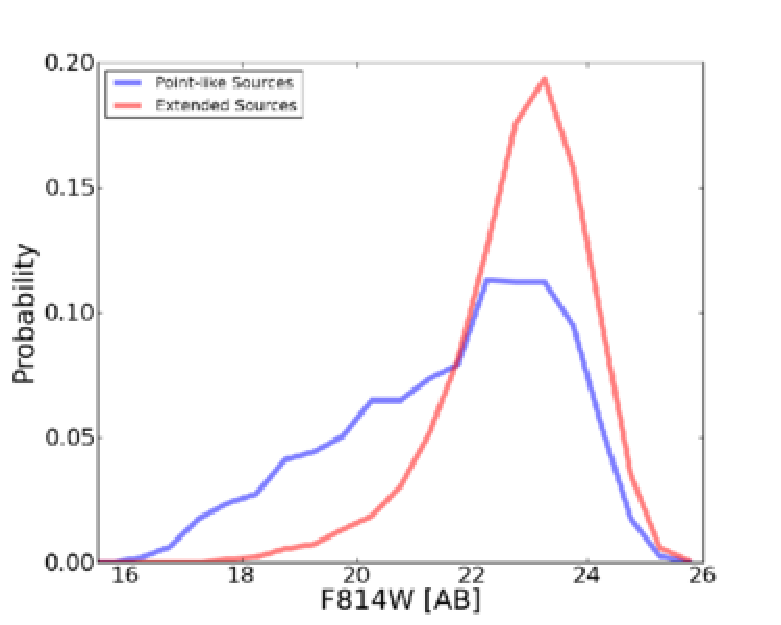}
\includegraphics[width=4.15cm]{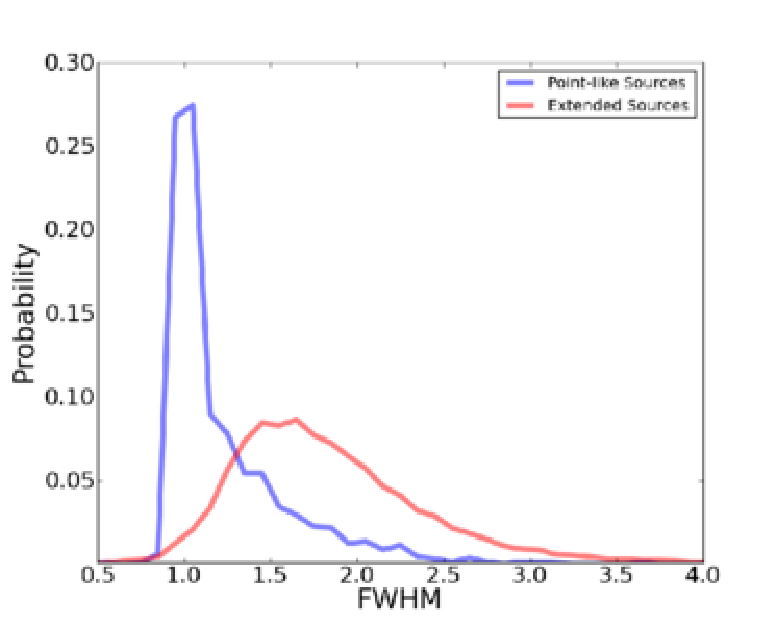}
\includegraphics[width=4.15cm]{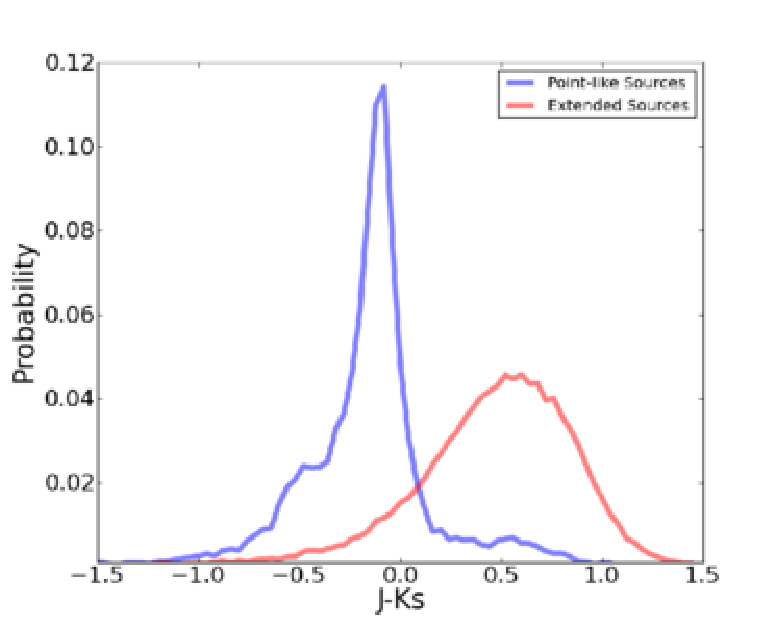}
\includegraphics[width=4.15cm]{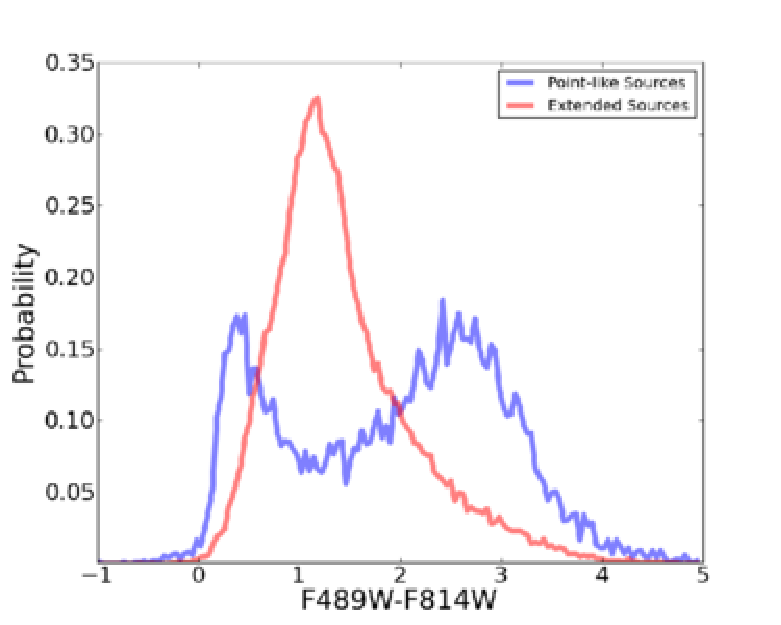}
\caption{Star/galaxy PDFs. The figure shows the four Probability Distribution Functions derived for a control sample of stars and galaxies selected from the ACS/F814W images. From left to right and top to bottom, the distribution of stars (blue line) and galaxies (red line) as a function of the apparent magnitude $F814W$, the apparent FWHM, the NIR and optical colors are shown. These PDFs were used to estimate the probability of a detection to be a star or a galaxy as explained in Section \ref{stFlag}.}
\label{stargalaxypdf}
\end{center}
\end{figure}

 Finally, we used the retrieved information from the star/galaxy geometry, F814W magnitude, optical and NIR colors to derive empirical PDFs as shown in Fig. \ref{stargalaxypdf}. Therefore we assigned a statistical classification to every detection given its observed information. Considering the level of both photometric and geometric uncertainties when deriving the PDFs, we excluded from the classification all detections with $m_{F814W}>22.5$, assigning them a $Stellar\_Flag$ value = 0.5.

\vspace{0.2cm}

\begin{figure*}
\begin{center}
\includegraphics[width=8.1cm]{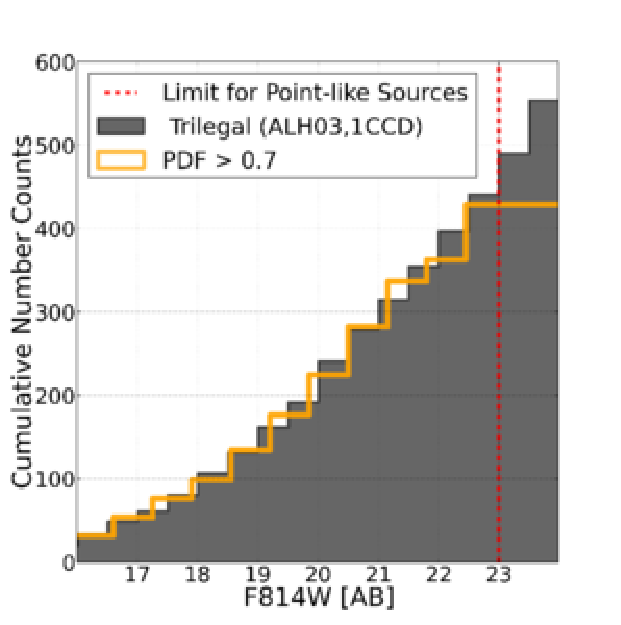}
\includegraphics[width=8.1cm]{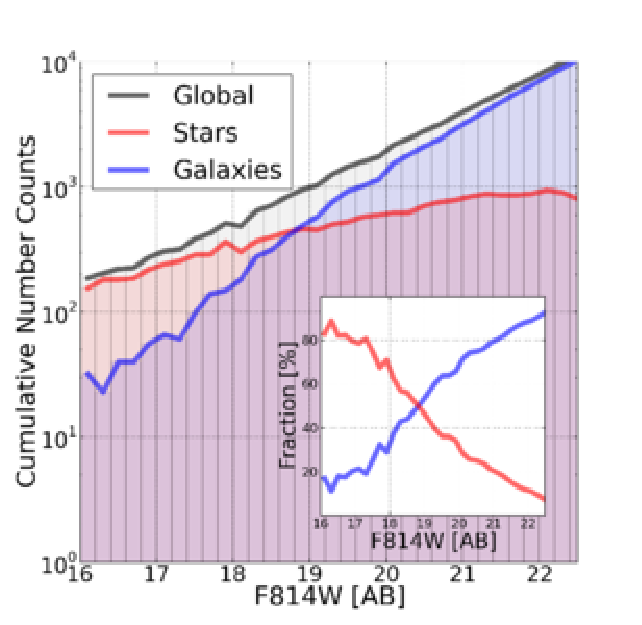}
\caption{Star number counts. For each field in ALHAMBRA, we compared the number of detections classified as stars based on our statistical criteria (solid yellow line) with that provided by the $Trilegal$ software (solid grey histogram), as seen in the left panel. The figure shows an example for a single CCD in the ALHAMBRA-03 field. The best match is reached for $Stellar\_Flag>0.7$. When applying this statistical criterium to the whole catalogue (right panel), we observe stars to dominate the sample down to a magnitude $m_{F814W}<19$. At fainter magnitudes, the fraction of stars rapidly declines with a contribution of $\sim10\%$ for magnitudes $m_{F814W}=22.5$, which quickly declines, as indicated in the \textbf{inset} panel, until it becomes almost negligible for $m_{F814W}>23.5$. Extending the analysis to the whole sample, we retrieved an average stellar density of $\sim$7000 stars per deg$^{2}$ ($\sim 450$ stars per CCD), for $m_{F814W}<22.5$}
\label{trilegalpdfs}
\end{center}
\end{figure*}

 We tested the goodness of this statistical classification by comparing the density of selected stars per unit of area with the numbers expected from the model of Girardi (2002,2005), implemented in the $Trilegal$ software, as a function of the area, the galactic position and the limiting magnitude. As seen in Fig. \ref{trilegalpdfs}, we find a very good agreement between measurements and predictions, which is optimized by using a threshold of  Stellar\_Flag$> 0.7$ to select our stars. 

\vspace{0.2cm}

 As shown in the right panel of Fig \ref{trilegalpdfs}, when applying this criterium, we observe that stars dominate the sample down to a magnitude $m_{F814W}<19$. For fainter magnitudes the fraction of stars rapidly declines with a contribution of $\sim$10\% for magnitudes $m_{F814W}=22.5$. As indicated in the \textbf{inset} panel, if we extrapolate the so-derived stellar number counts, the expected contamination for unclassified stars with magnitudes fainter than $m_{F814W}>22.5$ becomes negligible, with a contribution of stars of $\sim$1\% for magnitudes $m_{F814W}=23.5$. We retrieve an averaged stellar density in the galactic halo of $\sim$7000 stars per deg$^{2}$ ($\sim$450 stars per CCD) for sources brighter than $m_{F814W}=22.5$.

\vspace{0.2cm}

\subsection{Photometric errors}
\label{photoerr}

  Many of the steps involved in image processing introduce correlations between neighboring pixels making the background noise in images different from a Poissonian distribution. If these effects are not properly taken into account, they can lead to a severe underestimation of the real photometric uncertainties, critically affecting the photometric depth estimations (the survey photometric limiting magnitude) and the photometric redshift accuracy. We have therefore carefully estimated photometric errors using an empirical approach (similar to that described in Casertano et al. 2000, Labb\'e et al. 2003, Ben\'itez et al. 2004, Gawiser et al. 2006 and Quadri et al. 2007). 

\vspace{0.2cm}
 
 As explained in Section \ref{ColorPro}, \textit{ColorPro} was updated to automatically degrade every image which has a PSF narrower than that of the detection image. We also rescaled the original NIR images (from OMEGA-2000) to the LAICA pixel size (Crist\'obal-Hornillos et al. 2013, in prep.). Both procedures alter the properties of their original background distributions. Moreover, when deriving photometric uncertainties, SExtractor always assumes the image background to follow a  Poisson distribution with no correlation among pixels. This underestimates the real noise, as we will see below. 

\begin{figure}
\begin{center}
\includegraphics[width=8.75cm]{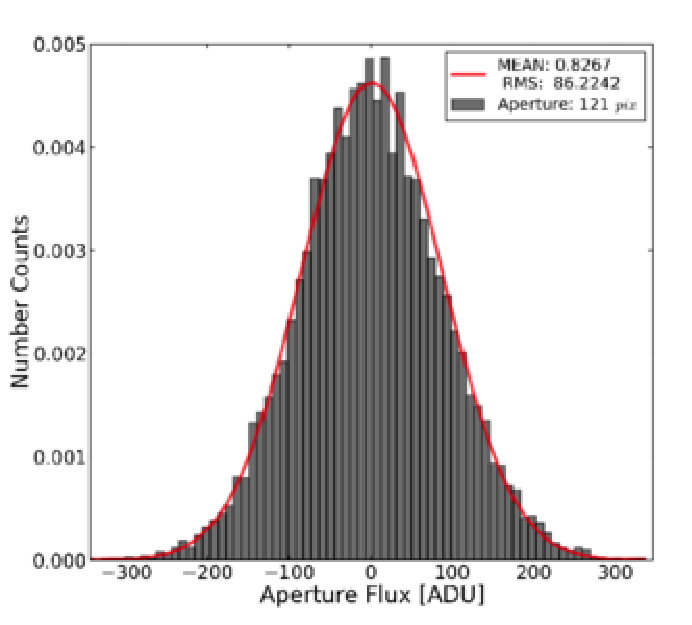}
\caption{The figures shows an example of the background distribution in 121-pixel apertures after drawing $\sim$50.000 apertures on blank regions. To properly estimate the empirical dependence between photometric apertures and the RMS, the procedure was repeated spanning a range of apertures between up to 250 pixels, covering the size distribution measured for ALHAMBRA objects.}
\label{s2a1p}
\end{center}
\end{figure}

 To derive empirical photometric uncertainties for each individual image, we mask out the objects detected by SExtractor using the segmentation map derived from the F814W detection image. Then we throw $\sim$50.000 apertures over the remaining area, measuring both the enclosed signal and the RMS inside it. The procedure is repeated for apertures in the 1-250 pixel range, correcting appropiately by the total exposure time of the pixels belonging to them using the weight maps. Fig. $\ref{s2a1p}$ shows an example of the typical measured background distribution for one pixel. The red line corresponds to the best Gaussian fit to the data.

\vspace{0.2cm}

  As expected, the ALHAMBRA images are accurately described by a Poisson distribution on small scales. However as apertures become larger, a second term starts dominating the distribution indicating the presence of large-scale correlations among pixels. In this case, the background distribution is described by the relation:

\begin{equation}
\label{backindexes}
\sigma(A) = {\sigma_{1}\sqrt{N}(C_{1}+C_{2}\sqrt{N}) \over \sqrt{w_{N}} }
\end{equation}

where coefficient $C_{1}$ indicates the Poisson contribution dominating on small scales, $C_{2}$ the contribution on large scales, $w_{N}$ the corresponding percent weight (from WEIGHT map) and $\sigma_{1}$ the background distribution measured for 1-pixel apertures.  

\vspace{0.2cm}
 
\begin{figure*}
\begin{center}
\includegraphics[width=7.67cm]{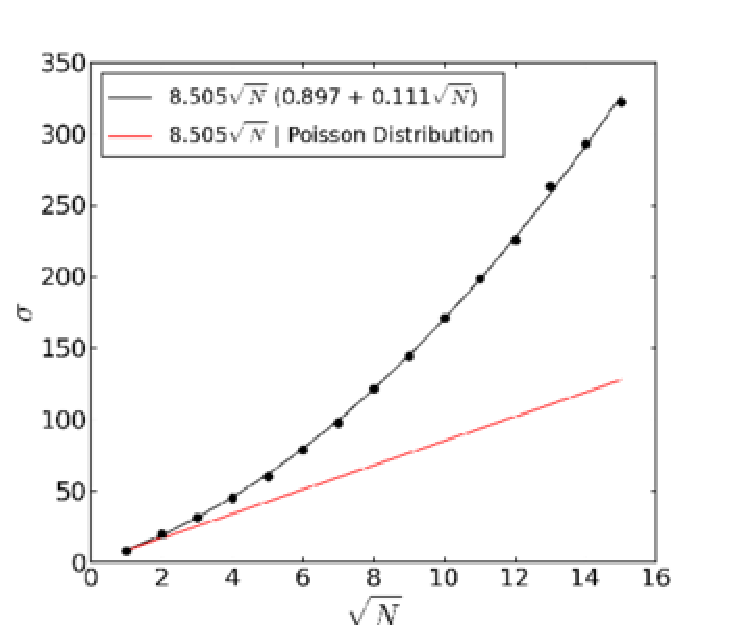}
\includegraphics[width=7.37cm]{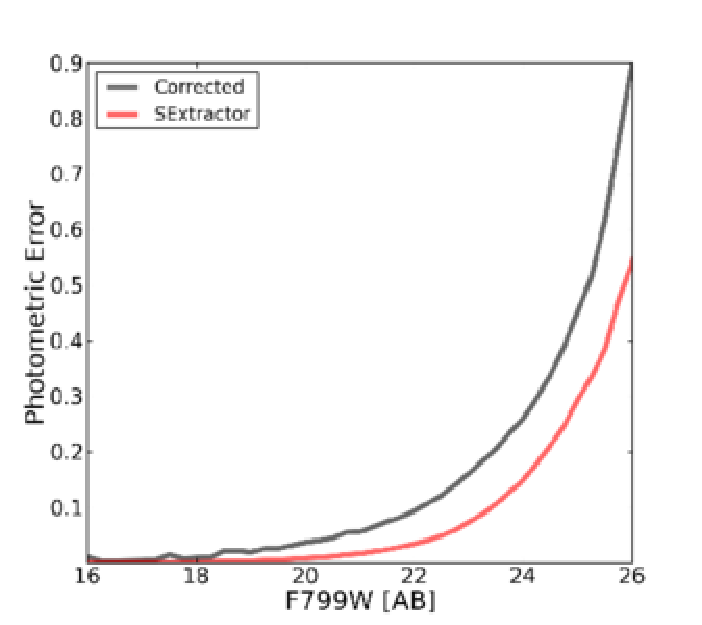}
\caption{Photometric uncertainties. The figure shows the differences between the photometric uncertainties yielded by $SExtractor$ (solid red line) assuming that the background follows a Poisson distribution and those empirically estimated (solid black line) using the methodology described in Section \ref{photoerr}. Left panel illustrates how the dependence between the RMS and the aperture-size ($\sqrt{N}$) becomes progressively underestimated by SExtractor due to the presence of large-scale correlations among pixels introduced during image processing. As described in equation \ref{backindexes}, the number outside the parenthesis in the legend corresponds to the background RMS derived for 1-pixel apertures and the number inside corresponds to the (Poisson) contribution, which dominate on small and large scales respectively.  The right panel compares the average photometric uncertainties as a function of magnitude, both derived by $SExtractor$ (solid red line) and using the empirical approach described in the text (solid black line). As expected, SExtractor underestimates the real photometric errors, what becomes especially significant at faint magnitudes.}
\label{apsigma}
\end{center}
\end{figure*}

  The relevance of this sort of corrections can be appreciated in Fig. $\ref{apsigma}$ where the differences between a Poisson-based treatment (solid red line) and an empirically estimated (solid black line) are shown. Whereas the left panel indicates the dependence of the expected RMS as a function of aperture size $\sqrt{N}$, the right panel shows the re-estimated mean photometric uncertainties as a function of magnitude. 

\vspace{0.2cm}

\subsection{Photometric verfication.}
\label{phintchecks}

As already mentioned in Section $\ref{general}$, we take advantage of the presence in LAICA of four CCDs which simultaneously image close regions on the sky under the same atmospheric conditions and with the same passband to carry out a statistical comparison among contiguous CCDs. 

\begin{figure*}
\begin{center}
\includegraphics[width=0.405\textwidth]{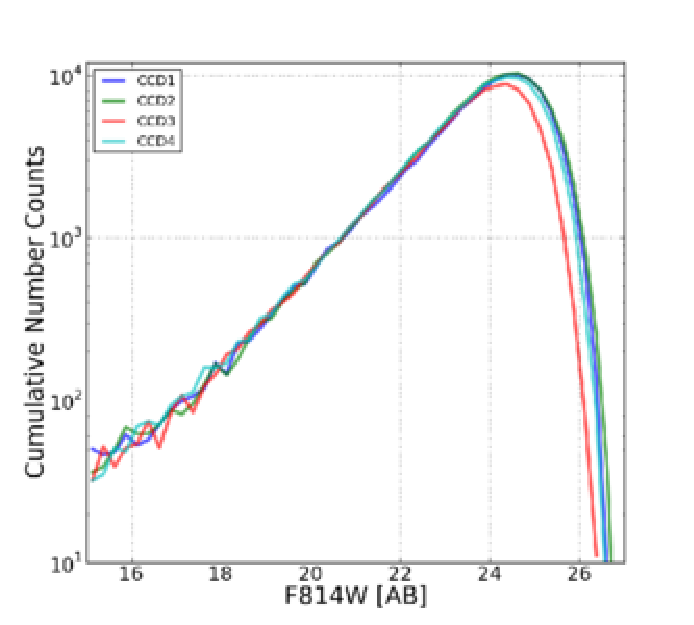}
\includegraphics[width=0.405\textwidth]{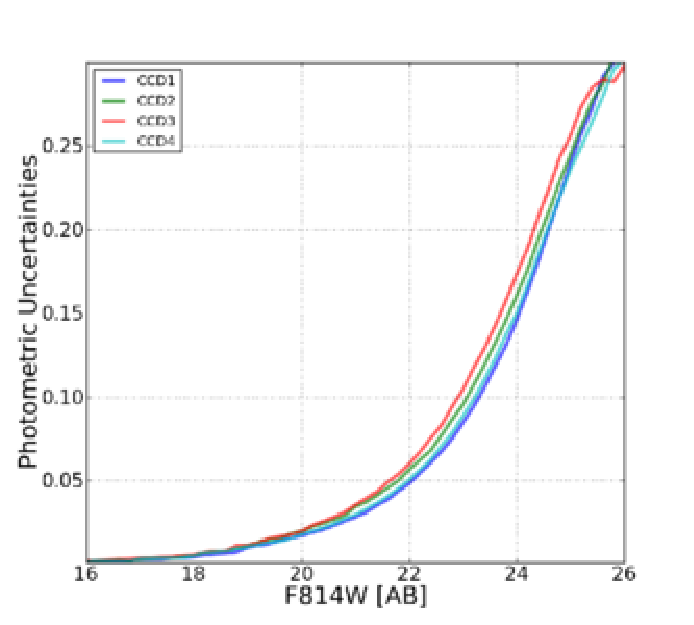}
\caption{Internal photometric verifications. Using the advantage that the four CCDs composing the LAICA optical system were simultaneously imaging (almost) the same regions of the sky, under equal atmospheric conditions and through the same pass-band, we performed internal photometric comparisons among the CCDs. As seen on the left panel, on the one hand, we compared number counts per magnitude range to ascertain the homogeneity during the detection process. As expected, whereas CCD1, CCD2 and CCD4 were similar, CCD3 behaved slightly worse with a shallower photometric depth. The presence of neither bumps nor horizontal shifts among CCDs indicated homogeneous detections and none photometric zeropoint offsets. On the other hand, we compared the photometric uncertainties as a function of the magnitude for the four CCDs, as seen on the right panel. As expected, CCD3 typically showed larger photometric uncertainties confirming its poorer performance.}
\label{intchecks}
\end{center}
\end{figure*}

 We looked at the number of detected sources per magnitude range, and as illustrated in the Fig. $\ref{intchecks}$, the results for the four CCDs were highly consistent for magnitudes $m_{F814W}<24$, where ALHAMBRA is photometrically complete. For fainter magnitudes CCD3 showed a decrease in the number of detections, probably due to its poorer efficiency. We also compared the photometric uncertainties between CCDs, and again,  CCD1, CCD2 and CCD4 showed a good agreement whereas CCD3 differed from the 
general trend showing larger photometric uncertainties.

\vspace{0.2cm}

 Finally, we did not observe any horizontal shifts among curves indicating no photometric bias at first order. This is illustrated in the right panel of Fig. $\ref{intchecks}$.

\section{PHOTOMETRIC REDSHIFTS}
\label{bpz}

 The BPZ code implements the Bayesian method of Ben\'itez 2000 to estimate photo-z. BPZ weights the redshift/type likelihood $L(C|z,T)$, obtained from the comparison of the redshifted template library with the observed galaxy magnitudes by a prior probability $p(z,T|I)$. $L(C|z,T)$ is often multimodal, due to color/redshift degeneracies, and the inclusion of prior information helps to eliminate unrealistic solutions and make $p(z,T)$ more compact, improving the photo-z accuracy and reducing the number of catastrophic outliers. In this work, we used an updated version of the code (BPZ2.0, Ben\'itez 2014) which includes several changes with respect to its original version.

\begin{figure*}
\begin{center}
\includegraphics[width=8.cm]{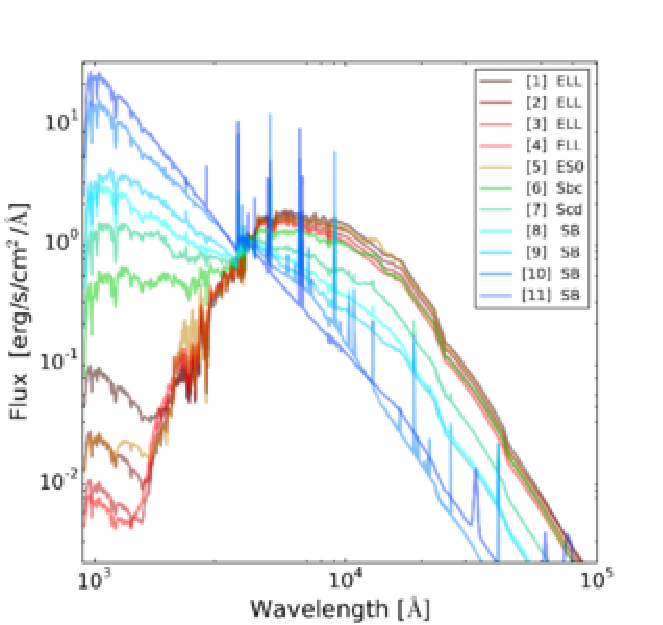}
\includegraphics[width=9.05cm]{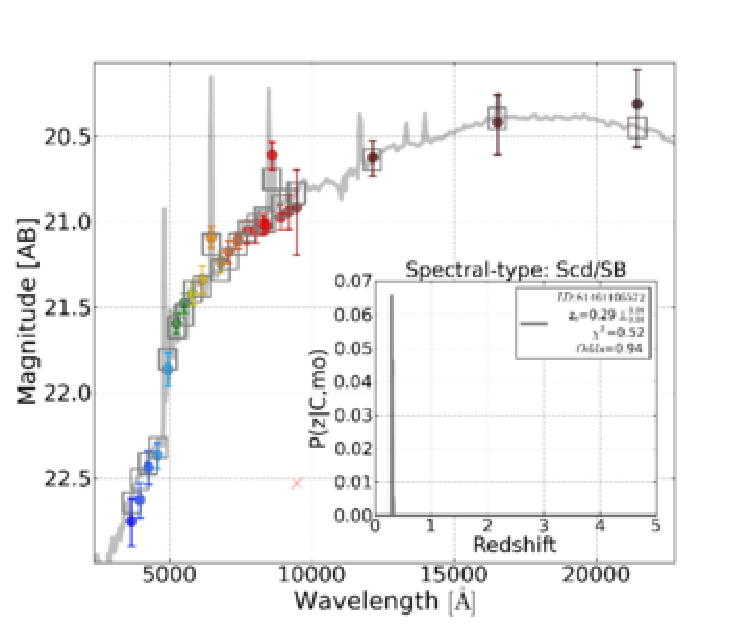}
\caption{We relied on an updated version of the BPZ code (BPZ2.0) to derive the photometric redshifts. Left: BPZ2.0 includes a new library composed by 11 galaxy templates: 5 originally drawn from PEGASE, 5 from GRASIL and 1 StarBurst. To make easier its visualization, the SEDs were arbitrarily normalized  to 4000$\AA$. The numerical notation used in the catalogues for the BPZ templates  is indicated in the legend. Right: shows an example of the typical spectral-fitting using the ALHAMBRA photometry, where the \textbf{inset panel} corresponds to its resulting redshift distribution function $p(z)$.}
\label{sedeb11}
\end{center}
\end{figure*}

 The BPZ2.0 uses a new library composed by six SED templates originally drawn from PEGASE (Fioc \& Rocca-Volmerange 1997) but then re-calibrated using the FIREWORKS photometry and spectroscopic redshifts (Wuyts et al. 2008) to optimize its performance. In addition to these basic 6 templates, four GRASIL and one Starburst template have been added. As seen in the left panel of Fig. $\ref{sedeb11}$, this new library includes five templates for elliptical galaxies, two for spiral galaxies and four for starburst galaxies along with emission lines and dust extinction. The opacity of the intergalactic medium was applied as described in Madau et al. (1995). An example of the typical spectral-fitting using the ALHAMBRA photometry is shown in the right panel of Fig. $\ref{sedeb11}$. The \textbf{inset} panel corresponds to the resulting redshift distribution function $p(z)$. The library and the procedure to obtain it will be described in detail in Ben\'\i tez (2014).    

\vspace{0.2cm}

 Likewise, the BPZ2.0 also includes a new prior which gives the probability of a galaxy with apparent magnitude $m_{0}$ having a certain redshift $z$ and spectral-type $T$. The prior has been empirically derived for each spectral-type and magnitude by the redshift distributions measured in the GOODS-MUSIC (Santini et al. 2009), COSMOS (Scoville et al. 2007) and UDF (Coe et al.2006) catalogs.  

\vspace{0.2cm}  

\begin{figure*}
\begin{center}
\includegraphics[width=8.25cm]{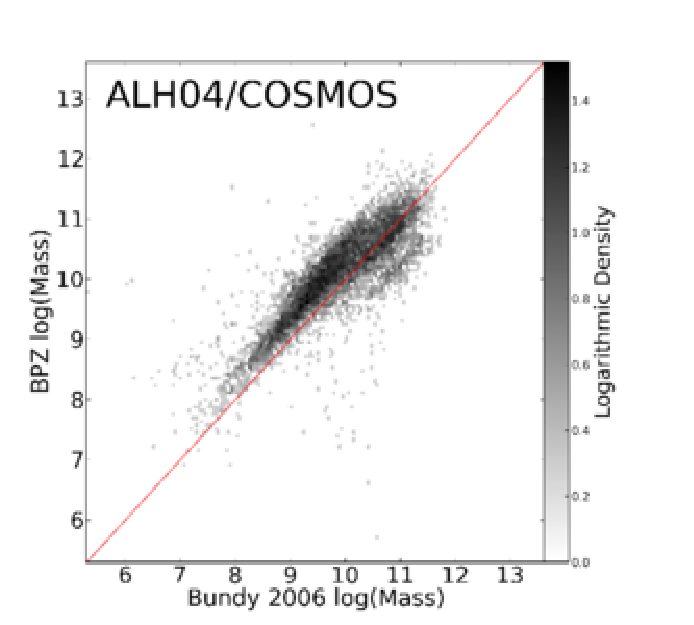}
\includegraphics[width=8.25cm]{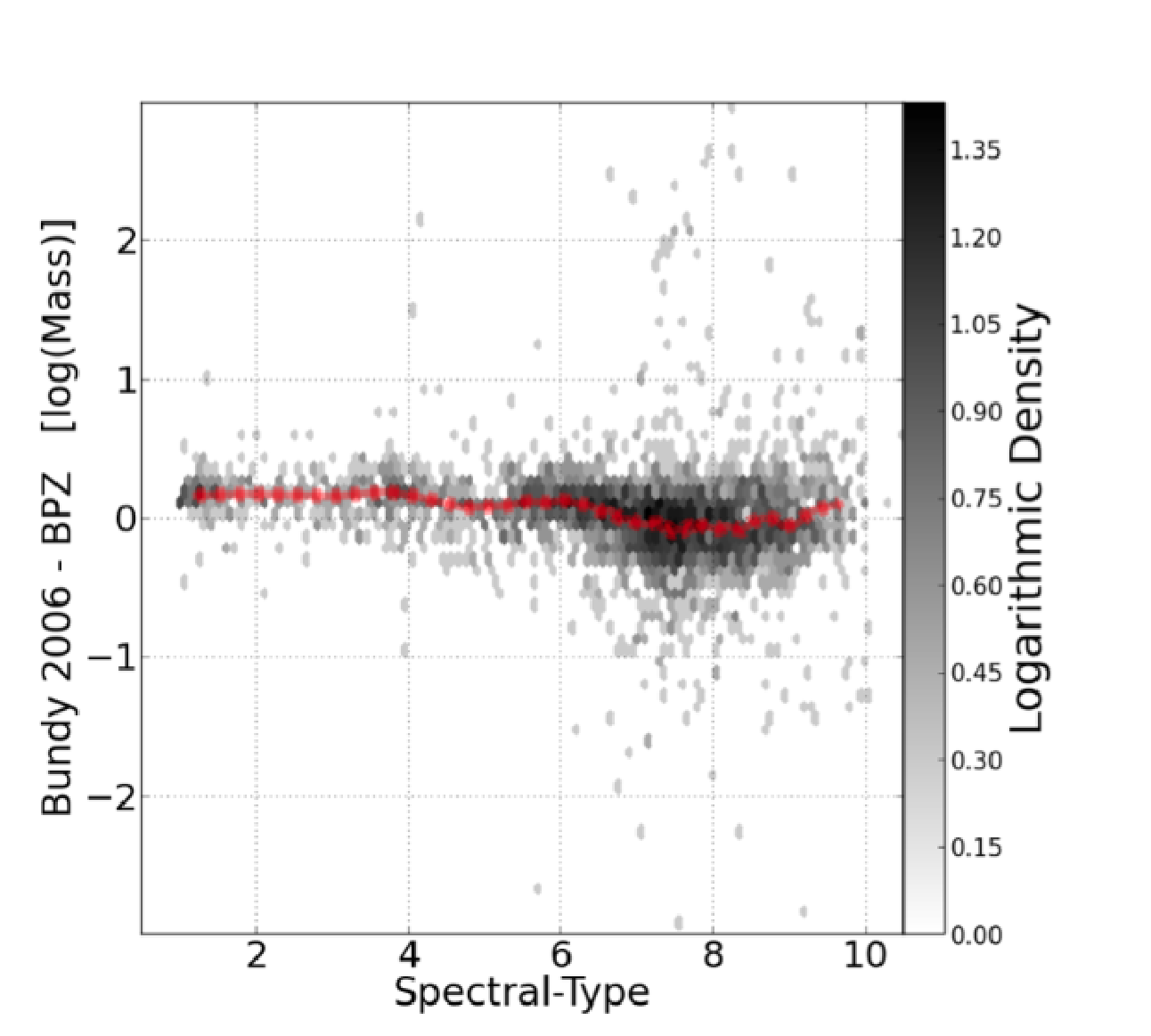}
\caption{The new version of BPZ provides an estimation of the galaxy stellar mass, calculated from the assigned interpolated spectrum, by applying the color-M/L ratio relationship established by Taylor et al. (2011) to the BPZ templates. Left: we show a comparison between the BPZ stellar masses with the masses measured by Bundy (2006a) on a sample of galaxies from the COSMOS field. We observe that the uncertainties (of about 0.1 - 0.2 $dex$) are within those expected by Bundy (2006a). Right: we represent the former comparison as a function of the spectral type. Again, a moderately dependence is observed with uncertainties within 0.1-0.2 $dex$.}
\label{bpzmasses}
\end{center}
\end{figure*}

  In addition, the BPZ2.0 also provides an estimation of the galaxy stellar mass, calculated from the assigned interpolated spectrum of the galaxy by applying the color-M/L ratio relationship established by Taylor et al. (2011) to the BPZ templates. For an in-depth discussion, we refer the reader to Ben\'itez 2014, in prep. We performed two different checks to show the robustness of the BPZ stellar masses.  First, when comparing the BPZ stellar masses with the masses measured by Bundy (2006a) on the COSMOS field, we observed that the uncertainties are within the expected by their analysis (of about 0.1 - 0.2 $dex$) with a moderately dependence on the spectral types, as seen in Fig. $\ref{bpzmasses}$. Secondly, we obtained BPZ stellar masses from a semi-analytical simulation (Merson et al. 2013, Ascaso et al. 2013 (in prep.)) and compared them with the input masses after correcting them by the effect of different IMFs (Bernardi et al. 2010). The mean value of the difference is $\sim 0.13 \pm 0.30$ dex which, as before, is consistent with the uncertainties reported (Mitchell et al. 2013), confirming the reliability of the stellar masses estimations.

\vspace{0.2cm}

BPZ2.0 provides an estimate for the galaxy redshift, generally defined as $z_{b}=\sum_{T}\int dz\, z\, p(z,T)$ and also the spectral type at that redshift, $T_{b}=[\sum_{T}\, p(T|z_{b})\,T]/[\sum_{T}\, p(T|z_{b})]$. To characterize the quality of the photo-z, BPZ provides the $Odds$ parameter (Ben\'itez 2000), $Odds=\sum_T\int_{-0.0125(1+z_{b})}^{0.0125(1+z_{b})}p(z-z_{b},T)$. It is worth emphasizing that the $Odds$ makes possible to derive high quality samples with very accurate redshifts and a very low rate of catastrophic outliers. We used a redshift resolution $DZ$  = 0.001 for the $0.001<z<7.0$ range. In order to fully cover the spectral-type space, we used an interpolation factor between templates of 7, i.e. generated seven models by linear interpolation in the flux space in between each of the original eleven models. The redshift confidence interval provided by $zb\_{min}$ and $zb\_{max}$ corresponds to the 68\% of the probability distribution function. Note that for multimodal probabilities, $p(z)$ can go to 0 within this range.

\subsection{Photometric redshift accuracy}
\label{photozaccuracy}

 The normalized median absolute deviation (NMAD) is a robust measurement of the accuracy reached by a sample of photometric redshifts (Brammer et al. 2008). A typical photometric redshift 
error distribution has fat tails, clearly departing from a pure Gaussian distribution, in addition to a relatively large fraction of outliers. NMAD estimator manages to get a stable estimate 
of the spread of the core of photo-z distribution without being affected by catastrophic errors. The NMAD is defined as:

\begin{equation}
\sigma_{NMAD} = 1.48 \times median({\left | \delta z - median(\delta z) \right | \over 1+z_{s}})
\end{equation}
\begin{equation}
and \,\,\, \delta z = z_{b} - z_{s} 
\end{equation}

  where $z_{b}$ corresponds to the Bayesian photometric redshift and $z_{s}$ to the spectroscopic redshift. Along with the scatter it is also important to quantify both the presence of any systematic bias $\mu$ and the fraction of catastrophic errors. In this work we user two different definitions for the outlier rate:

\begin{equation}
\eta_{1} = {\left | \delta z \right | \over 1+z_{s}} > 0.2
\end{equation}
\begin{equation}
\eta_{2} = {\left | \delta z \right | \over 1+z_{s}} > 5 \times \sigma_{NMAD} 
\end{equation}

\vspace{0.2cm}
 
\begin{figure}
\begin{center}
\includegraphics[width=8.85cm]{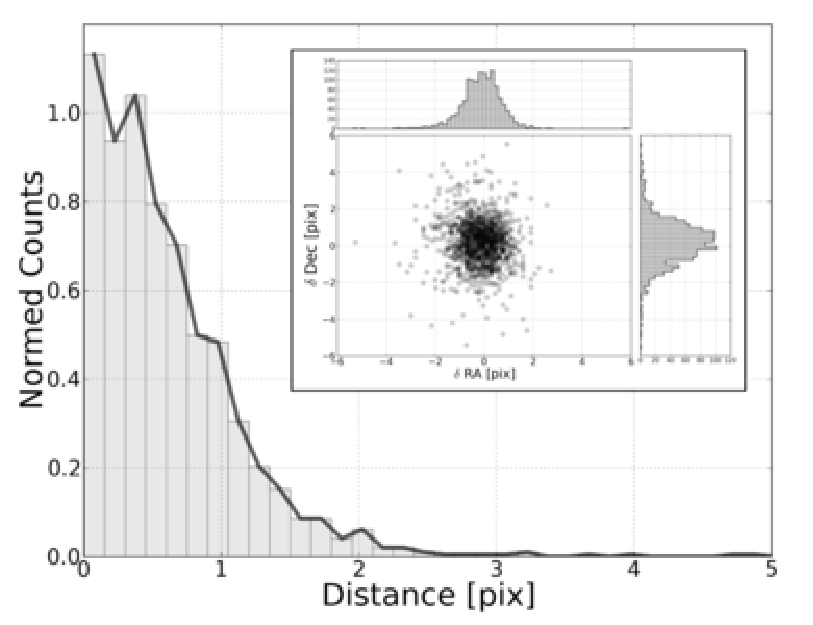}
\caption{Astrometric matching with spectroscopic samples. In order to reduce the fraction of potential mismatched galaxies, we initially performed second-order astrometric corrections between the ALHAMBRA fields and other surveys, to established a maximum distance of $\sim$3 pixels ($<$0.7") to match our detections. This maximum separation was manually set for each survey, being the distance at which the radial matching  distribution reached its first minimum. As seen in the main panel, $\sim$60\% of the selected spectroscopic sample is well accommodated within a 1-pixel distances. \textbf{Inset} panel illustrates the astrometric dispersion between ALHAMBRA and the overlapping surveys in terms of $\delta$RA (RA$^{ALH}$-RA$^{surv}$) and $\delta$Dec (Dec$^{ALH}$-Dec$^{surv}$) in units of pixel.}
\label{HSTmatching2D}
\end{center}
\end{figure}

\begin{figure}
\begin{center}
\includegraphics[width=9.cm]{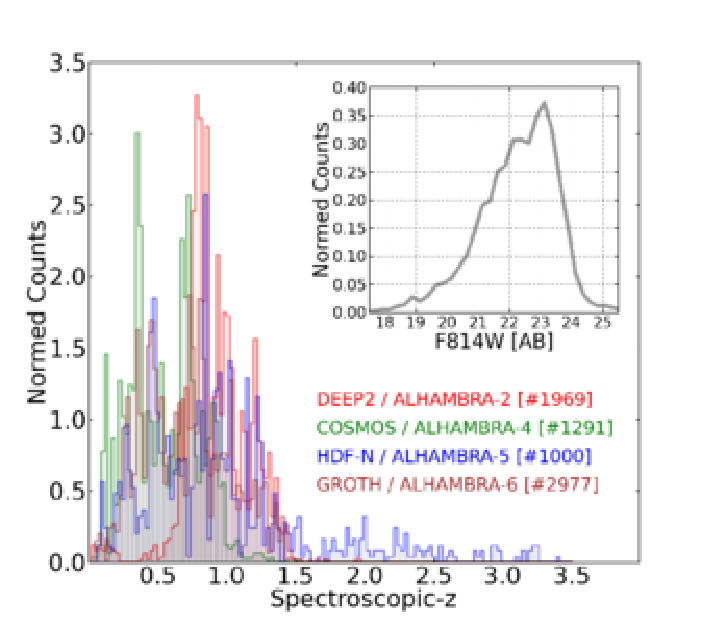}
\caption{Spectroscopic redshift compilation. Given the overlap between the ALHAMBRA fields and other existing spectroscopic surveys, we compiled a sample of $\sim7200$ galaxies with secure spectroscopic redshifts to quantify the accuracy for our photometric redshifts. Each survey contribution is color-coded (for visualization) as indicated in the legend. As seen in the figure, the compiled redshift sample mostly covering the ALHAMBRA parameter space, showing a redshift range 0$<z_{s}<$1.5 (with a mean redshift $<z_{s}>$$\sim$0.77) and a magnitude range (based on ALHAMBRA photometry) $18<m_{F814W}<25$ (with a mean magnitude $m_{F814W}\sim22.3$).}
\label{spzsample}
\end{center}
\end{figure}

  As explained in Section \ref{general}, ALHAMBRA was designed to partially overlap with fields observed by other surveys with extensive spectroscopic coverage. We compiled a sample of $\sim$7200 galaxies with spectroscopic redshifts from the publicly available data by selecting high quality (secure) objects with a good astrometric match to our data. The first condition is essential to accurately estimate of our outlier rate, even if it slightly biases our sample to brightest magnitudes, since as Fern\'andez-Soto et al. (2001) established, low quality spectroscopy can be much more unreliable than photo-z. To implement the second condition, we derived accurate astrometric corrections between samples (to avoid any offset) and then established a maximum matching distance of $\sim$3 pixels ($<$0.7"), as shown in Fig. \ref{HSTmatching2D}. This maximum separation was manually set for each survey being the distance at which the distribution of matching distances reached its first minimum. As seen in Fig. $\ref{spzsample}$, the compiled redshift sample shows a mean redshift $<z_{s}>\sim 0.77$ and a mean magnitude (based on ALHAMBRA photometry) $m_{F814W}\sim22.3$. In Table $\ref{tablespz}$ the contribution from each survey is specified, indicating the number of selected galaxies, the mean magnitude and redshift.

\vspace{0.2cm}
 
\begin{table}
\caption{{\small Spectroscopic Redshift samples.}}
\begin{center}
\label{tablespz}
\begin{tabular}{|l|c|c|c|c|c|c|c|}
\hline
\hline
\#    &   Survey  &  Reference  & $<m_{F814W}>$ & $<$z$>$ \\
\hline
1269       &   DEEP-2       &    Koo et al. 1995         &   22.64    &  0.92  \\
1291       &   COSMOS    &    Lilly et al. 2009          &   21.36    &  0.54   \\
1000       &   GOODS-N   &   Cooper et al. 2011     &   22.75    &  0.83  \\
2977       &   GROTH       &   Demian et al. 2011     &   22.21    &  0.70    \\
\hline
7237       &                       &                                      & 22.24      & 0.75 \\ 
\hline
\hline
\end{tabular}
\end{center}
\end{table}

\vspace{0.2cm}

\begin{figure*}
\begin{center}
\includegraphics[width=8.75cm]{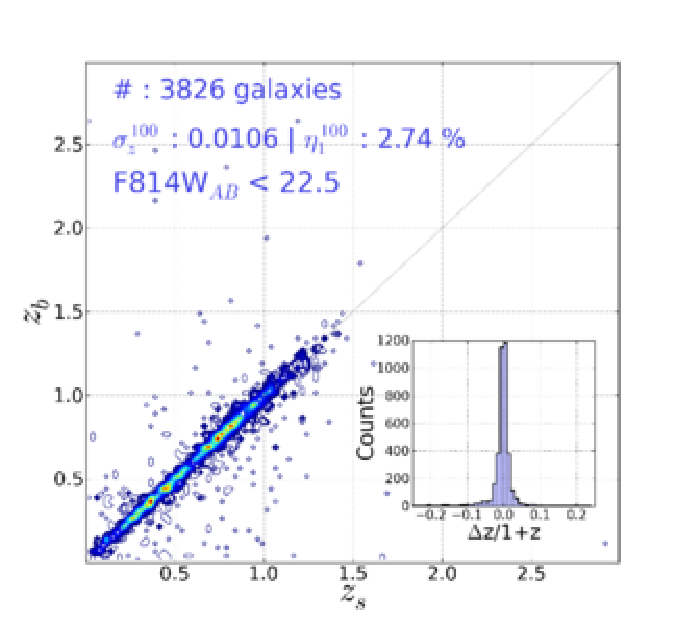}
\includegraphics[width=8.75cm]{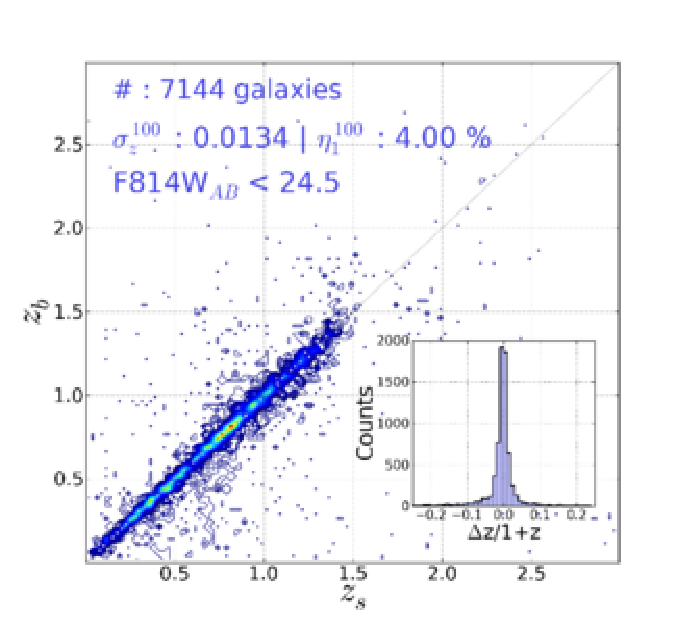}
\caption{Photometric redshift accuracy. The figures show the comparison between the ALHAMBRA photometric redshifts z$_{b}$ and the spectroscopic redshifts z$_{s}$ along with the error distribution $\Delta_{z}$/(1+z), for two different 
magnitude ranges. The  left plot shows the accuracy obtained for the bright sample ($m_{F814W}<22.5$) with a $\sigma_{z}$$<$0.0106 and a fraction of  catastrophic outliers $\eta_{1}$$\sim$2.7\%, the right plot shows a $\sigma_{z}$$<$0.0134 and a fraction of catastrophic  outliers $\eta_{1}$$\sim$4.0\% when including the entire sample. In both cases, the fraction of catastrophic outliers  (defined in Section \ref{photozaccuracy}) dramatically decreases when selecting galaxies with higher $Odds$, as indicated in Table \ref{dzvsoddstable}.}
\label{phzaccuracy}
\end{center}
\end{figure*}

\begin{figure*}
\begin{center}
\includegraphics[width=8.5cm]{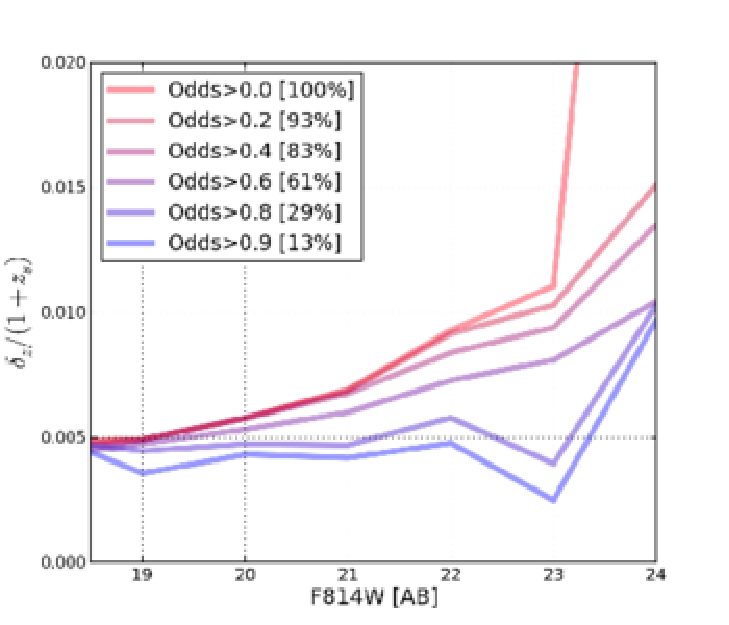}
\includegraphics[width=8.5cm]{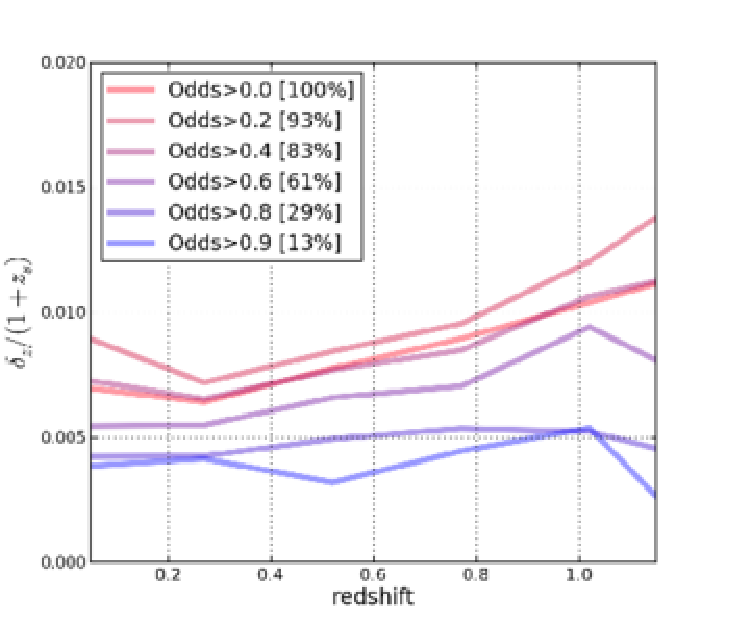}
\caption{Photometric redshift accuracy as a function of apparent magnitude F814W (left panel) and spectroscopic redshift (right panel). We explored the expected accuracy for our photometric redshifts in terms of a specific magnitude range and redshift range applying different $Odds$ intervals.}
\label{phzaccuracymz}
\end{center}
\end{figure*}

 As seen in Fig. \ref{phzaccuracy}, when compared with the spectroscopic sample, our photometric redshift estimations show a dispersion $\sigma_{z}$ = 0.0106 for $m_{F814W}<22.5$ with a fraction of catastrophic outliers $\eta_{1}$$\sim$2.7\%. For fainter magnitudes $m_{F814W}<24.5$ the accuracy observed is $\sigma_{z}$ = 0.0134 and the fraction of catastrophic outliers $\eta_{1}$$\sim$4.0\%. The fraction of catastrophic outliers dramatically decreases when selecting a more restricted sample (excluding X-ray emitters, AGNs or detections observed in only few bands). In addition, the photo-z error and the fraction of catastrophic outliers rapidly decreases as the $Odds$ interval increases. We show the expected accuracy for the photometric redshifts as a function of redshift, F814W magnitude and $Odds$ range in Fig. \ref{phzaccuracymz}. A more detailed analysis can be found in Tables \ref{dzvsztable} and \ref{dzvsmagtable}, respectively.  

\begin{table*}
\caption{{\small Photometric redshift quality vs spectroscopic redshifts.}}
\begin{center}
\label{dzvsztable}
\begin{tabular}{|l|c|c|c|c|c|c|c|c|c|c|c|c|c|c|c|c|c|c|c|}
\hline
\hline
Spectroscopic & $\sigma_{z}$    &  \#     &  $\eta_{1}$ & $\eta_{2}$ & $\sigma_{z}$  & \# & $\eta_{1}$ & $\eta_{2}$ & $\sigma_{z}$ & \# & $\eta_{1}$ & $\eta_{2}$  \\ 
   Redshift        & (Odds$>$0.0) & (\%)  &   (\%)         &  (\%)           & (Odds$>$0.5) & (\%)  &  (\%) &    (\%)   & (Odds$>$0.9) & (\%)  &    (\%)      &   (\%)          \\
\hline
0.00 $<$ z $<$ 0.25 & 0.0115 & 10.8    &  0.4  & 0.8  & 0.0086  &  3.8    &  0.1   &  0.2 & 0.0056  &  0.3   & 0.0  & 0.0 & \\ 
0.25 $<$ z $<$ 0.50 & 0.0101 & 21.0   &  0.6  & 1.5  & 0.0087  &  10.5  &  0.2  &  0.4 & 0.0062  &  1.6  & 0.0  & 0.1 & \\ 
0.50 $<$ z $<$ 0.75 & 0.0136 & 19.6   &  1.0  & 2.4  & 0.0107  &  11.5  &  0.3  &  0.8 & 0.0061  &  1.2  & 0.1  & 0.1 & \\ 
0.75 $<$ z $<$ 1.00 & 0.0135 & 21.4   &  0.7  & 2.3  & 0.0104  &  12.7  &  0.2  &  0.7 & 0.0066  &  1.4  & 0.0  & 0.1 & \\ 
1.00 $<$ z $<$ 1.25 & 0.0171 & 13.2   &  0.4  & 1.3  & 0.0125  &  7.0    &  0.1  &  0.4 & 0.0070  &  0.3  & 0.0  & 0.0 & \\ 
1.25 $<$ z $<$ 1.50 & 0.0194 & 7.2    &  0.3  & 0.9  & 0.0132  &  2.3     &  0.1    &  0.3  &    ---    &  ---  & ---  & --- & \\ 
1.50 $<$ z $<$ 1.75 & 0.0988 & 3.4    &  0.1  & 0.2  & 0.0567  &  0.3     &  0.0    &  0.0  &    ---     &  ---  & ---  & ---  & \\ 
1.75 $<$ z $<$ 2.00 & 0.1078 & 1.4    &  0.2  & 0.4  & 0.1620  &  0.0     &  0.0    &  0.1  &    ---     &  ---  & ---  & ---  & \\ 
\hline
\hline
\end{tabular}
\end{center}
\end{table*}

\begin{table*}
\caption{{\small Photometric redshift quality vs F814W magnitude.}}
\begin{center}
\label{dzvsmagtable}
\begin{tabular}{|l|c|c|c|c|clclclclc|c|c|clc|c|c|clc|c|c|clc|c|c|}
\hline
\hline
Magnitude  & $\sigma_{z}$    &  \#     &  $\eta_{1}$ & $\eta_{2}$ & $\sigma_{z}$  & \# & $\eta_{1}$ & $\eta_{2}$ & $\sigma_{z}$ & \# & $\eta_{1}$ & $\eta_{2}$  \\ 
 F814W           & (Odds$>$0.0) & (\%)  &   (\%)          &   (\%)         & (Odds$>$0.5) & (\%)  &   (\%)   &   (\%)       & (Odds$>$0.9) & (\%)  & (\%)   &      (\%)       \\
\hline
18.0 $<$ m $<$ 19.0 & 0.0081  & 0.8   & 0.0  & 0.1 & 0.0073 & 0.6 & 0.0   & 0.0 & 0.0055 & 0.1 & 0.0 & 0.0 \\
19.0 $<$ m $<$ 20.0 & 0.0083  & 2.2   & 0.1  & 0.3 & 0.0077 & 1.7 & 0.1   & 0.1 & 0.0056 & 0.3 & 0.1 & 0.1 \\ 
20.0 $<$ m $<$ 21.0 & 0.0095  & 5.3   & 0.3  & 0.7 & 0.0085 & 4.1 & 0.1   & 0.3 & 0.0059 & 0.7 & 0.0 & 0.0 \\ 
21.0 $<$ m $<$ 22.0 & 0.0101  & 11.9 & 0.4  & 1.1 & 0.0093 & 9.0 & 0.2 & 0.5 & 0.0058   & 1.3 & 0.0 & 0.0 \\ 
22.0 $<$ m $<$ 23.0 & 0.0140  & 26.0 & 0.7  & 2.1 & 0.0111 & 16.0 & 0.3 & 0.9 & 0.0065  & 1.5 & 0.0 & 0.0 \\ 
23.0 $<$ m $<$ 23.5 & 0.0182  & 22.8 & 0.6  & 2.1 & 0.0129 & 9.4 & 0.2 & 0.6 & 0.0045   & 0.5 & 0.0 & 0.0 \\ 
23.5 $<$ m $<$ 24.0 & 0.0263  & 30.7  & 0.9  & 2.3  & 0.0118 & 7.4 & 0.2 & 0.4 & 0.0038 & 0.3 & 0.0 & 0.0 \\ 
\hline
\hline
\end{tabular}
\end{center}
\end{table*} 

\vspace{0.2cm}

\begin{table}
\caption{{\small Photometric redshift accuracy vs Odds for the global sample.}}
\begin{center}
\label{dzvsoddstable}
\begin{tabular}{|l|c|c|c|c|clclclclc|c|c|}
\hline
\hline
Interval  &  Sample$^{1}$  &  $\sigma_{z}$  &  $\eta_{1}$  & $\eta_{2}$  \\ 
             &   (\%)        &                        &         (\%)      &  (\%)       \\
\hline
Odds $>$ 0.00	&   1.00   &  0.0137  & 3.04  &	8.64  \\ 
Odds $>$ 0.10	&   0.91   &  0.0131  & 2.56  &	7.36  \\ 
Odds $>$ 0.20	&   0.80   &  0.0123  & 2.13  &	6.01  \\
Odds $>$ 0.30	&   0.71   &  0.0116  & 1.71  &	5.00  \\ 
Odds $>$ 0.40	&   0.61   &  0.0109  & 1.46  &	4.09  \\
Odds $>$ 0.50	&   0.50   &  0.0102  & 1.08  &	2.89  \\
Odds $>$ 0.60	&   0.36   &  0.0093  & 0.72  &	1.86  \\
Odds $>$ 0.70	&   0.24   &  0.0082  & 0.51  &	1.13  \\
Odds $>$ 0.80	&   0.14   &  0.0069  & 0.30  &	0.57  \\
Odds $>$ 0.90	&   0.07   &  0.0062  & 0.15  &	0.23  \\
Odds $>$ 0.95	&   0.03   &  0.0057  & 0.08  &	0.14  \\
\hline
\hline
$^{1}m_{F814W}<24$ & & & & \\
\end{tabular}
\end{center}
\end{table}

\begin{figure}
\begin{center}
\includegraphics[width=9.0cm]{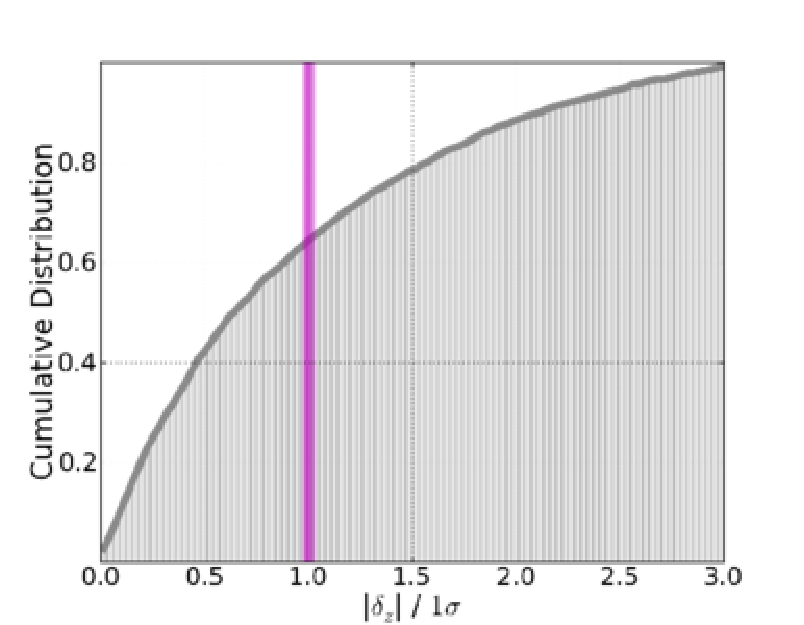}
\caption{Cumulative Distribution of the ratio $|$$\delta$z$|$/1$\sigma$. We observed that $\sim$64\% and $\sim$90\% of the photometric redshifts are well fitted within the formal 1$\sigma$ (magenta vertical line) and 2$\sigma$ confidence interval, respectively. This indicates that the photometric redshift uncertainties have been reliably established.}
\label{cumul1sigma}
\end{center}
\end{figure}

In order to verify that $\delta_{z}$/(1+z$_{s}$) is representative for the spectroscopic sample, the cumulative distribution of sources is represented in Fig. \ref{cumul1sigma}. We observed that $\sim$64\% and $\sim$90\% of the photometric redshifts are well fitted within the formal $1\sigma$ and $2\sigma$ confidence interval, respectively. This indicates that the uncertainties on $z_{b}$ are quite realistic. 

\vspace{0.2cm}

\begin{figure*}
\begin{center}
\includegraphics[width=8.5cm]{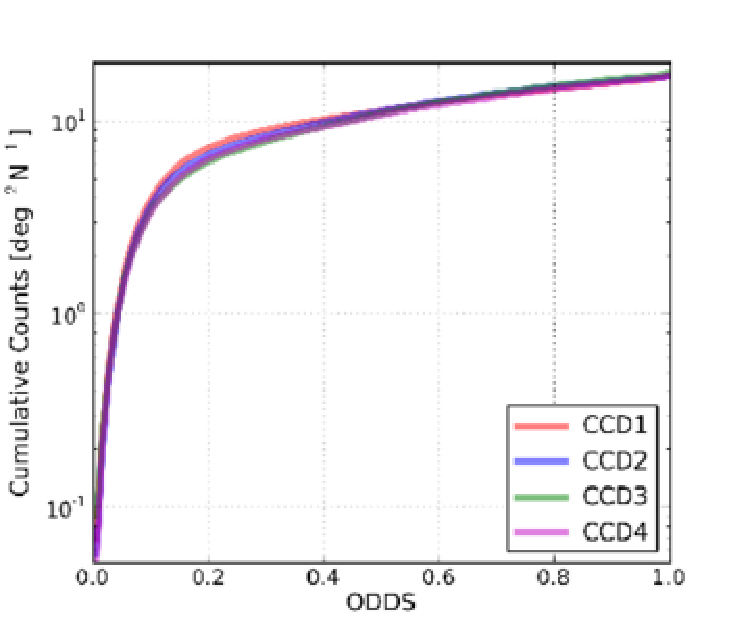}
\includegraphics[width=8.5cm]{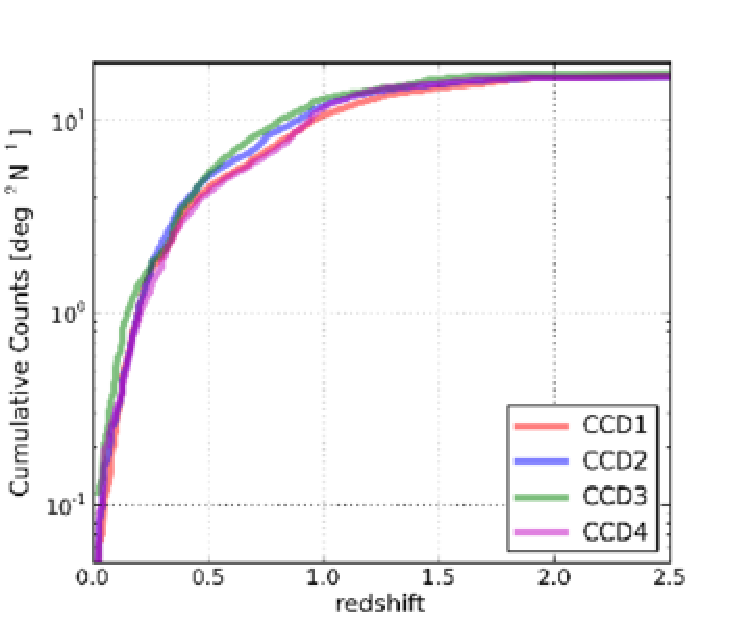}
\caption{Internal photometric redshift checks. Following the same approach as explained in Section \ref{phintchecks}, we systematically compared the $Odds$ (left panel) and photometric redshift $z_{b}$ (right panel) distributions among contiguous CCDs. The statistical results were consistent between each other with a scatter within the expected intrinsic variance for the sample imaged by each detector. These test served to ascertain the homogeneity within the different fields.}
\label{odds4distrib}
\end{center}
\end{figure*}

 Applying the same approach explained in Section \ref{phintchecks}, we performed internal photometric redshifts checks to compare our results among contiguous CCDs. As illustrated in Fig. \ref{odds4distrib}, the statistical results were consistent between CCDs, showing a scatter within the intrinsic variance for the sample observed by each CCD.  

\subsection{Photometric zeropoint recalibration.}
\subsubsection{Photometric ZP calibrations using spectroscopic redshifts.}
\label{PZPR}

  As it was shown in Coe et al. (2006), by comparing the observed colors of galaxies with spectroscopic redshifts against those expected from an empirically defined photo-z library, it is possible to calibrate photometric zeropoints to within a few percentage, similar or better than the accuracy reached by standard, stellar-based calibration techniques. This capability has been included in the BPZ software package from its initial release (Ben\'itez 2000) and has been applied successfully to several datasets (Capak et al. 2008, Hildebrandt, Wolf, \& Ben\'itez 2008).  

\begin{figure*}
\begin{center}
\includegraphics[width=8.8cm]{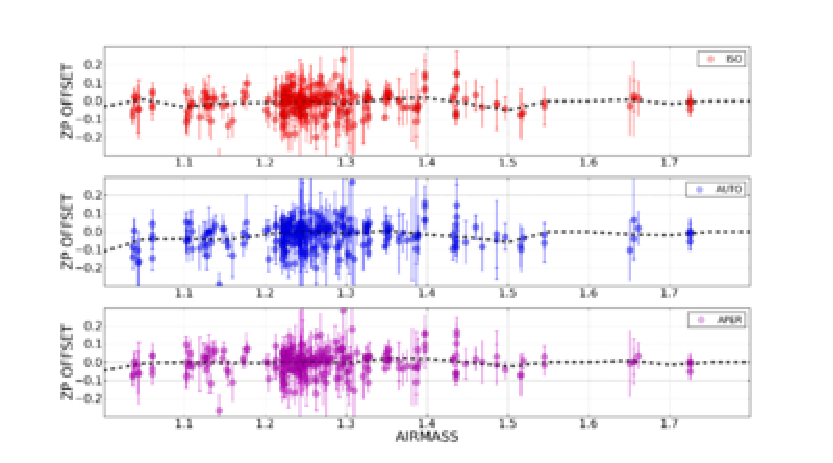}
\includegraphics[width=8.8cm]{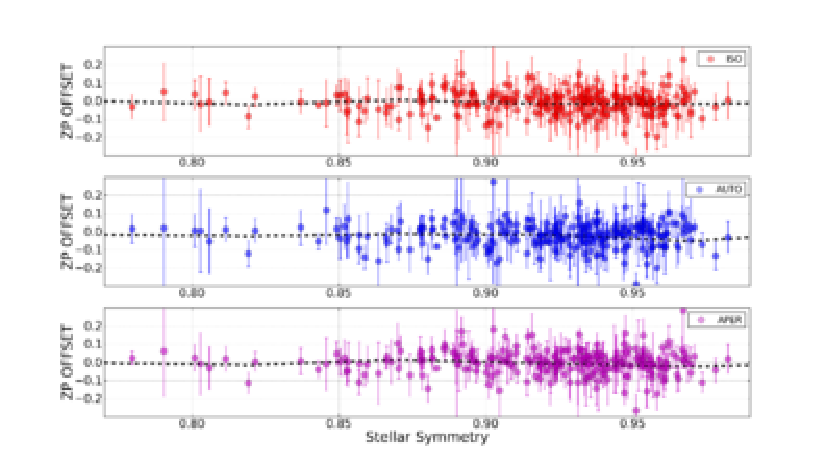} 
\includegraphics[width=8.8cm]{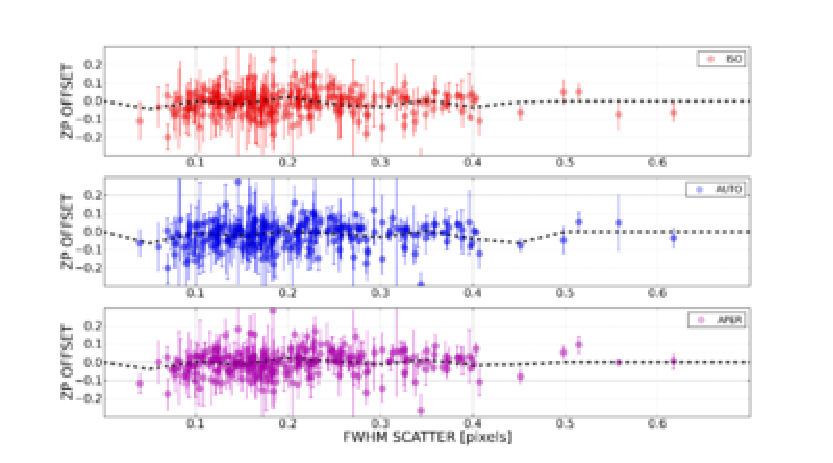}
\includegraphics[width=8.8cm]{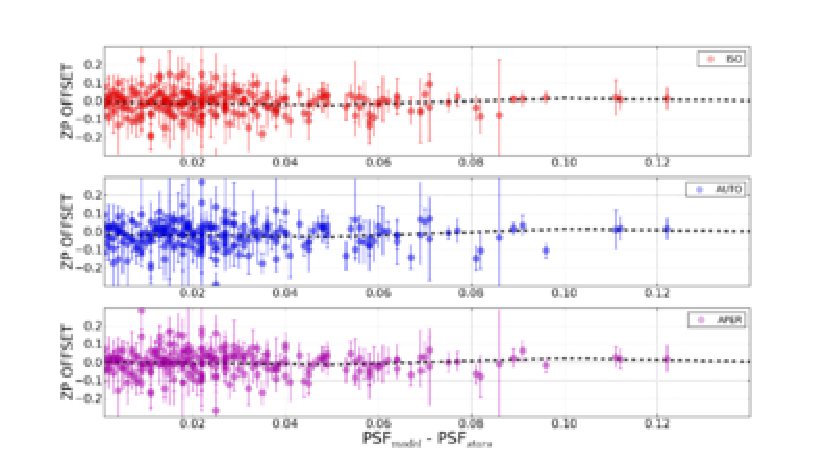}
\caption{Photometric zeropoint validations. We studied the source of the photometric zeropoint corrections (derived using SED-fitting algorithms) by comparing these quantities with several observational variables. Considering the possibility of a systematic effect during the data reduction, we represented globally the corrections for the $\sim$1100 individual images as a function of the AIRMASS (top left panel), the Stellar Symmetry  (top right panel), the FWHM Scatter  (bottom left panel) \& the differences between PSFs-models and stars (top left panel). The procedure was repeated using three different photometric apertures ($SExtractor\_ISOphotal$ as red circles, $SExtractor\_AUTO$ as blue circles and $SExtractor\_APER$ (3") as magenta circles) to discard any systematic effect due to the galaxy sampling regions. As indicated by the mean value of the distributions (dashed black lines), no clear correlations were observed, with fluctuations smaller than 1\%}.
\label{zpoffairmass}
\end{center}
\end{figure*}

\vspace{0.2cm}

 To calibrate the ALHAMBRA zeropoints, we followed this procedure for each individual CCD. First, we selected the spectroscopic redshift galaxies detected in all the 24 bands with a $S/N>10$, and chose the BPZ template which best fits its colors at their redshift. We then calculated the ratios between the fluxes predicted in each band by the templates and those observed; the median ratio, which converted to a magnitude represents the zeropoint offset (ZPO) required to match the observed magnitudes to the expected ones. We then corrected the fluxes by this value and iterated until the process  converged and the calculated correction was below $1\%$ in all the filters. Since all these changes are relative by nature, the synthetic F814W images were taken as anchor of the whole system. 

\vspace{0.2cm}

 Another useful quantity calculated by BPZ is the excess photometric scatter over the expected photometric error, what we call zeropoint error (ZPE). This noise excess could be due to two sources: a systematic mismatch between the templates and real galaxy colors, or a systematic error in the photometry. As Ben\'\i tez (2014) show, if enough spectroscopic redshifts are present, averaging over many templates and different SED rest frame locations, ensures that such residuals are typically due to flaws in the photometry. Including this factor allows for a much more realistic estimate of the error and significantly improves the photo-z precision. 
 
\vspace{0.2cm}

  We explored the dependence of the amplitude of these zeropoint corrections on several observational variables. Looking for possible systematic effects in the reduction, in Fig. $\ref{zpoffairmass}$ we plot globally the zeropoint corrections for the $\sim 1100$ individual images as a function of the airmass (top left panel),  Stellar Symmetry (top right panel; defined as the ratio of a/b parameters (Table $\ref{catalogdesc1}$), the FWHM Scatter  (bottom left panel) \& the  differences between PSFs-models and stars (top left panel). The procedure was repeated using three different photometric apertures ($SExtractor\_ISOphotal$ as red circles, $SExtractor\_AUTO$ as blue circles and $SExtractor\_APER$ (3") as magenta circles) to discard any dependence on the sampled area. As indicated by the mean value of the distributions (dashed black lines), we did not observe any clear correlations, with typical fluctuations smaller than 1\%.

\begin{figure}[h!]
\begin{center}
\includegraphics[width=9.2cm]{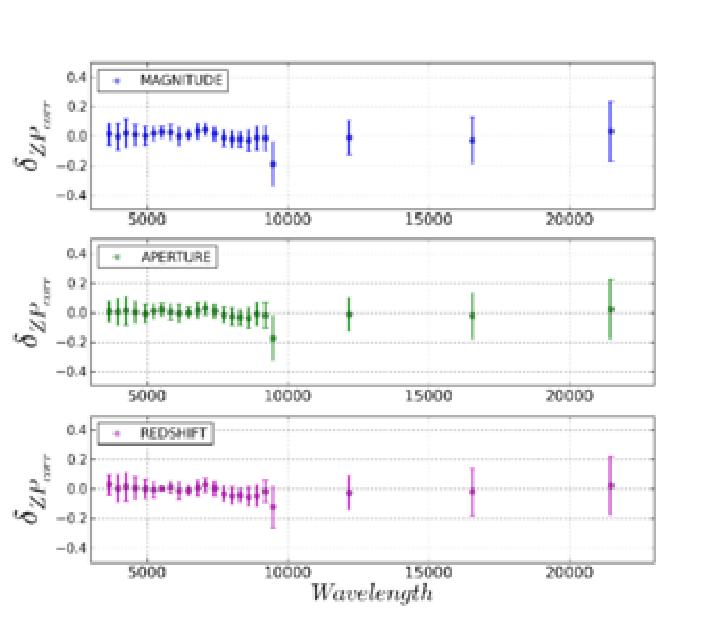}
\caption{Photometric zeropoint validations II. We also studied the robustness of the photometric zeropoint corrections using different samples of galaxies. We split the spectroscopic sample in equal-sized groups based on its magnitude ($m_{F814W}<22.5$ \& $m_{F814W}>22.5$), aperture size (area$<$125 pix \& area$>$125 pix) and redshift (z$<$0.81 \& z$>$0.81) and derive new photometric zeropoint corrections using BPZ. As observed in the figure, for all the three cases, the differences among samples ($\delta_{ZP}$) are quite small. This result shows that the zero-point corrections we derive do not strongly depend on the redshift range or spectral-type and therefore are indicative of true offsets in the zeropoints, most likely due to the differences between the calibrations obtained from traditional color transformations based on stars and the average colors of galaxies as defined by the BPZ template set, calibrated with HST observations.}
\label{zpcorrsMagAreaReds}
\end{center}
\end{figure}

 We explored whether zeropoint offsets depended on the magnitude. For that, we split the spectroscopic sample into two equal-sized groups with galaxies brighter and fainter than $m_{F814W}=22.5$. As seen in Fig. $\ref{zpcorrsMagAreaReds}$ (blue dots) the corrections derived for both samples are the same, within the typical level of photometric uncertainties. Even though filter F954W showed a clear disagreement among samples, its scatter was as large as $\sim$0.3 magnitudes indicating other sort of problems perhaps related to the reductions. To look for a dependence on the photometric aperture size, due to some effect related to the PSF corrections we again divided the spectroscopic sample into two equal-sized groups with photometric areas smaller (and larger, respectively) than 125 pixels. As seen in Fig. $\ref{zpcorrsMagAreaReds}$ (green dots) differences among samples were smaller than 1\%.

   Finally the dependence between redshift range and zeropoint offsets was also considered assuming a possible effect due to evolution in the galaxy populations (since the BPZ templates do not include any evolution). As seen in Fig. $\ref{zpcorrsMagAreaReds}$ (magenta dots) the differences obtained from both samples were smaller than $1\%$ and so within the error bars regime. We therefore conclude that the zeropoint offsets do not depend on the photometric treatment and represent real offsets between the zero-points defined by the colors of the BPZ templates, calibrated with the FIREWORKS HST photometry and those defined by the stellar-based color calibrations used for the primary ALHAMBRA photometric calibration. 

\subsubsection{Photometric ZP calibrations using photometric redshifts.}
\label{ZPCphz}

  Although ALHAMBRA was designed to overlap with other spectroscopic surveys, only $\sim$40\% of its fields had enough galaxies with spectroscopic redshifts to derive zeropoint corrections, as described above. As discussed in Section $\ref{PZPR}$ the absence of any clear dependence on the observational parameters  made unfeasible any extrapolations among different fields. Given the obvious improvement resulting from the zeropoint corrections, this lack of calibration spectroscopy created a serious sources of inhomogeneity across the survey. 

  We realized that the photometric redshifts obtained for emission line galaxies were quite robust to changes in the zeropoint calibration and therefore we could statistically treat those redshifts as spectroscopic for calibration purposes, obtaining an automatic source of zeropoint corrections for all our fields. Thus, we ran BPZ on the photometric catalogs with the original, stellar-based zeropoint estimations. Then we selected a sample formed by those galaxies observed in all 24 filters, large S/N ($m_{F814W}<23.0$), good fit to the SED ($Odds$$\ge$0.9 and $\chi^{2}$$\le$1) and classified by BPZ.2 as late-type galaxies ($tb$$>$7). We apply the procedure described in Section  \ref{PZPR}, using the photometric redshifts as spectroscopic values and iterating until convergence was reached. This is basically equivalent to calibrating the ZP using the continuum of the ELGs as defined by the BPZ.2 templates. 

\vspace{0.2cm}

\begin{figure*}
\begin{center}
\includegraphics[width=8.0cm]{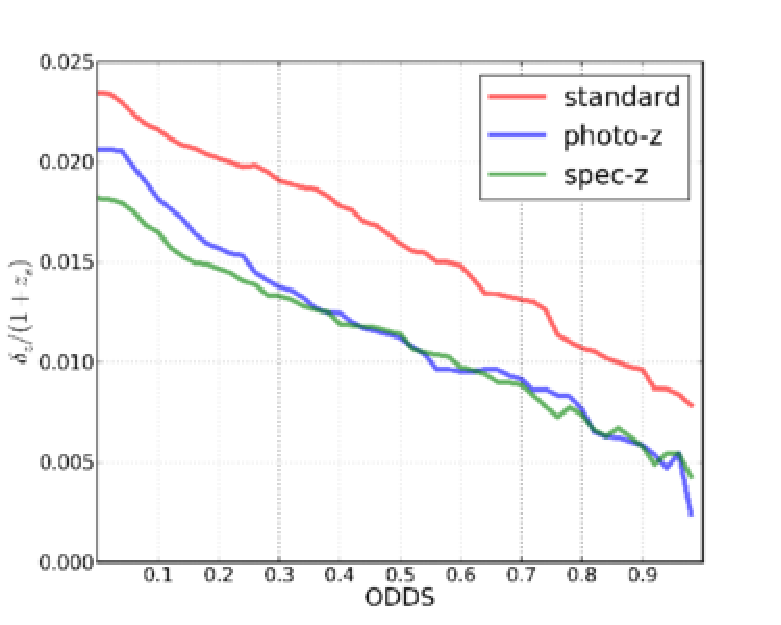}
\includegraphics[width=8.0cm]{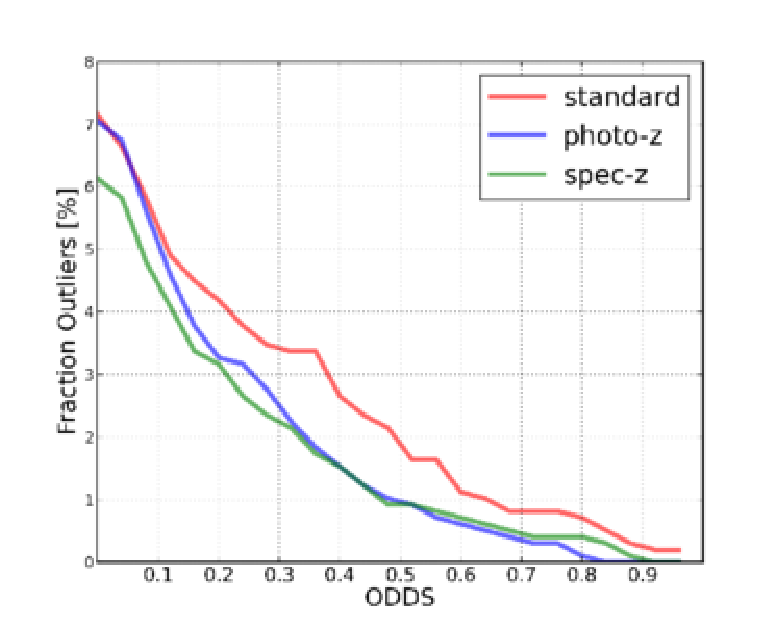}
\includegraphics[width=8.cm]{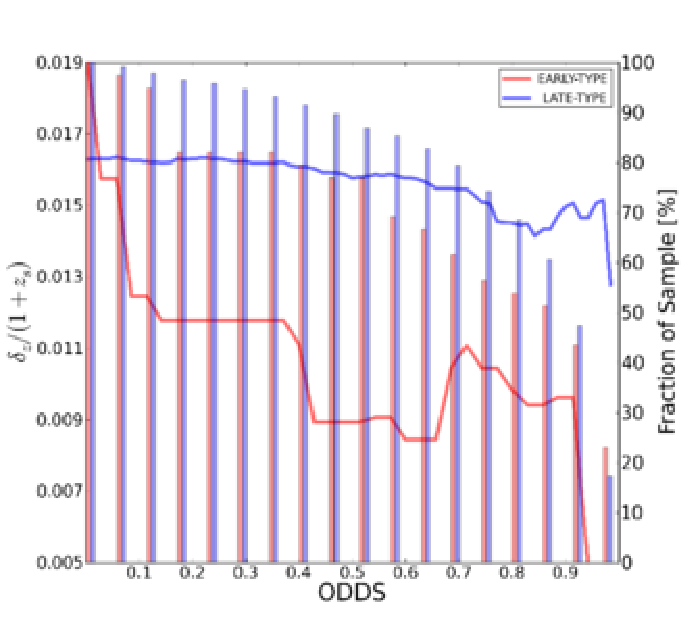}
\includegraphics[width=8.cm]{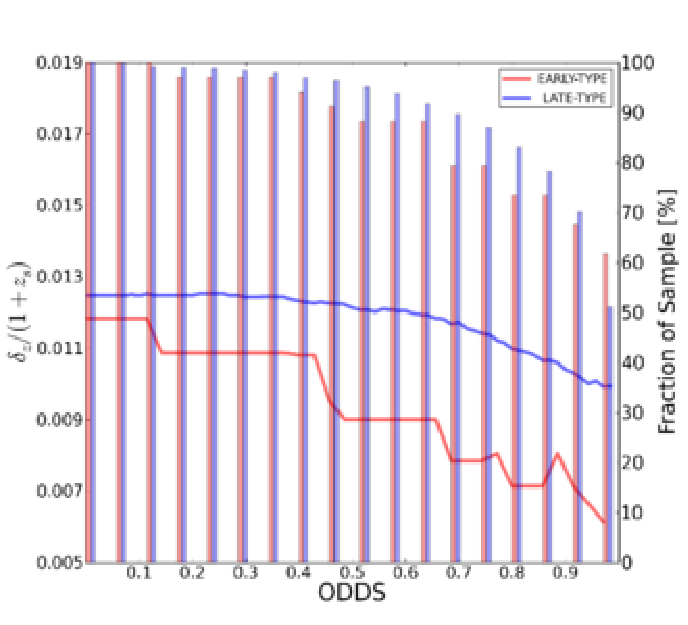}
\caption{We compare the spectroscopic redshifts with photo-z obtained after 
applying zero-point corrections with three different approaches: using the standard stellar-based method (red line), using photometric redshifts derived from emission line galaxies (blue line) and using a spectroscopic redshift sample (green line). As seen in the top left panel, photometric redshifts using emission line galaxies not only are a vast improvement with respect to the stellar-based method but also get quite close, in terms of accuracy, to the level of precision 
achieved by using the highest quality spectroscopic redshifts ($Odds$$>$0.3). In addition, the fraction of catastrophic outliers with high Odds was also significantly reduced, as shown in the  top right panel.  In the two bottoms panels we compare how the zero-point corrections derived with
the emission line galaxies (late-type) affect both main galaxy types. We show both the dependence of both the accuracy (left scale) and the outlier rate (right scale) before (left bottom panel) and after (bottom right panel) applying the zero-point corrections. Not only the accuracy for  the late-type galaxies (solid blue line) improved significantly with the corrections but also that of the early-type galaxies (solid red line). Meanwhile, the  fraction of galaxies per $Odds$ interval (vertical bars) increased homogeneously among spectral-types, indicating that the calibration also produced more galaxies with high quality photo-z-}
\label{zptcaltb}
\end{center}
\end{figure*}

 In the top panel of Fig. \ref{zptcaltb} we show the photometric redshift accuracy obtained with  three different calibration methods: the original zeropoints (red line), corrections derived from photometric redshifts (blue line) and corrections from spectroscopic redshifts (green line). The results indicate that the methodology presented here successfully improves the photometric redshifts accuracy almost up to the level provided by the spectroscopic sample, and it also dramatically reduces the fraction of catastrophic outliers. The bottom panel of the Fig. $\ref{zptcaltb}$ shows how the corrections derived with late-type galaxies worked very well for early-type galaxies, proving that they are independent of the particular choice of templates in the library. It is worth noting that even if the accuracy reached by this method was always slightly worse than that provided by a real spectroscopic sample, it was always much better than the standard stellar-based calibration. Therefore, we decided to apply this kind of zeropoint calibrations for all the fields without spectroscopy, what significantly improved the overall homogeneity of the ALHAMBRA sample. This kind of calibration promises to have wide application to future narrow-band surveys such as JPAS (Ben\'\i tez et al. 2009, Ben\'\i tez et al. 2014)

\subsection{Photometric Redshift Distributions}
\label{phzdepth}

  Despite having a relatively small FOV compared with other, much larger surveys, one of the main virtues of ALHAMBRA is that it includes 8 different lines of sight widely separated, which provides realistic estimation of both the typical redshift distribution of galaxies across cosmic time and its inherent variability (cosmic variance).

\vspace{0.2cm}   

  Photometric redshifts are probabilistic by nature, and the shape of the probability distribution is usually far from a well-behaved Gaussian. Therefore point estimates of the redshift and other parameters have a limited value, and it is much safer to work with the full probability distribution $p(z,T|C)$ (Ben\'itez 2000, Coe et al. 2006, Mandelbaum et al. 2008; Cunha et al. 2009; Wittman 2009; Bordoloi et al. 2010; Abrahamse et al. 2011; Sheldon et al. 2012). This is specially true for most faint galaxies, with noisy photometric information, where probability distributions usually become multimodal and completely asymmetric. Even in those cases, the distribution $p(z,T|C)$ obtained with a properly calibrated library and prior can be relied upon to produce accurate population properties like the redshift distribution. 

\vspace{0.2cm}

We therefore define the global photometric redshift distribution $P(z)$ as:

\begin{equation}
P(z)=\sum_{i=1}^{N_g} P_{i}(z) = \sum_{i=1}^{N_g} [\int dz \sum_T p_{i}(z,T|C)]
\end{equation}

\vspace{0.2cm}

where $p_i(z,T)$ represents the redshift/type probability distribution function for the $i^{th}$ galaxy. 

\begin{figure*}
\begin{center}
\includegraphics[width=8.5cm]{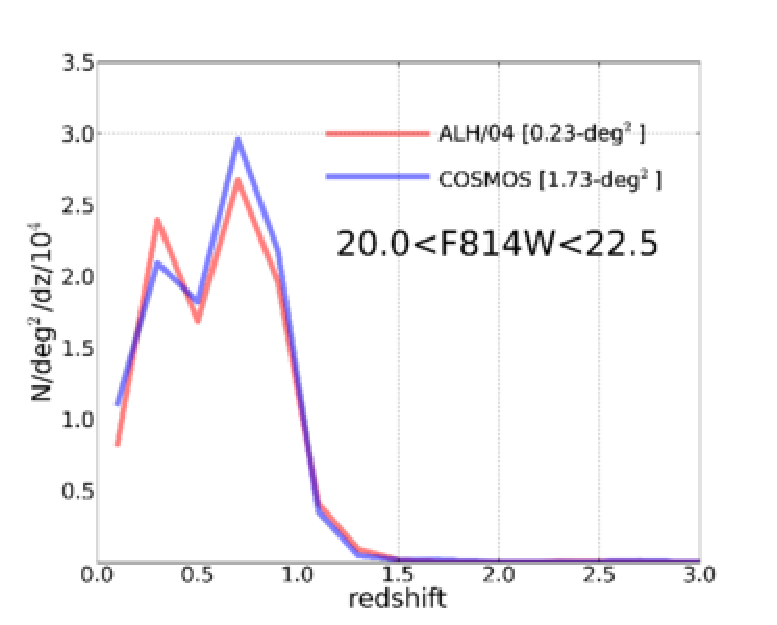}
\includegraphics[width=8.85cm]{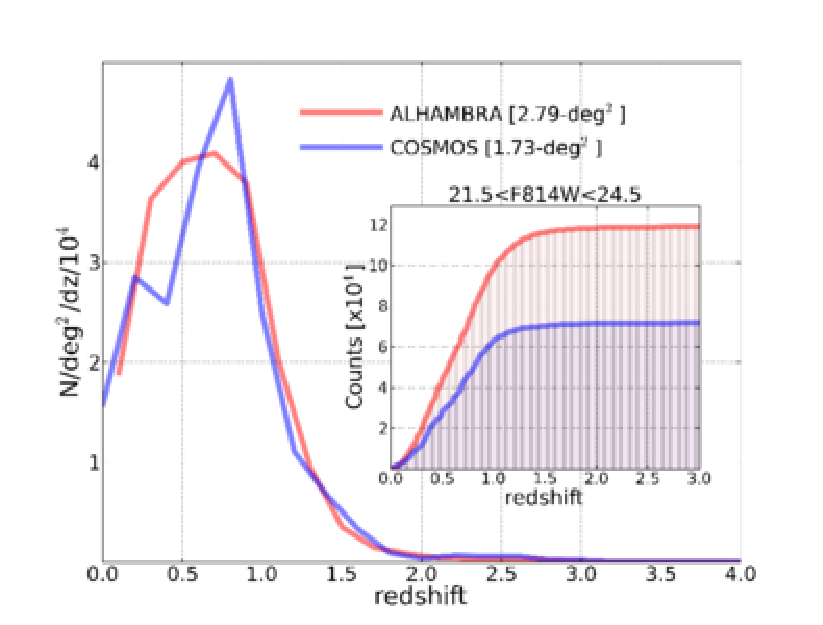}
\caption{Comparison of the redshift probability distribution function $P(z)$ for ALHAMBRA and the COSMOS field. We run BPZ on the Ilbert et al. (2009) catalog in the same way we did for our ALHAMBRA fields, and generate the corresponding $P(z)$ As seen in the left panel, the $P(z)$ for  the ALHAMBRA-04/COSMOS (red line) and COSMOS (blue line) are quite similar both distributions consistently reproduce a double peak at redshifts $z\sim0.3$ and $z\sim0.9$ The average redshifts are also similar, ALHAMBRA-4 has $\langle$$z$$\rangle$ = 0.60 for $m_{F814W}<22.5$ and $\langle$z$\rangle$ = 0.87 for $m_{F814W}<25.5$, whereas the COSMOS field shows a mean redshift $\langle$$z$$\rangle$ = 0.66 for $m_{F814W}<22.5$ and $\langle$$z$$\rangle$ = 0.96 for $m_{F814W}<25.5$. The global P(z) derived averaging the seven ALHAMBRA fields shows a mean redshift  $\langle$$z$$\rangle$ = 0.56 for $m_{F814W}<22.5$ and $\langle$$z$$\rangle$ = 0.85 for $m_{F814W}<25.5$, as seen in the right panel. It is obvious that the COSMOS field has a peculiar distribution, with a peak-through-peak structure with mimics a large evolution effect between $z=0.4$ and $z=0.9$}
\label{alh2cosPDZ}
\end{center}
\end{figure*}

\vspace{0.2cm}

\begin{figure*}
\begin{center}
\includegraphics[width=8.5cm]{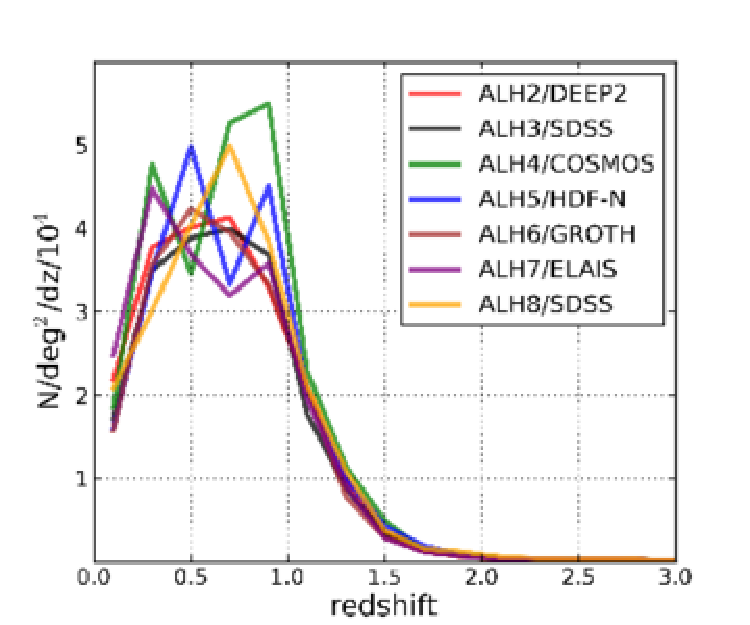}
\includegraphics[width=8.5cm]{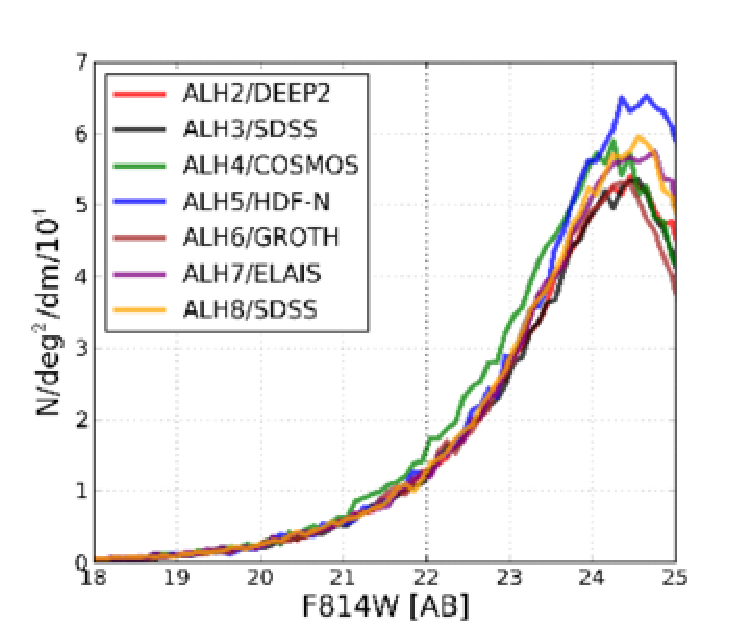}
\caption{Effect of the cosmic variance in the P(z). The left panel shows the redshift probability  distribution function $P(z)$ for all the seven ALHAMBRA fields, using a range in magnitudes between $19.0<m_{F814W}<23.5$. The different ALHAMBRA fields are color-coded as indicated in the legend. Once again the ALHAMBRA-04 field associated with the COSMOS fields (green line)  shows a peculiar distribution with a prominent peak at redshift z$\sim$0.86. The right panel shows the cumulative number counts for the seven fields. Again the ALHAMBRA-4 field (green line) shows a clear excess in the number of galaxies detected per magnitude range with respect to the other fields.}
\label{globalPDZ}
\end{center}
\end{figure*}

 We also ran BPZ on the photometric catalogue used by Ilbert et al. (2009) to derive the global redshift probability distribution function $P(z)$ for the COSMOS field and compare it consistently with our results. We first looked at the $P(z)$ derived using the ALHAMBRA-4/COSMOS data (red line) with the $P(z)$ derived using the COSMOS data (blue line), as seen in the left panel of Fig. \ref{alh2cosPDZ}, where both distributions consistently reproduce a double peak at redshifts z$\sim$0.3 and z$\sim$0.9, respectively. However, whereas the ALHAMBRA-4 field shows a mean redshift $\langle$$z$$\rangle$ = 0.60 for $m_{F814W}<22.5$ and $\langle$$z$$\rangle$ = 0.87 for $m_{F814W}<25.5$, the COSMOS field shows a mean redshift $\langle$$z$$\rangle$ = 0.66 for $m_{F814W}<22.5$ and $\langle$$z$$\rangle$ = 0.96 for $m_{F814W}<25.5$, as seen in the right panel of Fig. \ref{alh2cosPDZ}. Meanwhile the global photometric redshift distribution derived for all the seven  ALHAMBRA fields (excluding stars) shows a mean redshift $\langle$$z$$\rangle$ = 0.56 for $m_{F814W}<22.5$ and $\langle$$z$$\rangle$ = 0.85 for $m_{F814W}<25.5$, as seen in the right panel of Fig. \ref{globalPDZ}. This result indicates that as it is known, the COSMOS field shows a clear over-density with respect to the mean value derived averaging the seven ALHAMBRA fields. In fact, the average galaxy number in COSMOS goes up by a 60\% between z=0.4 and z=0.7, whereas no such effect is observed in our average. 

\vspace{0.2cm}

\begin{figure*}
\begin{center}
\includegraphics[width=8.5cm]{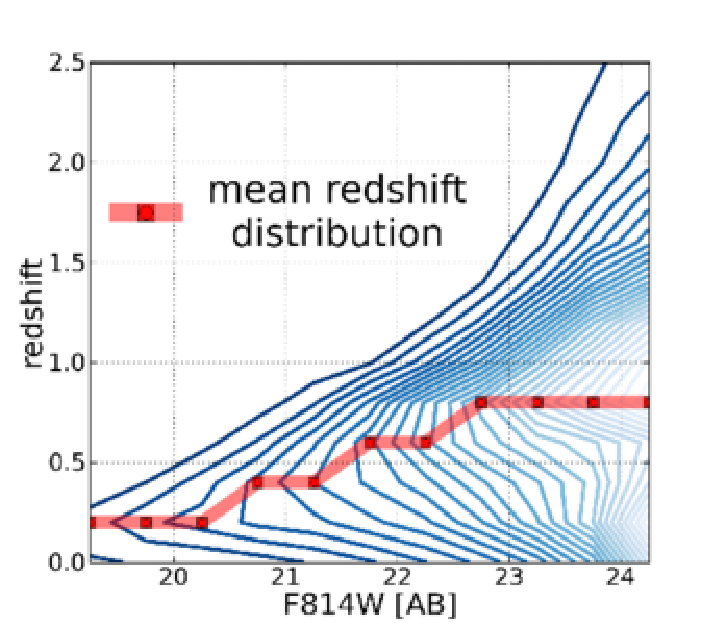}
\includegraphics[width=8.5cm]{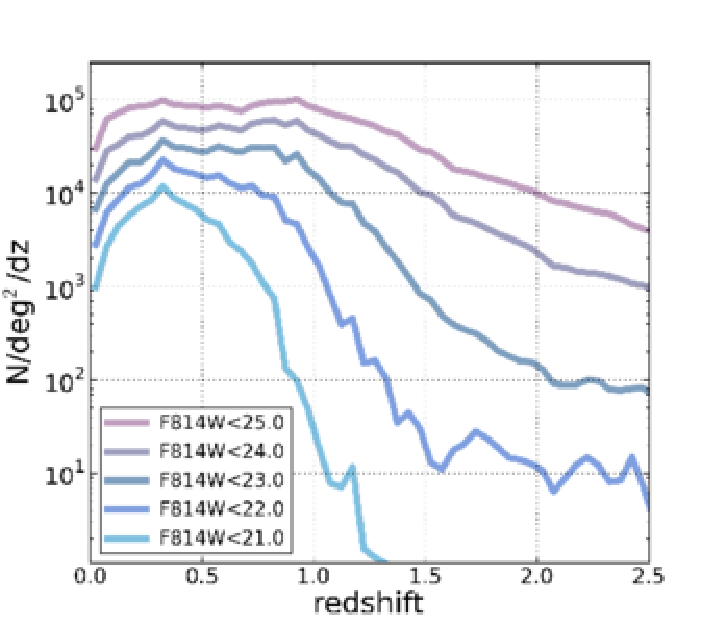}
\caption{Evolution of the redshift distribution. The left panel shows the evolution of the averaged redshift distribution for the ALHAMBRA fields, as a function of the magnitude F814W. The mean redshift distribution (solid red line) indicates a clear evolution moving from a $\langle$$z$$\rangle$$\sim$0.2 for $m_{F814W}<20.5$ to a $\langle$$z$$\rangle$$\sim$0.86 for $m_{F814W}>23.0$. The right panel shows the averaged distribution of galaxies for the ALHAMBRA fields, as a function of the redshift for different ranges in magnitude.}
\label{Pzvsmagfig}
\end{center}
\end{figure*}

  To study the evolution of the number counts as a function of the magnitude F814W and redshift, we derived the averaged redshift probability distribution function for the ALHAMBRA fields. As seen in the left panel of Fig. \ref{Pzvsmagfig}, the solid red line corresponding to the mean redshift  distribution (per bins of 0.5 mags) indicates a clear evolution moving from a  $\langle$$z$$\rangle$$\sim$0.2 for $m_{F814W}<20.5$ to a $\langle$$z$$\rangle$$\sim$0.8 for $m_{F814W}>23.0$. Inversely, the right panel of Fig. \ref{Pzvsmagfig} shows how the peak of the averaged distribution of galaxies increases as a function of the redshift for different magnitude ranges.

\begin{figure*}
\begin{center}
\includegraphics[width=18.cm]{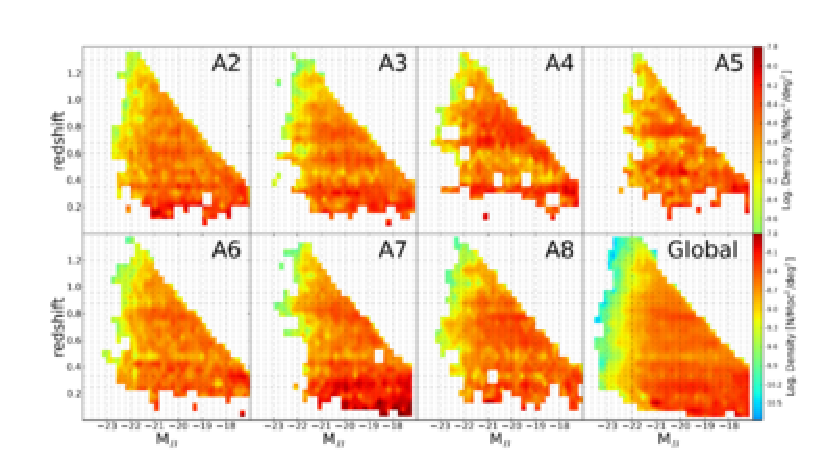}
\includegraphics[width=18.cm]{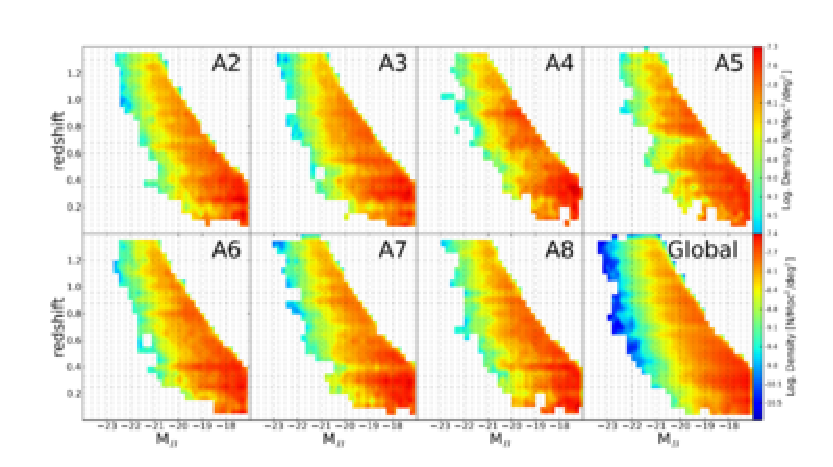}
\caption{Redshift distribution in the rest-frame per ALHAMBRA field. We explored the variance in the redshift-magnitude distribution of galaxies as a function of the absolute B magnitude and the spectral-type. After splitting the sample among early-type galaxies (top panel, defined as 1$<T_{b}<$5) and late-type galaxies (bottom panel, defined as 7$<T_{b}<$11), we compared the resulting distributions among each of the 7 individual ALHAMBRA fields (A$i$) + 1 averaged sample ($Global$). As observed from the panels, where the logarithmic density is color-coded, whereas each individual field shows clear and identifiable structures at different redshifts, the global samples show a more smooth distribution. In particular, we find that the well-known bimodal distribution in the COSMOS field (A4) is not systematically observed along the other fields, emphasizing the effects of cosmic variance on galaxy evolution studies.}
\label{redbluesequence}
\end{center}
\end{figure*}

  We explored the variance in the redshift-magnitude distribution of galaxies as a function of the absolute B magnitude and the spectral-type. As seen in Fig. \ref{redbluesequence}, we split the sample among early-type galaxies (top panel, defined as 1$<T_{b}<$5) and late-type galaxies (bottom panel, defined as 7$<T_{b}<$11). We plot the redshift distribution distributions for each of the 7 individual ALHAMBRA fields (A$i$) and 1 averaged ($Global$) sample. As observed from the figure, where the logarithmic density was color-coded, individual structures clearly seen in each individual field are smoothed over in the global distribution. In particular, the well-known bimodal distribution in the COSMOS field (A4) is not systematically observed along the other fields. 

\subsection{Photometric redshift depth}
\label{limmags}

  Due to the color/redshift degeneracies, it is possible to have galaxies which are detected at high $S/N$ in many filters but for which no unambiguous redshifts can be derived. One of the main practical ways of characterizing the effective completeness and depth of a photometric redshift catalog is by using the amount of galaxies with $Odds$ above a certain threshold, which basically tells us how many galaxies we can expect to have with meaningful, unambiguous photometric redshifts (Ben\'itez 2000, Ben\'itez 2009b).  

\vspace{0.2cm}

\begin{figure*}
\begin{center}
\includegraphics[width=8.6cm]{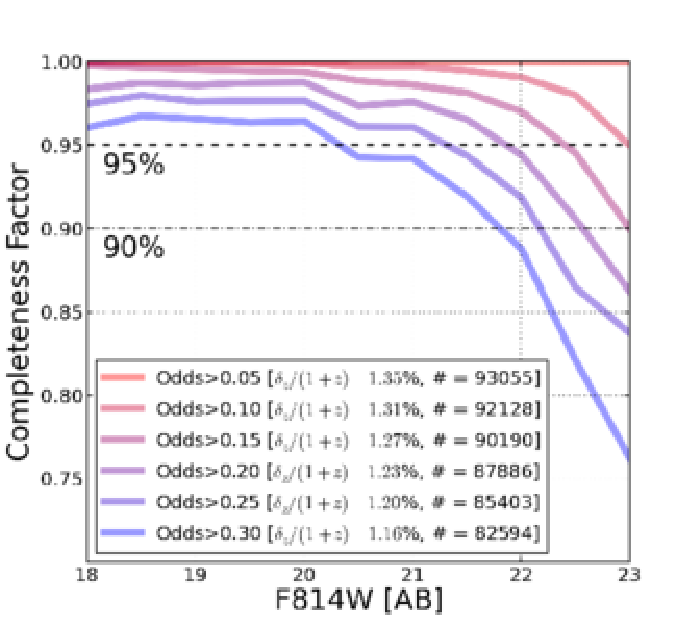}
\includegraphics[width=8.25cm]{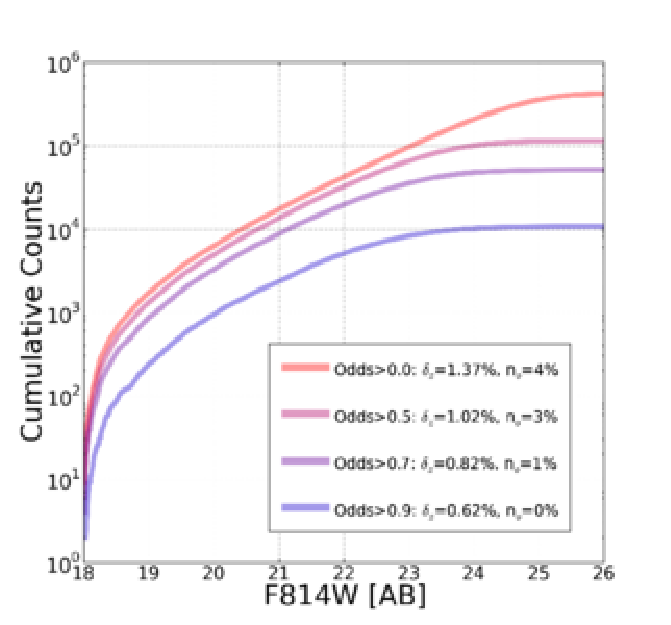}
\caption{Photometric Redshift Depth. In order to characterize the photometric redshift depth for the ALHAMBRA catalogues, we quantified the amount of galaxies per $Odds$ interval, which is equivalent to estimating the fraction and distribution of galaxies with a certain expected  photometric redshift accuracy. As seen in the left panel, we explored the expected Completeness Factor as a function of the magnitude F814W and $Odds$ interval. The total fraction of galaxies within each interval is specified in the legend. Similarly, the right panel shows the Cumulative distribution of galaxies as a function of the F814W magnitude for different $Odds$ intervals. The expected accuracy for the photometric redshifts $\delta_{z}$ and the fraction of catastrophic outliers $n_{o}$ (according to the spectroscopic sample) is indicated in the legend.}
\label{cnc}
\end{center}
\end{figure*}

  We therefore took into account the $Odds$ to carry out a set of analysis and evaluate the completeness and accuracy of our sample. For this, we set the interval to compute the $Odds$ parameter to $DZ$ = 2*0.0125*(1+z) since this quantity corresponds to twice the expected sigma. The completeness factor (fraction of galaxies per $Odds$ interval) as a function of F814W magnitude is presented in Fig. \ref{cnc}. 

\begin{figure}
\begin{center}
\includegraphics[width=8.5cm]{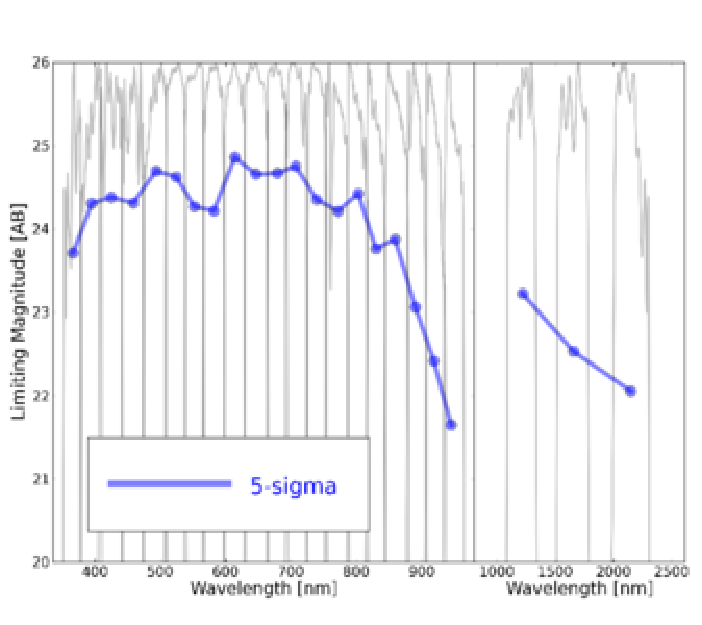}
\caption{Limiting magnitudes. We derived limiting magnitudes for every image, as they represent a very useful piece of information during the SED-fitting procedure. As required by BPZ, we replaced galaxies with measured fluxes equal or lower than the estimation of the background signal by an upper limit defined as 1-$\sigma$ above the background. Since limiting magnitudes depends on the photometric uncertainties, we computed limiting magnitudes after reestimating empirically the photometric errors. Meanwhile, we calculated the expected limiting magnitude using fixed apertures of 3" and 5-$\sigma$, as seen in the figure. These magnitudes correspond to the averaged values for the complete set of images.}
\label{maglim}
\end{center}
\end{figure}

 For sources detected only on the F814W detection image, an upper limit (defined as 1-$\sigma$ above the background) is provided. These limiting magnitudes represent the deepest magnitudes extractable from an image, providing useful information for the SED-fitting analysis. Limiting  magnitudes are applied whenever measured fluxes, inside a fixed aperture, are equal or lower than the estimation of the background signal. Since limiting magnitudes depends directly on photometric  errors, we computed limiting magnitudes after reestimating photometric errors, via empirical sigma estimation (Section \ref{photoerr}). Derived limiting magnitudes for each band can be found within the photometric catalogues. In Fig. \ref{maglim} we represent the averaged 5-$\sigma$ limiting magnitudes for all the 23 bands using fixed circular apertures of 3". 

\subsection{Emission-line galaxies}
\label{emlingal}

\begin{figure*}
\begin{center}
\includegraphics[width=9.cm]{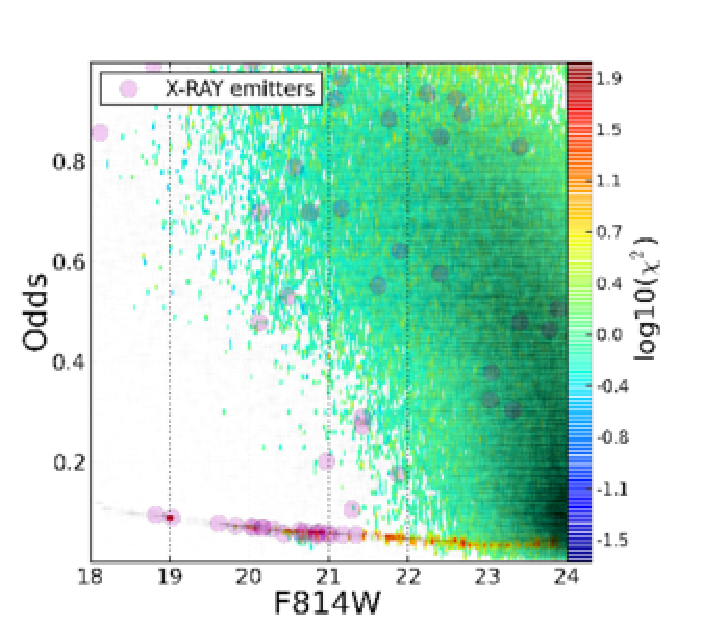}
\includegraphics[width=8.45cm]{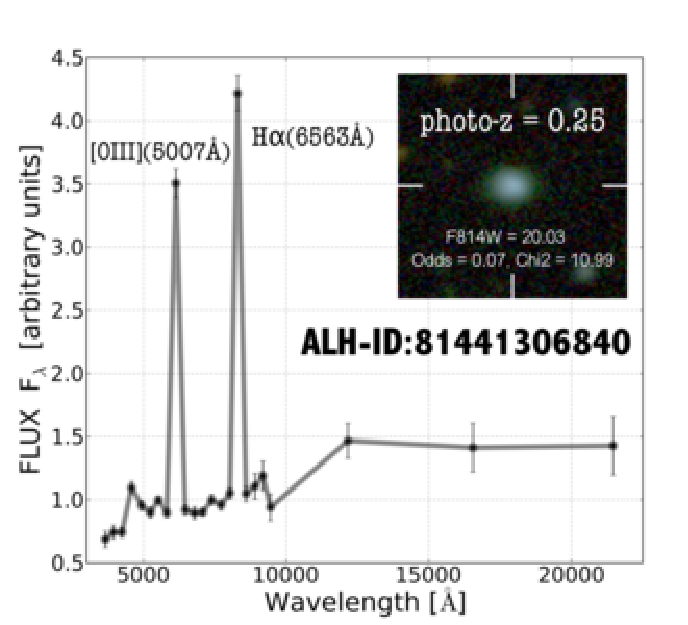}
\caption{Emission-line galaxies identification. The left panel shows the $Odds$ distribution as a function of magnitude F814W for the complete catalogue (stars excluded). As expected, there is a clear dependency between the $Odds$ and the magnitude, indicating how the photometric redshifts confidence decreases with the S/N. We observed an unexpected locus for magnitudes in between $18<m_{F814W}<24$ and $Odds$ in between 0.0$<$$Odds$$<$0.1. When plotting the logarithmic $\chi^{2}$ distribution, it revealed those detections to have extremely high $\chi^{2}$ values when SED-fitting the BPZ galaxy templates. We observed that those sources were mostly composed for very strong broad emission-line objects, AGNs and QSOs, galaxy types not included in the BPZ library of templates. The right panel shows an example of an intense emission-line galaxy within that horizontal sequence.}
\label{elg}
\end{center}
\end{figure*}

 When plotting the $Odds$ distribution as a function of F814W magnitude for all galaxies (Fig. \ref{elg}) we find a concentration of objects with low odds at magnitudes in between $18<m_{F814W}<23$ and 0.0$<$$Odds$$<$0.1. When plotting the logarithmic $\chi^{2}$ distribution for those objects, it immediately reveals those detections to have very high $\chi^{2}$ values (since BPZ.2 will automatically assing low odds to those objects) indicative of a poor fit, as the right panel in Fig. \ref{elg} illustrates.    

\vspace{0.2cm}

  Detections with unexpected poor $\chi^{2}$ fitting (given its magnitude) could be due to incorrect photometry or to an incomplete library of templates. After purging the sample from objects with high SExtractor flags, we saw that the remaining galaxies in that locus could be classified into  two different groups: 1. Unresolved stellar pairs (identified as a single detection by SExtractor) with clearly asymmetric morphologies (despite their photometric colors) and 2. Very strong broad emission-line objects. As mentioned in Section \ref{bpz}, neither AGN nor QSO templates were included in the BPZ library and therefore these sources might be expected to show poor fits to any BPZ template. In fact, this is a good way of finding active galaxies, as illustrated in the left panel of Fig. \ref{elg}. Given the high value of $\chi^{2}$, the new version of BPZ generates for them a redshift distribution which (in most cases) assigned by the prior  probability, favoring  solutions near the prior mode z$_{b}$$\sim$0.4 for  intermediate spectral-types $Tb$ (E0/Scd/Sbc). 

\section{PHOTOMETRIC CATALOGS}
\label{catalogs}

  As explained in Section \ref{f814wimages}, we have generated homogeneous F814W-based catalogs across the $\sim 3.0$deg$^{2}$ combining the available information from our 23 + 1 photometric filters. The full area is then divided in 48 individual catalogues containing the information listed in Appendix $\ref{catalogdesc1}$ \& $\ref{catalogdesc2}$.

\vspace{0.2cm}

 Unique IDs are given to every detection according to the following criteria: ID = 81442100119 stands for 814 (detection image) +  4 (field) + 2 (pointing) + 1 (CCD) + 00119 (ColorPro\_ID). Both astrometric and geometrical information is therefore derived from its corresponding F814W detection image: f04p02\_F814W\_1.swp.fits (following former example). SExtractor detection parameters were chosen differently for every detection image as discussed in Section \ref{SExconf}. Kron apertures (RK) and fraction-of-light radii (RF) were settled according to the aperture parameters defined in Table \ref{detconftable}. Total magnitudes and empirically corrected uncertainties (Section \ref{photoerr}) are given by all the 24 filters. Along with this, F814W\_3arc represent F814W magnitudes measured on a 3" circular apertures and ditto F814W\_3arc\_corr, but corrected to match COSMOS/F814W photometry (Section \ref{f814wimages}). As every detection in the ALHAMBRA fields was covered by all the 24 filters, \textit{nfd} indicates the number of filters a source was detected. Whenever a source was not detected, its magnitude was set to 99. and its photometric uncertainty replaced by a 1-${\sigma}$ upper limit (Section \ref{limmags}) suitable for BPZ. 

\vspace{0.2cm}

The final catalogues contain several quality flag. \textit{PhotoFlag} corresponds to the standard SExtractor photometric flag (Bertin \& Arnouts 1996), \textit{Satur\_Flag} indicates a possible saturated source (typically stars with magnitudes brighter than $m_{F814W}=16$), \textit{Stellar\_Flag} represents a source-by-source statistical classification among stars and galaxies (Section \ref{SG}), \textit{irms\_OPT\_Flag} and \textit{irms\_NIR\_Flag} indicates the number of optical and NIR bands a detection was observed with a normalized exposure time below its 80\%, respectively. The \textit{DupliDet\_Flag} indicates either a source was detected twice (within the overlapping area among consecutive detectors). If so, the detection with the poorest S/N is set to a value \textit{DupliDet\_Flag} = 1. Therefore, selecting detections with \textit{DupliDet\_Flag} = 0 removes all duplicated detections when combining  catalogues.

\vspace{0.2cm}

 The best photometric redshift estimate for every source is $z_b$. Additionally, $z_{b}^{min}$ and $z_b^{max}$ represent the lower and upper limits for the first peak within a $1\sigma$ interval. Spectral-type classification is given by $t_b$ where its number refers to the selected template as indicated in Fig. \ref{sedeb11}. $Odds$ gives the amount of redshift probability enclosed around the main peak (see Section \ref{bpz}) and $\chi^{2}$ the reduced chi-squared from the comparison between observed and predicted fluxes according to the selected template and redshift. An estimation of each detection stellar mass content (in units of log10($M\odot$)) is given by $Stell\_Mass$. $M\_ABS$ corresponds to the absolute (AB) magnitude for the Johnson B-band. $MagPrior$ corresponds to the F814W magnitude used to derived the BPZ Prior. Finally $z_{ml}$ and $t_{ml}$ represent respectively the maximum likelihood photometric redshift and spectral-type.

\section{The ALHAMBRA ``gold'' catalogue}
\label{alhambragold}

 The ALHAMBRA gold catalogue corresponds to a subsample of $\sim$100k galaxies ($17<m_{F814W}<23$) with very accurate and reliable photometric redshifts, an expected error $\sigma_{z}$$<$0.012 and redshift probability distribution functions $P(z)$ well-defined by a single peak. The catalogue also includes PSF-corrected photometry for $\sim$20.000 stars in the galactic halo (identified according to the methodology described in Section \ref{SG}) along with $\sim$1000 AGN candidates found with the method discussed in Section \ref{emlingal}. 

\vspace{0.2cm}

The ALHAMBRA gold catalog can be downloaded from the website: http://cosmo.iaa.es/content/alhambra-gold-catalog

\section{SUMMARY}

 The ALHAMBRA (Advance Large Homogeneous Area Medium Band Redshift Astronomical) survey has observed 8 different regions of the sky, including sections of the COSMOS, DEEP2, ELAIS, GOODS-N, SDSS and Groth fields using a new photometric system with 20 contiguous $\sim$300$\AA$ filters covering the optical range, combining them with deep $JHKs$ imaging. The observations, carried out with the Calar Alto $3.5$m telescope using the wide field (0.25 deg$^{2}$ FOV) optical camera LAICA and the NIR instrument Omega-2000, correspond to $\sim$700hrs of on-target science images. The photometric system was specifically designed to maximize the effective depth of the survey in terms of accurate spectral-type and photometric redshift estimation along with the capability of identification of relatively faint emission lines. 

\vspace{0.2cm}

Synthetic HST/ACS F814W detection images were generated to be able to define a constant and homogeneous window for all the ALHAMBRA fields. These images, photometrically complete down to a magnitude $m_{F814W}\le25.5$ AB, served not only to improve the quality of the photometric detections but also to carry out systematic comparisons with the COSMOS-survey. In order to improve the source detection efficiency, we masked every saturated star, stellar spike, ghost and damaged area. To minimize the variation of the image RMS, all flagged pixels were replaced with background noise.

\vspace{0.2cm}

To deal with the observed PSF-variability across filters, \textit{ColorPro} (Coe et al. 2005) was used to perform accurate aperture-matched PSF-corrected photometry retrieving robust photometric colors ideal for photometric redshift estimations. For this purpose PSF models were generated for individual images by manually selecting several hundred of non saturated and well isolated stars across the field. Using a compilation of $\sim$20.000 stars, the mean radial PSF variation was smaller than 5\%, enabling the usage of a single PSF model per image.  

\vspace{0.2cm} 
 
We carried out several simulations to test the accuracy of ColorPro retrieving precise photometry, across images with varied PSF. We degraded ACS/HST images (from COSMOS) to the typical ALHAMBRA conditions (in terms of PSF and background noise) and run ColorPro on it expecting retrieve null colors (equal magnitudes). We found that simulated colors showed a dispersion of $\sigma$$\sim$0.03 which marks a photometric precision floor, for sources brighter than magnitude $m_{F814W}=23.0$ and, as expected from the uncertainties arising from the photometric noise, an increasing error for fainter magnitudes. For most of the magnitude range, there are negligible biases.

We also studied the expected photometric completeness for the ALHAMBRA fields in terms of the PSF and background level, using the previous simulations. So, we derived the statistical probability of detecting a sample of faint galaxies when observed under the typical ALHAMBRA conditions. The result indicates that ALHAMBRA is photometrically complete down to a magnitude of $m_{F814W}\sim24$. At fainter magnitudes, the number of detections decreases rapidly, with $\sim$40\% of the galaxies lost at $m_{F814W}\sim25$. 
 
\vspace{0.2cm}
 
To decontaminate extragalactic sources from field stars, we followed a statistical approach where every detection was classified in terms of the probability of being a star or a galaxy, given its apparent geometry, F814W magnitude, optical $F489W-F814W$ and NIR $J-Ks$ colors. We tested the goodness of our statistical classification by comparing the density of finding stars against that predicted by the Trilegal software (Girardi 2002, 2005). We found a very good agreement between both samples. When this statistical criteria is applied to the complete catalogue, we observed that stars dominate the sample down to a magnitude $m_{F814W}<19$. For fainter magnitudes, the fraction of stars rapidly declines with a contribution of $\sim$1\% for magnitudes $m_{F814W}=22.5$. We retrieve an averaged stellar density in the galactic halo of $\sim$7000 stars per deg$^{2}$ ($\sim$450 stars per CCD) for sources brighter than $m_{F814W}=22.5$. 

\vspace{0.2cm} 
 
Given the correlation among pixels introduced during image processing, we empirically re-calculated photometric uncertainties for every detection following a similar approach as that described in Labb\'e et al. (2003), Gawiser et al. (2006) and Quadri et al. (2007). Spanning a range of radius between 1-250 pixels, we thrown $\sim$50.000 apertures on blank areas across the images to measure both the enclosed background signal and its scatter. The methodology served to properly estimate the empirical dependence between any galaxy photometric aperture and its photometric uncertainty.

\vspace{0.2cm}

We calculated photometric redshifts with the BPZ2.0 code (Ben\'itez 2013, in prep). This new version includes a new prior empirically derived for each spectral-type and magnitude by fitting luminosity functions provided by GOODS- MUSIC (Santini et al. 2009), COSMOS (Scoville et al. 2007) and UDF (Coe et al.2006), a new empirically calibrated library of galaxy templates and an estimation of the galaxy stellar mass based on the color-M/L ratio relationship established by Taylor et al. (2011).

\vspace{0.2cm}

Given the overlap between the ALHAMBRA fields and other existing spectroscopic surveys, we compiled a sample of $\sim$7200 galaxies with high quality (secure) spectroscopic redshifts from the publicly available data, mostly covering the ALHAMBRA parameter space, i.e., with a redshift range 0$<$$zs$$<$1.5 ($<$$zs$$>$$\sim$0.77) and a magnitude range (based on ALHAMBRA photometry) $18<m_{F814W}<25$ ($<m_{F814W}>\sim22.3$). Based on this spectroscopic sample, our photometric redshifts have a precision of $\delta_{z}$/(1+$z_{s}$)=1$\%$ for I$<$22.5 and $\delta_{z}$/(1+$z_{s}$)=1.4$\%$ for 22.5$<$I$<$24.5. Precisions of $\delta_{z}$/(1+$z_{s}$)$<$ 0.5 $\%$ are reached for the brighter spectroscopic sample, showing the potential of medium-band photometric surveys. 

\vspace{0.2cm}

We refined photometric zeropoints derived using standard stellar-based calibration techniques, by comparing the observed colors of galaxies (for which spectroscopic redshifts were available) with those expected by the BPZ library of templates. We found that the so-derived corrections improved not only the photometric redshifts accuracy but also reduced the fraction of catastrophic outliers. Considering the possibility of a systematic effect during the data reduction, we represented globally the zeropoint corrections for the all the individual images as a function of the AIRMASS, the Stellar Symmetry, the FWHM Scatter, differences between PSFs-models and stars, magnitude ranges, redshift ranges or aperture sizes. No clear correlations were observed with typical fluctuations smaller than 3\%. We therefore conclude that the zeropoint offsets do not depend on the photometric treatment and represent real differences between the calibration obtained from traditional color transformations based on stars and the average colors of galaxies as defined by the BPZ template set, calibrated with HST observations.

\vspace{0.2cm}

For those fields without spectroscopic coverage, a new methodology (described in this work) was applied to calibrate photometric zeropoint estimations using photometric redshifts. Essentially, we realized that the photometric redshifts obtained for emission line galaxies were quite robust to changes in the zeropoint calibration and therefore could be treated as spectroscopic for calibration purposes, obtaining an automatic and self-contained zeropoint correction for all our fields. This methodology not only successfully improved the photometric redshifts accuracy (almost up to the level provided by the spectroscopic sample), but also dramatically reduced the fraction of catastrophic outliers, avoiding serious problem of inhomogeneity among fields.

\vspace{0.2cm}

Considering the probabilistic nature of the photometric redshift estimations, we worked with the full probability distribution P (z, T$|$C) (Ben\iõtez 2000, Coe et al. 2006, Mandelbaum et al. 2008; Cunha et al. 2009; Wittman 2009; Bordoloi et al. 2010; Abrahamse et al. 2011; Sheldon et al. 2012). This approach represents a more convenient estimator for most faint galaxies as the p(z) usually becomes multimodal and completely asymmetric, so not well represented by a single and symmetric distribution. Using the complete probability distribution functions we found that the global photometric redshift distribution shows a mean redshift $<$z$>$=0.56 for I$<$22.5 AB and $<$z$>$=0.86 for I$<$24.5 AB. In particular, comparison with our average $n(z)$ shows that the COSMOS field has a rather peculiar redshift distribution, with a large spike a z$\sim$0.7 and an underdensity for z$<$0.5 which mimics a significant redshift density evolution effect. 

\vspace{0.2cm}

Despite having a relatively small FOV compared with other surveys, one of the main virtues of ALHAMBRA is that it includes 8 different lines of sight widely separated, providing a realistic estimation of both the typical redshift distribution of galaxies across cosmic time and its inherent variability. We explored the cosmic variance effect in the redshift distribution of galaxies as a function of the absolute B magnitude and the spectral-type. We found that the well-known bimodal distribution in the COSMOS field was not systematically observed along the other fields, emphasizing the usefulness of the ALHAMBRA data on galaxy evolution studies.

\vspace{0.2cm}

We discovered a new methodology to identify potential AGN candidates using BPZ. When plotting the Odds distribution as a function of F814W magnitude for all galaxies, we found a concentration of objects with low odds at magnitudes in between $18<m_{F814W}<23$. When plotting the logarithmic $\chi^{2}$ distribution for these objects, it immediately revealed those detections to have the highest $\chi^{2}$ values indicative of a poor fit. After purging the sample from objects with high SExtractor flags, we saw that the remaining objects could be classified into two different groups: 1. unresolved stellar pairs with clearly asymmetric morphologies (spite of its photometric colors) and 2. very strong broad emission-line objects, AGNs or variable sources. Precisely the templates not included in the BPZ library.

\vspace{0.2cm}

The PSF-corrected multicolor photometry and photometric redshifts for $\sim$438,000 galaxies presented in this work, covers an effective area of $2.79$ deg$^{2}$, split into 14 strips of 58.5'x15.5' and represents a $\sim$32 hrs of on-target exposure time. Given its depth, multiband coverage and a much smaller cosmic variance than other similar projects, ALHAMBRA is a unique dataset for galaxy evolution studies. Several of the techniques presented here will have a wide applicability to future large scale narrow-band photometric redshift surveys like JPAS (Javalambre Physics of the Accelerated Universe, Ben\'itez 2009a, Ben\'itez 2013, in prep.).

\section*{Acknowledgments}

We acknowledge support from the Spanish Ministerio de Educaci\'on y Ciencia through grant AYA2006-14056  BES-2007-16280. We acknowledge the financial support from the Spanish MICINN under the Consolider-Ingenio 2010 Program grant CSD2006-00070: First Science with the GTC. Part of this work was supported by Junta de Andaluc\'ia, through grant TIC-114 and the Excellence Project P08-TIC-3531, and by the Spanish Ministry for Science and Innovation through grants AYA2006-1456, AYA2010-15169, AYA2010-22111-C03-02, AYA2010-22111-C03-01 and Generalitat Valenciana project Prometeo 2009/064. We acknowledge the excellent work carried out by the CAHA observatory. This project was based on observations collected at the Centro Astron\'omico Hispano Alem\'an (CAHA) at Calar Alto, operated jointly by the Max-Planck Institut f\"ur Astronomie and the Instituto de Astrof\'isica de Andaluc\'ia (IAA-CSIC).

\appendix
\section{The ALHAMBRA fields nomenclature.}
\label{LAICAscheme}
In this appendix we illustrate an example of the Pointing layout for the ALHAMBRA fields. The combination of two contiguous pointing yields a final layout composed by two strips of 58.5'x15.5' (comprising four individual CCDs) with a separation of $\sim$13.0'.

\begin{figure*}
\begin{center}
\includegraphics[width=17.cm]{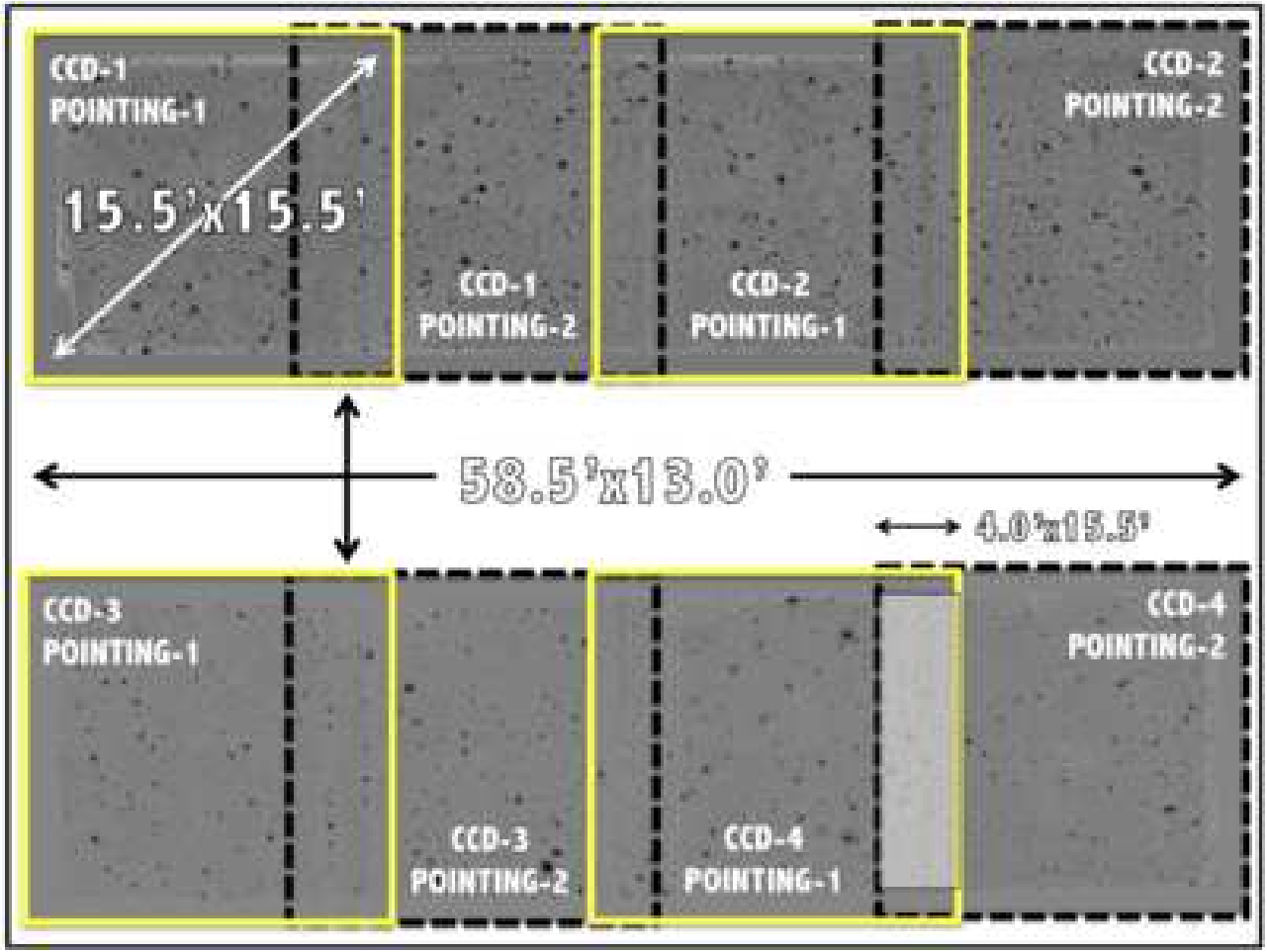}
\caption{Pointing layout for the ALHAMBRA fields. Given the geometrical configuration of the optical imager LAICA, each pointing is composed by four CCDs (as marked with the yellow squares) with an internal gap of $\sim$13.0'. The combination of two contiguous pointing yields a final layout composed by two strips of 58.5'x15.5' (comprising four individual CCDs) with a separation of $\sim$13.0'. Contiguous CCDs within each strip show a maximum overlap of 4.0'x15.5'.}
\label{LAICA}
\end{center}
\end{figure*}

\section{Photometric redshift surveys comparison}
\label{surveycomparison}
The figure represents the photometric redshift accuracy versus the covered area for several surveys (see Table \ref{surveyaccuracy}). The number of photometric passbands is color-coded as described in the inset panel. The marker-size represents logarithmically the number of detections.

\begin{figure*}
\begin{center}
\includegraphics[width=17.cm]{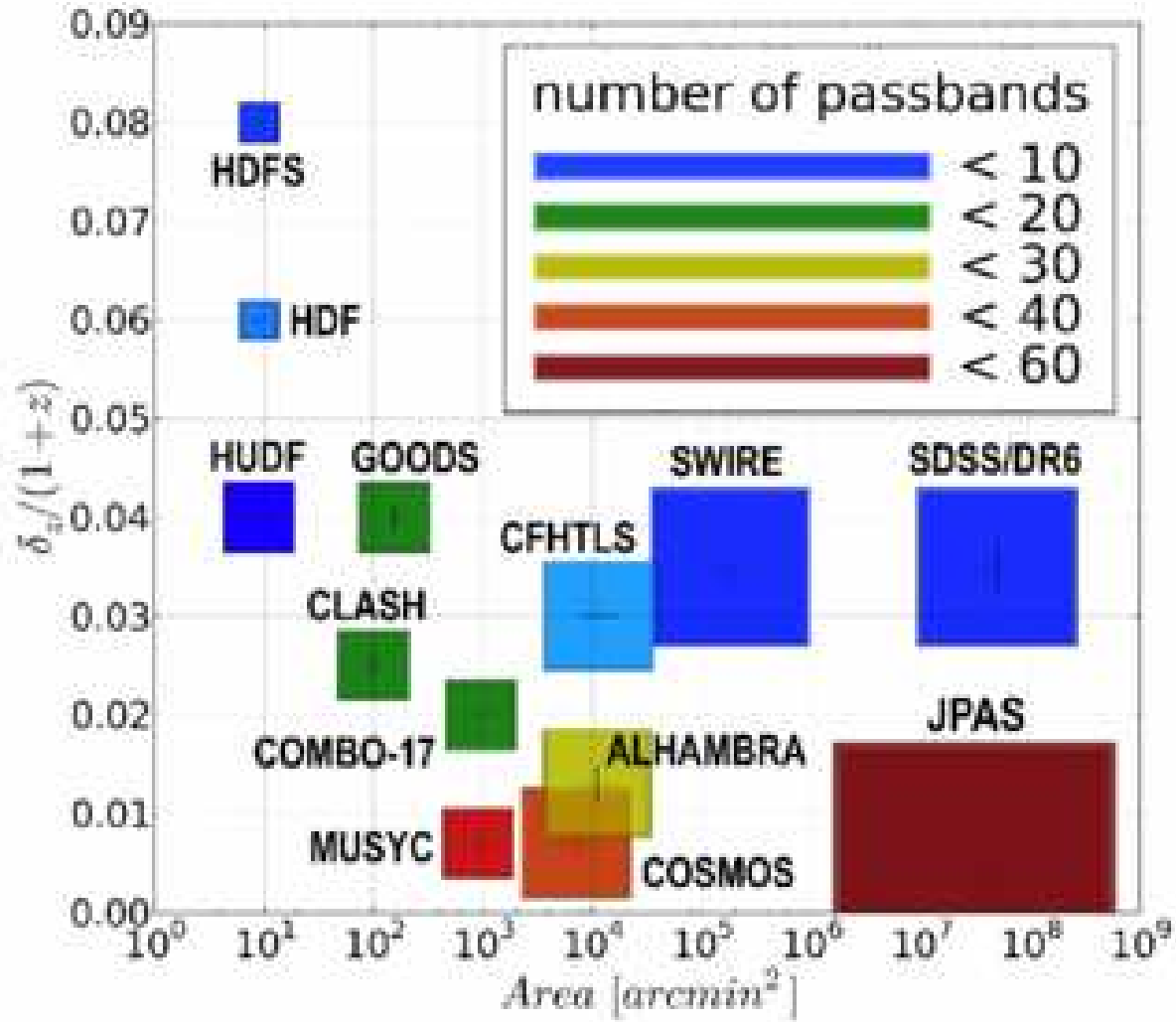}
\caption{Photometric redshift surveys comparison. The figure represents the photometric redshift accuracy versus the covered area for several surveys (see Table \ref{surveyaccuracy}). The number of photometric passbands is color-coded as described in the inset panel. While the marker-size represents logarithmically the number of detections, the position of the internal plus sign indicates to its photometric redshift accuracy.}
\end{center}
\end{figure*}

\section{Photometric Redshift Catalogues description.}
\label{descriptcatalogues}
In this appendix we include the description of the photometric redshift catalogues content in more detail.

\begin{table*}
\caption{The ALHAMBRA Photometric Redshift Catalogs Content. Part I}
\begin{center}
\label{catalogdesc1}
\begin{tabular}{|l|c|c|c|c|c|c|c|}
\hline
\hline
& COLUMN  &   PARAMETER  &  DESCRIPTION \\
\hline
& 1        		  &      ID\_ColorPro	                  &  Object ID Number	 \\
& 2                      &      Field	                           &  ALHAMBRA field   \\
& 3                      &      Pointing                             &  Pointing within the field   \\
& 4                      &      CCD	                           &   Detector within the pointing   \\
& 5                      &      RA	                                    &  Right Ascension in decimal degrees [J2000]   \\
& 6        		  &      DEC	                           &  Declination in decimal degrees [J2000]   \\
& 7                      &      XX 	                                    &  x pixel coordinate	 \\
& 8                      &      YY 	                                    &  y pixel coordinate	 \\
& 9                      &      AREA	                           &  Isophotal aperture area (pixels)	 \\
& 10                    &      FWHM	                           &  Full width at half maximum (arcsec)	 \\
& 11                    &      STELL                             &    SExtractor 'stellarity' (1 = star; 0 = galaxy) \\
& 12                    &      ELL                                  &   Ellipticity = 1 - B/A \\
& 13                    &      a                                      &   Profile RMS along major axis (pixels) \\
& 14                    &      b                                       &  Profile RMS along minor axis (pixels) \\ 
& 15			  &      THETA                             &   Position Angle (CCW/x)  \\
& 16                    &      RK                                   &   Kron apertures in units of A or B (pixels) \\ 
& 17                    &      RF                                   &    Fraction-of-light radii (pixels)  \\
& 18                    &      S/N                  		 &   Signal to Noise (SExt\_FLUX\_AUTO/SExt\_FLUXERR\_AUTO) \\
& 19                    &      PhotoFlag                        &   SExtractor Photometric Flag \\
& 20,21              &       F365W, dF365W 	          &  F365W total magnitude \& uncertainty \\
& 22,23              &       F396W, dF396W               &  F396W total magnitude	\& uncertainty \\
& 24,25              &       F427W, dF427W 	           &  F427W total magnitude \& uncertainty	 \\
& 26,27              &       F458W, dF458W               &  F458W total magnitude \& uncertainty	 \\
& 28,29              &       F489W, dF489W 		   &  F489W total magnitude \& uncertainty	 \\
& 30,31              &       F520W, dF520W 		   &  F520W total magnitude \& uncertainty	 \\
& 32,33              &       F551W, dF551W 		   &  F551W total magnitude \& uncertainty	 \\
& 34,35              &       F582W, dF582W 		   &  F582W total magnitude \& uncertainty	 \\
& 36,37              &       F613W, dF613W 		   &  F613W total magnitude \& uncertainty	 \\
& 38,39              &       F644W, dF644W 		   &  F644W total magnitude \& uncertainty	 \\
& 40,41              &       F675W, dF675W 		   &  F675W total magnitude \& uncertainty	 \\
& 42,43              &       F706W, dF706W 		   &  F706W total magnitude \& uncertainty	 \\
& 44,45              &       F737W, dF737W 		   &  F737W total magnitude \& uncertainty	 \\
& 46,47              &       F768W, dF768W 		   &  F768W total magnitude \& uncertainty	 \\
& 48,49              &       F799W, dF799W 		   &  F799W total magnitude \& uncertainty	 \\
& 50,51              &       F830W, dF830W 		   &  F830W total magnitude \& uncertainty	 \\
& 52,53              &       F861W, dF861W 		   &  F861W total magnitude \& uncertainty	 \\
& 54,55              &       F892W, dF892W 		   &  F892W total magnitude \& uncertainty	 \\
& 56,57              &       F923W, dF923W 		   &  F923W total magnitude \& uncertainty	 \\
& 58,59              &       F954W, dF954W 		   &  F954W total magnitude \& uncertainty	 \\
& 60,61              &       J, dJ 	                            &  NIR-J total magnitude \& uncertainty	 \\
& 62,63              &       H, dH 	                            &  NIR-H total magnitude \& uncertainty	 \\
& 64,65              &       KS, dKS 	                            &  NIR-KS total magnitude \& uncertainty	 \\
& 66,67              &       F814W, dF814W               &  F814W total magnitude \& uncertainty \\
& 68                   &      F814W\_3arcs                   &  3arcsec Circular Aperture magnitude [AB]     \\
& 69                   &      dF814W\_3arcs                 &  3arcsec Circular Aperture magnitude uncertainty [AB] \\
& 70                   &       F814W\_3arcs\_corr         &  Corrected 3arcsec Circular Aperture Magnitude [AB]  \\
& 71                   &       nfd                                     &     Number Filters Detected (out of 24)  \\
& 72                   &       xray                                   &     X-Ray Source [0:NO,1:YES] (2XMM;Watson et al. 2009) \\
& 73                   &       PercW                               &     Percentual Photometric Weight (on detection image). \\
& 74                   &     Satur\_Flag                         & Photometric Saturation-Flag [0:Good Detection, 1:Saturated Detection] \\
& 75                   &     Stellar\_Flag                       & Statistical STAR/GALAXY Discriminator [0:Galaxy,0.5:Unknown,1:Star] \\
& 76                   &     DupliDet\_Flag                   & Duplicated Detection Flag [0:Non duplicated, 1:Duplicated] \\
\hline
\hline
\end{tabular}
\end{center}
\end{table*}

\begin{table*}
\caption{Photometric Redshift Catalogs Content. Part II}
\begin{center}
\label{catalogdesc2}
\begin{tabular}{|l|c|c|c|c|c|c|c|}
\hline
\hline
&  COLUMN  &   PARAMETER  &  DESCRIPTION \\
\hline
&  77                         &     zb                     & BPZ most likely redshift  \\
&  78                         &     zb\_min             &  Lower limit (95p confidence) \\
&  79                         &     zb\_max             & Upper limit (95p confidence) \\
&  80                         &     tb                      &  BPZ most likely spectral type \\
&  81                         &     Odds                & P(z) contained within zb +/- 2*0.01*(1+z) \\
&  82                         &     z\_ml                    &  Maximum Likelihood most likely redshift \\
&  83                         &     t\_ml                    &  Maximum Likelihood most likely spectral type \\
&  84                         &     Chi2                   &  Poorness of BPZ fit: observed vs. model fluxes \\ 
&  85                         &     Stell\_Mass     &     Stellar Mass (log10($M_{\odot}$))  \\
&  86                         &     M\_ABS           &    Absolute Magnitude [AB] (B\_JOHNSON) \\
&  87                         &     MagPrior            &  Magnitude Used for the Prior (F814W) \\
&  88                         &     irms\_F365W     &     Percentual Weight on F365W 1/RMS image (within ISOphotal Area). \\  
&  89                         &     irms\_F396W    &     Percentual Weight on F396W 1/RMS image (within ISOphotal Area). \\
&  90                         &     irms\_F427W     &     Percentual Weight on F427W 1/RMS image (within ISOphotal Area). \\
&  91                         &     irms\_F458W     &     Percentual Weight on F458W 1/RMS image (within ISOphotal Area). \\
&  92                         &    irms\_F489W     &     Percentual Weight on F489W 1/RMS image (within ISOphotal Area). \\
&  93                          &    irms\_F520W     &     Percentual Weight on F520W 1/RMS image (within ISOphotal Area). \\
&  94                          &    irms\_F551W     &     Percentual Weight on F551W 1/RMS image (within ISOphotal Area). \\
&  95                          &    irms\_F582W     &     Percentual Weight on F582W 1/RMS image (within ISOphotal Area). \\
&  96                          &    irms\_F613W     &     Percentual Weight on F613W 1/RMS image (within ISOphotal Area). \\ 
&  97                          &    irms\_F644W     &     Percentual Weight on F644W 1/RMS image (within ISOphotal Area). \\
&  98                          &    irms\_F675W     &     Percentual Weight on F675W 1/RMS image (within ISOphotal Area). \\
&  99                          &    irms\_F706W     &     Percentual Weight on F706W 1/RMS image (within ISOphotal Area). \\ 
&  100                        &     irms\_F737W     &     Percentual Weight on F737W 1/RMS image (within ISOphotal Area). \\
&  101                        &    irms\_F768W     &     Percentual Weight on F768W 1/RMS image (within ISOphotal Area). \\
&  102                        &    irms\_F799W     &     Percentual Weight on F799W 1/RMS image (within ISOphotal Area). \\Ê
&  103                        &    irms\_F830W     &     Percentual Weight on F830W 1/RMS image (within ISOphotal Area). \\ 
&  104                        &    irms\_F861W     &     Percentual Weight on F861W 1/RMS image (within ISOphotal Area). \\
&  105                        &    irms\_F892W     &     Percentual Weight on F892W 1/RMS image (within ISOphotal Area). \\
&  106                       &     irms\_F923W     &     Percentual Weight on F923W 1/RMS image (within ISOphotal Area). \\ 
&  107                       &     irms\_F954W     &     Percentual Weight on F954W 1/RMS image (within ISOphotal Area). \\
&  108                       &     irms\_J               &  Percentual Weight on J  1/RMS image (within ISOphotal Area).   \\
&  109                       &     irms\_H              &  Percentual Weight on H 1/RMS image (within ISOphotal Area).   \\
&  110                        &    irms\_KS            & Percentual Weight on KS 1/RMS image (within ISOphotal Area).   \\
&  111                        &    irms\_F814W     &      Percentual Weight on F814W 1/RMS image (within ISOphotal Area). \\
&  112                       &     irms\_OPT\_Flag    &     Optical-Quality-Flag. Number of Optical Filters with PercW $<$ 0.8  \\
&  113                       &     irms\_NIR\_Flag      &   NIR-Quality-Flag. Number of NIR Filters with PercW $<$ 0.8  \\
\hline
\hline
\end{tabular}
\end{center}
\end{table*}

\section{SExtractor configuration and the effective area.}
\label{SExconfApp}
In this appendix we present an example of the SExtractor configuration used to derive the F814W detections. Along with this, we also includes several tables containing statistical information concerning the observations.

\begin{table*}
\caption{Example of the typical SExtractor configuration used to derive the ALHAMBRA photometric catalogs. Asterisked parameters may vary among CCDs.}
\begin{center}
\label{detconftable}
\begin{tabular}{|l|c|c|c|c|c|c|c|}
\hline
\hline
PARAMETER            &  SETTING  &  COMMENT \\
\hline
ANALYSIS\_THRESH 	& 1.3*                            	& Limit for isophotal analysis $\sigma$\\
BACK\_SIZE       		& 256                            	& Background mesh in pixels \\
BACK\_FILTERSIZE 	& 5                              	& Background filter\\
BACKPHOTO\_THICK 	& 102                            	& Thickness of the background LOCAL annulus\\
BACKPHOTO\_TYPE  	& LOCAL                          	& Photometry background subtraction type\\
CATALOG\_NAME    	& STDOUT                       	& Output to pipe instead of file \\
CATALOG\_TYPE    		& ASCII                          	& Output type\\
CLEAN           			& Y                              	& Clean spurious detections\\
CLEAN\_PARAM     		& 1                              	& Cleaning efficiency\\
CHECKIMAGE\_TYPE	& SEGMENTATION          & Output Image Type \\
DETECT\_MINAREA  	& 8*                              	& Minimum number of pixels above threshold\\
DEBLEND\_MINCONT 	&0.0002                            	& Minimum contrast parameter for deblending\\
DEBLEND\_NTHRESH 	& 64                             	& Number of deblending sub-thresholds\\
DETECT\_THRESH   	& 1.35*                            	& Detection Threshold in $\sigma$\\
DETECT\_TYPE     		& CCD                            	& Detector type\\
FILTER          			& Y                              	& Use filtering \\
FILTER\_NAME		& tophat\_3.0\_3x3.conv	& Filter for detection image\\
GAIN            			& 57.68*				& Gain is 1 for absolute RMS map\\
MAG\_GAMMA                  & 4.0                                   &  Gamma of emulsion \\
MAG\_ZEROPOINT   	& 0.*                             	& Magnitude zero-point\\
MEMORY\_PIXSTACK 	& 2600000                        	& Number of pixels in stack\\
MEMORY\_BUFSIZE  	& 4600                           	& Number of lines in buffer\\
MEMORY\_OBJSTACK 	& 15000                          	& Size of the buffer containing objects\\
MASK\_TYPE       		& CORRECT                     & Correct flux for blended objects\\
PARAMETERS\_NAME 	& ColorPro.param		& Fields to be included in output catalog\\
PHOT\_APERTURES  	& 14.0                                & MAG\_APER aperture diameter(s) in pixels\\
PHOT\_AUTOPARAMS 	& 2.5,3.5                      	& MAG\_AUTO parameters: $<$Kron\_fact$>$,$<$min\_radius$>$\\
PIXEL\_SCALE     		& 0.221                         	& Size of pixel in arcseconds\\
SATUR\_LEVEL     		& 50000                         	& Level of saturation\\
SEEING\_FWHM     		& 0.86*                           	& Stellar FWHM in arcseconds\\
STARNNW\_NAME    	& default.nnw                	& Neural-Network\_Weight table filename \\
WEIGHT\_TYPE 		& MAP\_WEIGHT                     & Set Weight image type\\
\hline
\hline
\end{tabular}
\end{center}
\end{table*}

\begin{table*}
\caption{{\small Effective Surveyed Area. The definition of the effective area is provided in section \ref{Flagimages}.}}
\begin{center}
\label{effectivearea}
\begin{tabular}{|l|c|c|c|}
\hline
\hline
Field   & Eff. Area  & Field   &  Eff. Area \\
Name & [deg$^{2}$]        & Name  & [deg$^{2}$]       \\
\hline
ALHAMBRA\_F02P01C01 & 0.0580 & ALHAMBRA\_F06P01C01 & 0.0593 \\
ALHAMBRA\_F02P01C02 & 0.0584 & ALHAMBRA\_F06P01C02 & 0.0583 \\
ALHAMBRA\_F02P01C03 & 0.0540 & ALHAMBRA\_F06P01C03 & 0.0585 \\
ALHAMBRA\_F02P01C04 & 0.0582 & ALHAMBRA\_F06P01C04 & 0.0582 \\
ALHAMBRA\_F02P02C01 & 0.0596 & ALHAMBRA\_F06P02C01 & 0.0587 \\
ALHAMBRA\_F02P02C02 & 0.0506 & ALHAMBRA\_F06P02C02 & 0.0587 \\
ALHAMBRA\_F02P02C03 & 0.0538 & ALHAMBRA\_F06P02C03 & 0.0572 \\
ALHAMBRA\_F02P02C04 & 0.0586 & ALHAMBRA\_F06P02C04 & 0.0589 \\
ALHAMBRA\_F03P01C01 & 0.0586 & ALHAMBRA\_F07P03C01 & 0.0587 \\
ALHAMBRA\_F03P01C02 & 0.0589 & ALHAMBRA\_F07P03C02 & 0.0590 \\
ALHAMBRA\_F03P01C03 & 0.0578 & ALHAMBRA\_F07P03C03 & 0.0576 \\
ALHAMBRA\_F03P01C04 & 0.0592 & ALHAMBRA\_F07P03C04 & 0.0587 \\
ALHAMBRA\_F03P02C01 & 0.0592 & ALHAMBRA\_F07P04C01 & 0.0589 \\
ALHAMBRA\_F03P02C02 & 0.0577 & ALHAMBRA\_F07P04C02 & 0.0566 \\
ALHAMBRA\_F03P02C03 & 0.0569 & ALHAMBRA\_F07P04C03 & 0.0580 \\
ALHAMBRA\_F03P02C04 & 0.0590 & ALHAMBRA\_F07P04C04 & 0.0590 \\
ALHAMBRA\_F04P01C01 & 0.0589 & ALHAMBRA\_F08P01C01 & 0.0588 \\
ALHAMBRA\_F04P01C02 & 0.0590 & ALHAMBRA\_F08P01C02 & 0.0590 \\
ALHAMBRA\_F04P01C03 & 0.0569 & ALHAMBRA\_F08P01C03 & 0.0577 \\
ALHAMBRA\_F04P01C04 & 0.0589 & ALHAMBRA\_F08P01C04 & 0.0587 \\
ALHAMBRA\_F05P01C01 & 0.0595 & ALHAMBRA\_F08P02C01 & 0.0585 \\
ALHAMBRA\_F05P01C02 & 0.0594 & ALHAMBRA\_F08P02C02 & 0.0583 \\
ALHAMBRA\_F05P01C03 & 0.0588 & ALHAMBRA\_F08P02C03 & 0.0558 \\
ALHAMBRA\_F05P01C04 & 0.0594 & ALHAMBRA\_F08P02C04 & 0.0576 \\
\hline
\hline
\end{tabular}
\end{center}
\end{table*}

\label{lastpage}
\end{document}